\documentclass[manuscript,screen]{acmart}

\usepackage{booktabs}
\usepackage{multirow}
\usepackage{enumitem}
\usepackage{subfigure}
\usepackage[ruled]{algorithm2e}
\usepackage[flushleft]{threeparttable}
\usepackage{bbm}
\usepackage{mathrsfs}
\usepackage{longtable}
\usepackage{lscape} 
\usepackage{makecell}

\AtBeginDocument{%
  \providecommand\BibTeX{{%
    \normalfont B\kern-0.5em{\scshape i\kern-0.25em b}\kern-0.8em\TeX}}}

\usepackage{xcolor}
\usepackage{soul}
\usepackage{colortbl}

\begin{document}

\title{All Roads Lead to Rome: Unveiling the Trajectory of Recommender Systems Across the LLM Era}

\author{Bo Chen}
\affiliation{%
	\institution{Noah's Ark Lab, Huawei}
	\country{China}}
\email{chenbo116@huawei.com}

\author{Xinyi Dai}
\affiliation{%
\institution{Noah's Ark Lab, Huawei}
\country{China}}
\email{daixinyi3@huawei.com}

\author{Huifeng Guo}
\affiliation{%
\institution{Noah's Ark Lab, Huawei}
\country{China}}
\email{huifeng.guo@huawei.com}

\author{Wei Guo}
\affiliation{%
\institution{Noah's Ark Lab, Huawei}
\country{Singapore}}
\email{guowei67@huawei.com}

\author{Weiwen Liu}
\affiliation{%
	\institution{Noah's Ark Lab, Huawei}
	\country{China}}
\email{liuweiwen8@huawei.com}

\author{Yong Liu}
\affiliation{%
	\institution{Noah's Ark Lab, Huawei}
	\country{Singapore}}
\email{liu.yong6@huawei.com}

\author{Jiarui Qin}
\affiliation{%
	\institution{Noah's Ark Lab, Huawei}
	\country{China}}
\email{qinjiarui3@huawei.com}

\author{Ruiming Tang$^\dagger$}
\affiliation{%
\institution{Noah's Ark Lab, Huawei}
\country{China}}
\email{tangruiming@huawei.com}

\author{Yichao Wang}
\affiliation{%
	\institution{Noah's Ark Lab, Huawei}
	\country{China}}
\email{wangyichao5@huawei.com}

\author{Chuhan Wu}
\affiliation{%
	\institution{Noah's Ark Lab, Huawei}
	\country{China}}
\email{wuchuhan1@huawei.com}

\author{Yaxiong Wu}
\affiliation{%
\institution{Noah's Ark Lab, Huawei}
\country{Singapore}}
\email{yx.wu@huawei.com}

\author{Hao Zhang}
\affiliation{%
	\institution{Noah's Ark Lab, Huawei}
	\country{Singapore}}
\email{zhang.hao3@huawei.com}

\thanks{$^*$Authors are listed in alphabetical order.}
\thanks{$^\dagger$Ruiming Tang is the corresponding author.}


\renewcommand{\shortauthors}{B. Chen et al.}

\begin{abstract}

\end{abstract}

\begin{abstract}
Recommender systems (RS) are vital for managing information overload and delivering personalized content, responding to users' diverse information needs. 
The emergence of large language models (LLMs) offers a new horizon for redefining recommender systems with vast general knowledge and reasoning capabilities. 
Standing across this LLM era, we aim to integrate recommender systems into a broader picture, and pave the way for more comprehensive solutions for future research.
Therefore, we first offer a comprehensive overview of the technical progression of recommender systems, particularly focusing on language foundation models and their applications in recommendation.
We identify two evolution paths of modern recommender systems---via list-wise recommendation and conversational recommendation.
These two paths finally converge at LLM agents with superior capabilities of long-term memory, reflection, and tool intelligence.
Along these two paths, we point out that the information effectiveness of the recommendation is increased, while the user's acquisition cost is decreased.
Technical features, research methodologies, and inherent challenges for each milestone along the path are carefully investigated---from traditional list-wise recommendation to LLM-enhanced recommendation to recommendation with LLM agents.
Finally, we highlight several unresolved challenges crucial for the development of future personalization technologies and interfaces and discuss the future prospects.
\end{abstract}

\maketitle

\section{Introduction}

We are facing an unprecedented era of information explosion, that profoundly affects our perception, cognition, and actions in both virtual and physical worlds~\cite{bawden2020information}. 
Thus, picking helpful and reliable information with acceptable effort is important for humans and the society~\cite{bobadilla2013recommender}.
Over the past decades, researchers and practitioners have made tremendous efforts in building information filtering systems (a.k.a, recommender systems) to accommodate individual's needs by choosing relevant information pieces from a potentially overwhelming number of candidates~\cite{koren2009matrix, guo2017deepfm}.
Accompanied by the rise or decline of the World Wide Web, the development of recommender systems empowers a number of great products and companies, such as YouTube, Facebook, and TikTok~\cite{covington2016deep, shapira2013facebook, wang2022recommendation}.
As an underlying technique, modern recommender systems are de facto shaping and leading our views and decisions through the penetration of these personalized applications~\cite{ko2022survey}.

An ultimate goal of recommender systems is fully understanding users and satisfying their needs by providing accurate and effective information under minimal user effort in the loop of human-system interactions~\cite{ko2022survey}.
A common form of recommendation services is displaying candidate items such as products and news on lists or streaming layouts for information delivery and feedback collection.
To support this paradigm of applications, recommender systems generally need to generate a ranked list of items based on their relevance to user preference inferred from various genres of user feedback, such as clicks, comments, shares, and consumptions.
Different from conventional search engines, recommender systems free users from writing search queries to explicitly explain their intent and can discover their potential interest by analyzing the latent patterns behind their behaviors~\cite{wang2021dcn, guo2017deepfm, covington2016deep}.
Within minor efforts of the clicks made by our fingers, recommender systems can roughly understand our preferences and mitigate our burden in information seeking.

Nonetheless, simple user feedback like clicks is far from reflecting the complexity of human intention~\cite{jannach2021survey}.
In fact, language is a more natural and powerful information carrier to express our requirements, opinions, feelings, and experiences~\cite{sun2018conversational}.
Thus, conversational recommender systems that can receive and understand natural language instructions have the potential to capture user interest signals that cannot be evoked by other user feedback~\cite{gao2021advances}.
During interactive and engaging conversations to guide the recommender systems, they are expected to find items that exactly match user intention and even provide complementary information through natural language explanations~\cite{christakopoulou2016towards}.
However, writing textual messages in multi-round conversations significantly increases the effort of users in seeking desired information, though the final results may be more accurate and effective than the guess of non-conversational recommender systems.
In addition, how to fully understand and generate natural languages during interactions poses huge challenges to recommender systems.

Fortunately, the task of understanding and applying natural languages is not so daunting in recent years, owing to the emergence of large language models (LLMs).
By self-supervised learning on the huge unlabeled corpus, LLMs can obtain rich general knowledge about the physical world and human society through the lens of human languages.
Equipped by such knowledge, LLMs bring new opportunities to various fields in recommendation research, including user modeling, item understanding, result explanation, conversation generation, and even pipeline coordination~\cite{lin2023can}.
These application forms of LLMs can empower recommender systems in different aspects, aiming to optimize the accuracy of information delivery and explore new forms of interactions between users and systems~\cite{gao2023chat}.
Due to the human-like thought mode of LLMs that can quickly understand user motivation through world knowledge and commonsense reasoning, LLM-empowered conversational recommender systems have the potential to assist users without verbose and rigid chats~\cite{friedman2023leveraging}.
In the flooding era of LLMs, how to introduce, embrace, and enjoy LLMs has triggered an emerging revolution of fundamental technical and business paradigms of personalized recommendation~\cite{fan2023recommender}.

Standing at the crossroads of this technical reform, there have been several surveys that retrospect some methodologies and taxonomies of LLM-empowered recommender systems~\cite{liu2023pre,wu2023survey, fan2023recommender,lin2023can}.
However, most of them focus on utilizing LLMs to enhance or replace the modules in conventional recommender systems in list- or stream-based information delivery, while the crucial trend of conversational recommendation led by LLMs is overlooked.
Moreover, there lacks a big picture of the technical position of LLMs in recommender systems and the paradigm shift of recommendation in the dawn of artificial general intelligence.
Thus, a timely overview and outlook of the connections between LLMs and recommender systems can bring useful insights to the design and implementation of next-generation recommendation applications.

In this article, we present a panoramagram of the technical evolution of recommender systems in the context of language foundation models and their applications.
We discover two paths in the paradigm evolution of modern recommender systems in improving the effectiveness of information acquirement, and their trajectories converge at the same point that can maximally elicit and exploit the potential of LLMs (Fig.~\ref{fig:trend}).
We systematically review the milestones on both paths in the aspects of their technical characteristics, research practices, and scientific limitations. 
Furthermore, we analyze the possible forms of LLM applications in recommender systems and raise several open problems in this area that are core to the design of future personalization products and interaction interfaces.
Our survey is expected to stimulate discussions on the future technical and commercial shapes of recommender systems in the coming era of super artificial intelligence.

The rest of this paper is organized as follows. In Section 2, we will give a brief overview of the development path of recommender systems and introduce five different types of recommendation techniques, including list-wise recommendation in the deep learning age, LLM-enhanced list-wise recommendation, conversational recommendation before the LLM era, conversational recommendation in LLM era, and LLM-powered recommendation agents. In Sections 3-7, we will elaborate on each technique in detail. We then propose several open problems and raise potential future directions in Section 8. Finally, we conclude the survey in Section 9.

\section{Overview}
\begin{figure}[t]
    \centering
    \includegraphics[width=\textwidth]{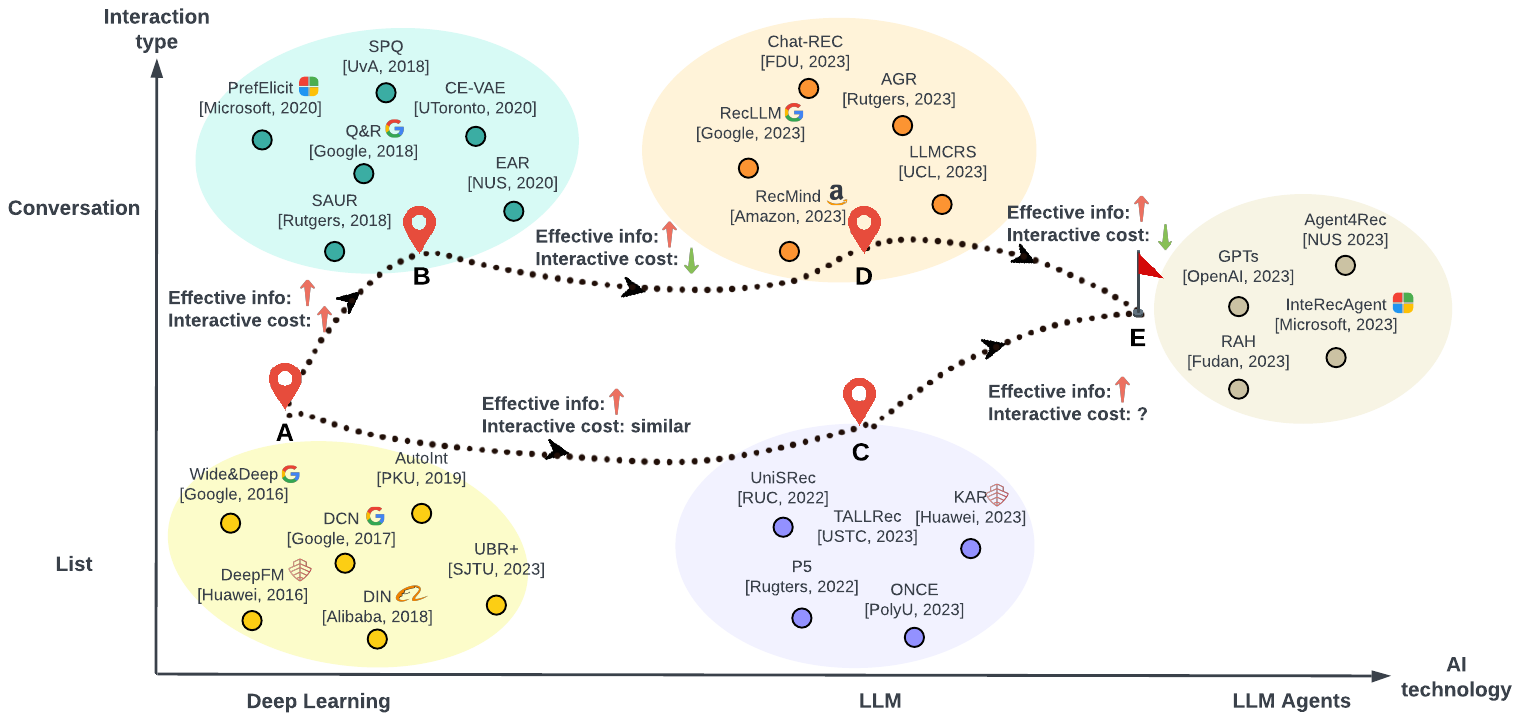}
    \caption{Development trajectory of the modern recommender systems.
    The $x$-axis denotes key technical milestones in artificial intelligence, while the $y$-axis represents various interaction types, leading to five different recommendation paradigms (marked by different colors). 
    Arrows present the general paradigm shift: recommender systems evolve from the lower left corner to the upper right corner, with a decrease in interactive cost and an increase in effective information.
    }
    \label{fig:trend}
\end{figure}

Throughout the history of the Internet, we have experienced distinct feelings when facing information explosion at different ages, i.e., from panic and overwhelming to enjoying and immersing.
Accompanied by the ever-deepening connection between the virtual and physical worlds, acquiring and digesting proper information becomes a fundamental requirement and right of people in social life.
Thanks to the rapid development of information-seeking and filtering technologies such as recommender systems, we can easily touch and affect the information we are concerned about, no longer drowning in the flood of information.
By living in the world through the assistance of personalized recommendations, we are characterizing the web and also characterized by the web unconsciously.

Enlightened by the spark of general intelligence typified by GPT-4 and Sora in this age~\cite{bubeck2023sparks}, new technical and business orders have been brought to the world.
We have unprecedented opportunities to turn massive information into new knowledge and even wisdom by exploiting the universal and human-like abilities of large foundation models.
Different from deterministic recommender systems in the early times, these powerful models not only perceive the world but also create new ideas and content.
Thus, the collision between two technique forms is igniting new sparks that can draw forth novel discipline paradigms and business models in the personalization, media, and information management areas. 
However, both academia and industry are still struggling to figure out the correct direction of this revolution.

\textit{``The farther back we can look, the farther forward we are likely to see.''}
By carefully reviewing the passing history of recommender systems in this article, we analyze the underlying inherent impetus behind their technical development trajectory.
From thirty years ago to now, the general goal of recommender systems, i.e., finding and providing useful and relevant information to users, still remains unchanged.
Given limited time and space, recommender systems are pursuing increasing the amount of effective information that can really attract and help users by optimizing data, algorithms, and interaction interfaces.
From the view of users, we are hoping to be satisfied with minimal effort due to the nature of humans, though we may tolerate the cost if the information is sufficiently valuable.
This naturally forms a dilemma between information effectiveness and acquisition cost: recommender systems need more feedback from users to calibrate and improve their mechanisms, but users often mean to provide feedback and educate the systems if their requirements are not accommodated~\cite{he2016interactive}.
This dilemma actually implies two essential directions to optimizing recommender systems, i.e., \textbf{improving the amount of effective information} by making more accurate and helpful recommendations, and \textbf{decreasing the user interaction cost} by freeing users from frequent and time-consuming operations.

In fact, we find the evolution of recommendation techniques tacitly follows the two philosophies, as shown in Fig.~\ref{fig:trend}.
Transferring from the lower left to the upper right, there are two paths that change the technical forms of recommender systems in different aspects.
The upper path mainly enhances effective information by introducing additional feedback such as natural languages, but sacrifices the interaction efficiency due to the low convenience of text inputs compared to simple clicks.
The lower path focuses on raising effective information by exploiting the potential of large language models to facilitate and automate different modules and stages in recommendation.
Both paths converge at the same point, i.e., converting recommender systems into personalized agents.
Driven by the perception and reasoning abilities provided by large language models, intelligent recommendation agents can fully understand users and actively interact with the world to acquire and aggregate rich knowledge without extensive user instructions and guidance.
The roads to this ideal form are difficult and roundabout due to the temporary sacrifice of user experience (upper path) and the huge paradigm leap between conventional list-oriented recommender systems and conversation-dominated recommendation agents (lower path). 
This long journey is remarked by a series of milestones and breakthroughs, tackling the numerous challenges in approaching and breaking the limitation of personalized recommendation.

Over the past decade, recommender systems have greatly enjoyed the bloom of deep learning and achieved huge business success~\cite {zhang2021deep}. 
The universal approximation property of neural networks makes it possible to train neural models to mine complex interest patterns from user behaviors and model their relevance to candidate items~\cite{cybenko1989approximation}.
Numerous neural recommendation methods have been developed to address diverse recommendation tasks and have found extensive application in industrial products. 
As shown in area A of Fig.~\ref{fig:trend}, notable contributions such as DCN~\cite{wang2021dcn}, DeepFM~\cite{guo2017deepfm}, and DIN~\cite{zhou2018deep} facilitate the success of many online customer-centric enterprises. 
These models usually aim to capture the collaborative signals encoded by user feedback like clicks or ratings, which can be sparse and noisy.
Without further feedback and external knowledge, it is difficult for recommender systems to understand the exact user intents in many scenarios.
Thus, the intrinsic limitations of neural recommenders in this generation stimulate the paradigm shift toward two directions.

An intuitive way to improve the effectiveness of recommender systems is incorporating richer-information feedback such as natural languages (area B).
By parsing and processing user instructions in a multi-turn conversation, recommender systems can better understand the real requirements of users, and make accurate recommendations by filling dialog templates~\cite{zhao2013interactive,christakopoulou2016towards} or generating language responses~\cite{ren2023kecr}.
Moreover, users can actively guide the system to adjust the recommendation results through iterative queries and feedback until they are satisfied~\cite{gao2021advances}.
Conversational recommendation enables users to interact with machines like human-human chatting, thereby providing a novel and engaging experience for its customers.
However, the input of textual messages significantly increases the effort of users to complete a round of information acquisition, which is against the product design principles of many online applications on swift and convenient interaction.

The emergence of large language models (LLMs) opens another direction to improve recommender systems by boosting effective information (area C).
By exploiting the universal language understanding and generation capabilities of LLMs, many major modules and stages in practical recommender systems can be benefited and even replaced.
For example, LLM can enhance the user/item representations and generate auxiliary textual features~\cite{xi2023towards, torbati2023recommendations}.
In addition, the rich world knowledge condensed by LLMs is complementary to conventional recommender systems in comprehending users and items, and unifying their knowledge can collaboratively their capabilities in making recommendations~\cite{li2023ctrl}.
Since the interaction interface of LLM-assisted recommenders usually remains unchanged, users can enjoy more targeted and accurate recommendation services without any additional effort.
However, the limited information disclosed by simple user feedback is still a barrier that hinders machines from truly understanding user interest.

With the human alignment process of LLMs, they are usually specialized in generating natural language conversations.
Different from traditional dialogue expert systems and small generative models, LLMs not only master strong language skills but also better comprehend complicated contexts and capture user interest based on their commonsense knowledge.
Therefore, they naturally become the chatting engine in conversational recommendations to aggregate and explain the results.
Furthermore, equipped with external tools such as databases, search engines, and recommendation models, LLMs can touch and fuse heterogeneous knowledge and incorporate it into recommendations.
By intelligently understanding user intent, digesting multi-source information, and providing convincing explanations, LLMs can not only improve recommendation accuracy and informativeness but also accommodate user requirements with fewer inquiries to reduce the interaction cost~\cite{friedman2023leveraging,liu2023conversational}. 
Nonetheless, the accuracy of many LLM-centric conversational recommenders is restricted by the capabilities of available tools, such as the accuracy of integrated recommendation models.
In addition, many high-level skills including complex planning, reasoning, and action execution are difficult to incorporate, which are essential to many genres of information acquisition and processing tasks in personalized recommendations.

In the two paradigms mentioned above, recommender systems evolve in two distinct ways to provide more effective information in user-machine interactions.
In LLM-centric conversational recommendations, LLM acts as a coordinator to arrange the execution of recommendation tasks and command various external tools including conventional recommendation models.
By contrast, in LLM-empowered systems for list-based recommendation, LLMs play auxiliary roles and are utilized in certain processes such as feature engineering and item ranking~\cite{xi2023towards, torbati2023recommendations,li2023taggpt}, which enables partial automation of recommendation processes to lessen the need for human intervention.
Both paradigms come to the same destination that converts recommender systems into recommender agents.
With the ultra-strong capabilities of LLM agents in terms of long-term memory, reflection, and tool intelligence, personalized assistants can become indispensable intimates that provide users with thoughtful and tailored services.
Without wordy conversations and mechanical interactions, humans and personal agents can reach an unvoiced pact and finally build harmony and solid trust.
By scrutinizing the recent history of recommender systems, the outline of their future shape is becoming clear.

In the following sections, we will walk along the ``all roads lead to Rome'' and draw lessons from the revolutionary history of personalized recommendation in detail.
Starting from the conventional list-wise recommendation in the deep learning age in Section~\ref{dl}, we then introduce LLM-enhanced list-wise recommendation in Section~\ref{sec:llm_enhanced_rec}, followed by the conversational recommendation before and in the LLM era in Section~\ref{sec:pre_llm_crs} and Section~\ref{sec:llm-crs}, respectively. 
Finally, the LLM-powered recommendation agents are described in Section~\ref{sec:recagent}.

\section{Conventional List-wise Recommendation}
\label{dl}
\begin{figure}[t]
    \centering
    \includegraphics[width=0.8\textwidth]{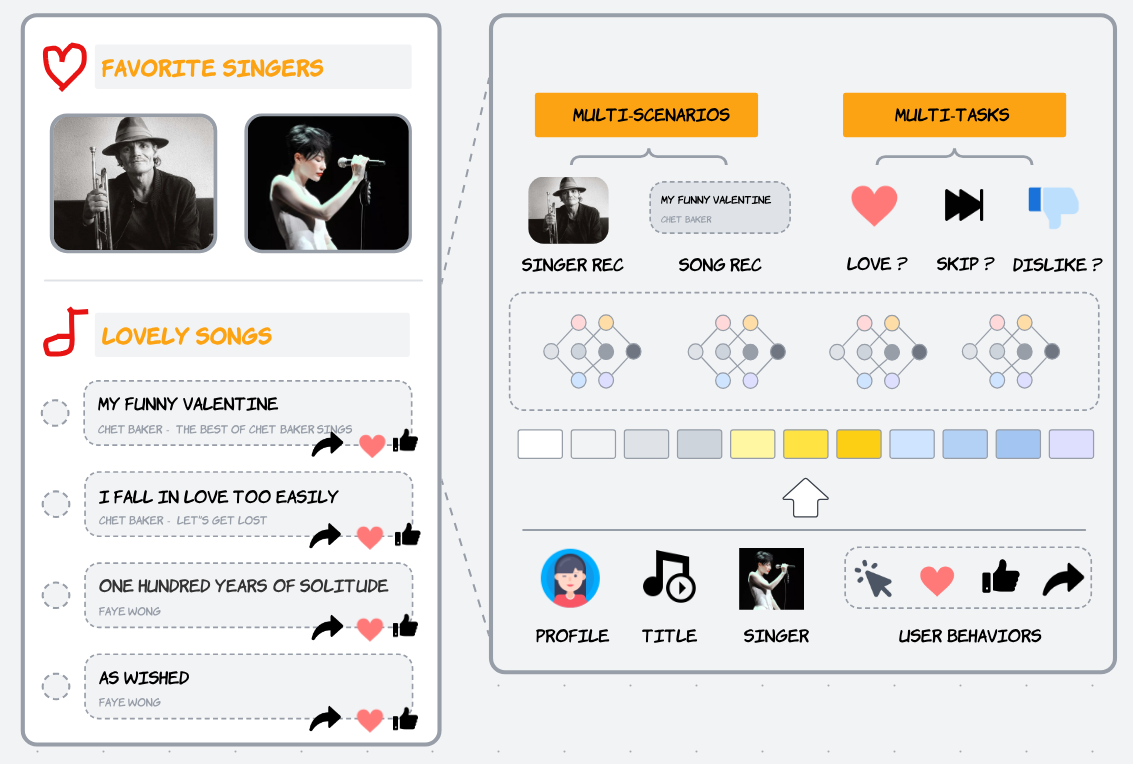}
    \caption{Example of a conventional list-wise recommendation system.
    }
    \label{fig:list-wise}
\end{figure}

Recommendation systems in the past decades benefited from the bloom of deep learning and usually presented the recommendation results to users with a ranked list. 
The representative commercial products include YouTube, Facebook, TikTok, etc. 
These recommendation systems generate a ranked list of items based on their users' preferences. 
The preferences are inferred with well-trained deep-learning models exploiting the user feedback, such as clicks and downloads, from the interaction logs generated over these products. 
For example, Fig.~\ref{fig:list-wise} presents a typical process of conventional list-wise music recommender systems.
Recommendation lists for singers or songs are generated according to the user features (e.g., the user profile), the item feature (e.g., song title, singer), and the user's past behaviors (e.g., clicked songs, liked songs).

In recommender systems, the number of candidate items can be millions or even billions.
To balance the effectiveness and efficiency, \textbf{a multi-stage cascade ranking system} is usually conducted.
Specifically, the ranking system is divided into mainly three stages including \textit{recall, ranking, and reranking}, taking into consideration efficiency and diversity requirements \cite{rankflow}.
Section~\ref{subsec:stages} introduces related studies for different recommendation stages.
Moreover, to exploit the complex and heterogeneous user feedback, the rankers in each stage encounter many challenges such as \textit{feature interaction, user behavior modeling, multi-task and multi-scenario learning}. Recent work focusing on these challenges will be discussed in Section~\ref{subsec:challenges}.

\subsection{Stages in list-wise recommendation}\label{subsec:stages}
To build a large-scale real-world recommender system capable of finding a list of items out of millions to billions of available options, a multi-stage system is deployed for ranking items efficiently, which usually contains recall, ranking, and reranking.
The recall stage seeks to retrieve a small set of candidate items users may be interested in from millions or even billions of available items.
After the recall stage, candidate items from multiple channels are fused and sorted by the ranking model according to the estimated scores to select dozens of items.
To meet certain requirements and further improve the recommendation results, a reranking model is applied to adjust the ranking of the item list according to specific metrics such as diversity and fairness.
Through these cascaded stages, the number of candidates is reduced to tens, which are finally exposed to users.

\subsubsection{Recall Stage}
To support the multiple interests of users and the diverse content of the platform, the traditional recall structure generally contains multiple sources, which can be classified into non-personalized recall and personalized recall.
Non-personalized recall algorithms don't exploit the information about a specific user, but utilize general information like item popularity, operating policy, city, and request time for recommendation.
Typical non-personalized recall algorithms include \textit{hot item recall} based on the item popularity, \textit{new item recall} based on the operating policy, and \textit{high accuracy recall} based on the click rate.
Without considering the diverse characteristics of different users, these non-personalized methods can hardly capture the users' unique needs and real-time preferences.
To pursue customized recall results based on users' different profiles and historical behaviors, many personalized recall algorithms like collaborative-based methods and embedding-based methods are proposed.
YoutubeDNN \cite{covington2016deep} uses deep neural networks with categorical features and average pooled user behavior sequential features to learn informative user representation for relevant item recall.
Bert4Rec \cite{sun2019bert4rec} improves YoutubeDNN by adopting a transformer layer to learn the complicit item dependencies and interest evolution.
As a powerful tool for learning user-item relations, GNN-based methods are widely adopted to the recall stage.
Pinsage \cite{ying2018graph} constructs a pin-board bipartite graph, then uses the graph convolutional networks to mine the collaborative signals and learn high-quality representations.
RippleNet \cite{wang2018ripplenet} introduces an auxiliary knowledge graph to complement additional relations to improve the recommendation.
To achieve faster computational efficiency, contextual features and cross features are often discarded, which greatly limits the effect of the recall model.
As a result, we usually design a complex ranking model with numerous hand-crafted features to enhance the final result.

\subsubsection{Ranking Stage}
The recall stage provides a preliminary filtering of items, which are delivered to the ranking models for a fine-grained selection based on users' preferences.
Since there are usually dozens of items to be ranked, the algorithms in this stage are typically much more complicated combined with rich features for precise user interest mining and item relevance prediction.
To achieve this goal, click-through rate (CTR) prediction and post-click conversion rate (CVR) prediction are the two most critical tasks for personalized ranking systems, which estimate the probability of a user click or a conversion over a given item.
Then these items are sorted by the combination of CTR, CVR, and item bidding price, where the combination strategies are often different in different platforms. 

The ranking models are typically trained using historical exposure logs, following an \textit{Embedding \& Representation learning \& Prediction paradigm}.
They firstly transform the input features (including categorical features, sequential features, numerical features, and combinational features) into embedding vectors, then use an advanced network structure to learn high-level feature representation via modeling feature interaction or mining user behavior, finally fed into fully connected layers to get the prediction score.
Since a single model is usually applied in this stage, many ingenious network structures are proposed, and different companies may have different base models with customized optimization to their different application scenarios. 
Representative works include DeepFM~\cite{guo2017deepfm} proposed by Huawei for CTR prediction, DCN~\cite{wang2017deep} proposed by Google for CTR prediction, DIN\cite{zhou2018deep} proposed by Alibaba for CTR prediction, and ESMM\cite{ma2018entire} proposed by Alibaba for CVR prediction.

\subsubsection{Reranking Stage}
The reranking stage plays a crucial role as the last controller of the recommendation list, it amends the final recommendation results by considering the item relationships and list-wise context~\cite{liu2022neural}.
Early works usually optimize the single accuracy objective, where many deep neural network (DNN) based models have been proposed, such as multi-layer perceptron (MLP) based \cite{jiang2018beyond,liu2021variation}, recurrent neural
networks (RNN) based \cite{ai2018learning,zhuang2018globally}, multi-head self-attention (MHSA) based \cite{pei2019personalized,huang2020personalized}, and graph neural network (GNN) based \cite{liu2020personalized}.

Later, \citet{xi2021context} points out that the feedback used in earlier reranking models depends heavily on the displayed ranking list, while different permutations of the ranking list may yield different observation results.
Hence, some recent works seek to provide unobserved signals with an extra evaluator assuming the counterfactual permutations.
Generally, there is a generator to produce possible permutations followed by an evaluator to assess the utility of the corresponding permutation \cite{wang2019sequential,song2020co,huzhang2021aliexpress}.
Moreover, multi-objective optimization is proposed to improve the single accuracy objective by providing more supervision signals.

Though most existing models focus on accuracy metrics, several other metrics like diversity and fairness are attracting more and more attention due to the long-term benefits and regulatory requirements.
To increase the diversity, some models propose to decrease the similarity among items \cite{feng2018greedy,yan2021diversification}, while some other models choose to promote the coverage over some specific categories \cite{qin2020diversifying,abdool2020managing,liu2023personalized}.

Besides the above-mentioned works about different loss functions and supervision signals, there are also some new directions---mixed reranking tries to reorder the list of mixed sources of items \cite{xie2021hierarchical,zhao2020jointly}; edge reranking considers the reranking with the collaboration of cloud and edge \cite{gong2020edgerec}.

\subsection{Major challenges in list-wise recommendation}\label{subsec:challenges}
With the rapid development and great success of deep learning for recommendation, there emerges a large variety of research problems, including feature interaction modeling, user behavior modeling, and multi-task and multi-scenario learning.
In this section, we pick some representative models to have a brief view of the development of deep learning models for recommendation.
\subsubsection{Feature Interaction in Recommendation}
Early deep learning based models focused on the adequate modeling of feature interactions, which can be divided into three categories according to the different operators: \textit{product based, convolutional based and
attention based} \cite{zhang2021deep}.

\textbf{Product based models} use different product operators like the inner product, outer product, or Hadamard product to capture relevance between vectors for effective feature interaction modeling.
DeepFM \cite{guo2017deepfm} proposes to utilize an FM layer in parallel with an additional DNN to model feature interactions.
Deep \& Cross Network (DCN)~ \cite{wang2017deep} recursively applies vector-wise Hadamard product layer by layer, to capture feature interactions of different orders.
DCN V2~\cite{wang2021dcn} enhances DCN by introducing a matrix to replace the feature crossing vector to improve its modeling capability. 

\textbf{The convolutional based models} explore the convolutional neural networks
(CNN) and graph convolutional networks (GCN) for feature interaction modeling.
FGCNN \cite{liu2019feature} exploits the fine-grained local patterns mining capacity of the CNN layer to generate high-quality features, then feeds the combination of these new features with existing features into IPNN ~\cite{qu2016product} for final prediction.
FiGNN \cite{li2019fi} considers different features as graph nodes and designs a fully connected graph with all node pairs fully linked, then conducts the graph propagation for feature interaction modeling.

\textbf{The attention based models} exploit the different attention architectures for feature interaction modeling.
AutoInt~\cite{song2019autoint} proposes to utilize the self-attention layer to learn the relevance between the central feature and all other features, thus learning a contextualized representation for better prediction. 
FibiNet \cite{huang2019fibinet} applies the SENET \cite{hu2018squeeze} to the input features to learn dynamical feature weights and alleviate the noise inside feature interactions.
Though many deep models have been proposed for feature interaction modeling in the last 10 years, the performance difference is indistinguishable and some new models obtain even worse results than previous SOTAs \cite{zhu2021open}.
One reason is that the recommendation system research field lacks unified datasets and baselines.
Another possible reason is that seldom work pays attention to the basic theory of these models, which makes the conclusions solely depend on the experimental results, influenced by many factors, such as dataset splits, dataset size, hyper-parameters tuning, etc.

\subsubsection{User Behavior Modeling in Recommendation}
In addition to the modeling of feature interaction, user behavior features that record the user's historical interaction in a given time range are also critical for recommendation.
To mine the various and crucial interest patterns of users, existing user behavior modeling methods can be summarized into three research trends according to the different characteristics of input user behavior features: \textit{traditional sequential modeling, long-sequence modeling, and multi-behavior modeling} \cite{he2023survey}.

\textbf{Traditional sequential modeling} seeks to extract item co-occurrence and sequentially dependency from a single short behavior sequence.
DIN~\cite{zhou2018deep} proposes a target-aware attention network to learn different weights for different items, thus obtaining the user's real interests and removing the noise.
CAN~\cite{zhou2020can}  designs a co-action network with dynamic network parameter generation to capture the interaction between the target item and the historical behaviors.

\textbf{Long-sequence modeling} attempts to use a longer sequence to model users' lifelong behaviors and mine more accurate user interests.
SIM \cite{pi2020search} proposes to search a subset of behaviors most similar to the target item at first, then uses target-aware attention to extract user interests.
ETA \cite{chen2021end} leverages locality-sensitive hashing (LSH) and Hamming distance to retrieve relevant items with an end-to-end architecture.

\textbf{Multi-behavior modeling} aims to explicitly consider the different behavior types and rich behavior attributes for more accurate prediction.
MBSTR \cite{yuan2022multi} regards multi-type behaviors as a unified sequence, then proposes a new heterogeneous transformer layer to capture multi-behavior dependencies and behavior-specific semantics.
SC-CNN \cite{zhang2022deepvt} concatenates side information with the item identifications as a 3D cube,
then uses a CNN to capture the relations among side features and item identification. 

User behavior modeling is similar to the token sequence modeling in the NLP field with the common sequential characteristic.
Inspired by the great success of large language models, it's very promising to exploit the scaling law of user behavior modeling for training large recommendation models.
HSTU \cite{zhai2024actions} proposed by META regards all the interacted items and behavior types as a unified sequence, then applies a modified transformer architecture to scale the model to a maximum of 1.5 trillion parameters.

\subsubsection{Multi-task and Multi-scenario learning in Recommendation}
As a pivotal tool to provide online information and services, there are usually multiple optimization tasks (e.g., click, like, purchase, etc.) to reflect the various actions of users.
Besides, there are also a large number of scenarios (e.g., video, music, reading, etc.) to satisfy the diversified requirements of users.

To jointly optimize multiple tasks, the \textbf{multi-task learning} paradigm~ \cite{caruana1997multitask} which uses a unified structure to optimize multiple targets is widely used.
Representative models are Shared Bottom~ \cite{caruana1997multitask}, MMOE~\cite{mmoe}, and PLE~\cite{PLE}.
Shared Bottom feeds a shared bottom layer to multiple prediction layers for multi-task learning.
MMOE~\cite{mmoe} learns different gating networks to assign different weights over different experts for more flexible prediction.  
PLE~\cite{PLE} further proposes to use shared and specific expert networks to mitigate information conflicts among different tasks. 

Considering the various recommendation scenarios, \textbf{multi-scenario learning} seeks to learn a unified model for different scenarios to avoid the sparsity of cold scenarios and reduce the cost of training multiple models.
STAR~\cite{sheng2021one} proposes a star topology structure, which contains the shared centered parameters and scenario-specific parameters to handle the problem of scenario distribution discrepancy.
SASS~\cite{zhang2022scenario} utilizes a multi-layer domain adaptive transfer module with a two-stage training process to facilitate the multi-domain information transfer.
APG~\cite{yan2022apg} designs a new learning paradigm where input features are used to generate model parameters to achieve domain adaptation. 
M2M \cite{zhang2022leaving} introduces the meta unit to incorporate scenario knowledge to learn inter-scenario correlations.

There are also works that try to use a single framework to handle multi-task learning and multi-scenario learning simultaneously.
HiNet \cite{zhou2023hinet} proposes a domain-aware attentive network on top with a hierarchical information extraction network to optimize multiple objectives in multiple domains.
The seesaw phenomenon (i.e., models improve the performance of some tasks or scenarios, but harm the others) widely exists.
Though many deep models have been proposed, the deep understanding of how and when the negative transfer happens is still under-explored.

\subsection{Discussion}
The list-wise recommendation system is a mature recommendation paradigm, it can provide abundant information to users at one time, making it convenient for users to quickly obtain the information they want.
However, the conventional list-wise recommendation systems do have some problems. 

\textbf{Implicit signals can be vague and noisy.} 
The user feedback on a recommended list (e.g., browse, click) is implicit, which can be vague and noisy to learn user preference. 
For example, song completion in a music recommendation system can represent a positive signal because the user did not change the song. 
But it can also be a negative signal cause the user might not actually pay attention to the song. 
Moreover, the position of an item in a list can influence the user's attention.
Specifically, items appearing in the first position are more likely to be seen and clicked by users even if the users might prefer the item in the middle.

\textbf{Lack semantic knowledge}
Merely based on the ID-based collaborative signals, the recommendation models can hardly know about the user and item thoroughly. 
A music recommendation system may contain millions of songs and lyrics, but they do not include the stories behind the songs and the social connections of the singers. This information is in fact informative for a better representation understanding. What's more, most of the list-wise recommendation systems currently do not take into consideration the explanation for the recommendation results, which makes the system ambiguous to the users and the users accordingly lose control of the personalized recommendation system.

Fortunately, two emerging recommendation paradigms have been proposed to address the problems mentioned above. The first one is the conversational recommendation system (CRS), in which users can engage in multiple turns of interactions with the system, and the system will in return guide the users to clarify their needs and preferences gradually and finally provide the suitable items. In most cases, the CRS will provide a corresponding explanation according to the user history or previous conversations.
The other is the LLM-enhanced recommendation systems, which can utilize the open-world knowledge acquired from the pre-training stage to help the recommendation models better understand the item and user portrait. Besides, the logical reasoning ability and tool-calling ability can be further exploited to optimize the user behavior understanding and recommendation procedure.

\section{LLM-enhanced Recommendation}\label{sec:llm_enhanced_rec}
Thanks to the development of deep learning, list-wise recommender systems have achieved remarkable progress over the past decades as depicted in Section~\ref{dl}.
However, their recommendation performance is still limited, hampered by several major drawbacks as follows: 
1) The list-wise recommendation models based on deep learning generally capture the user-item occurrence relationship with discrete ID features by feature interaction and user behaviors modeling~\cite{guo2017deepfm,zhou2018deep}, thus ignoring their original semantic information.
2) The scope of involved knowledge for recommendation modeling is limited within the domain. The lack of open-domain world knowledge obstructs the ability to obtain the latest information and understand the relationship between items and users.
3) These recommendation models might lack recommendation explainability.

\begin{figure}[t]
    \centering
    \includegraphics[width=0.95\textwidth]{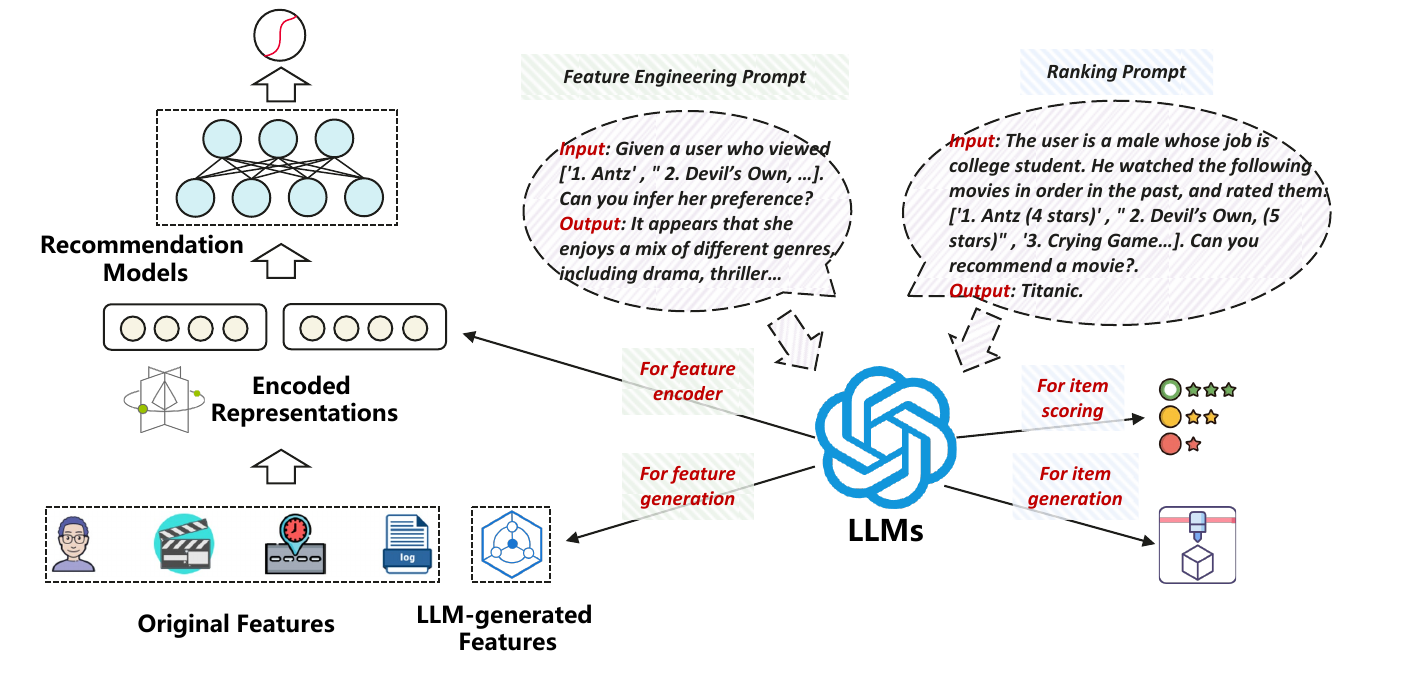}
    \caption{The illustrative dissection of LLM-empowered recommendation. LLM-empowered recommender systems can be divided into two categories: LLMs for feature engineering and LLMs for ranking.
    }
    \label{fig:llm-rs}
\end{figure}

Recently, with the emergence of pretrained language models (PLM), especially large language models (LLMs) such as GPT-3~\cite{brown2020language}, LLaMA~\cite{touvron2023llama}, ChatGLM~\cite{du2022glm}, these drawbacks are expected to be solved. LLMs have showcased powerful energy in general intelligence, such as memorizing vast amounts of open-world factual knowledge, excellent understanding of content and context, as well as logical and commonsense reasoning ability, which have been applied in various tasks.
By involving LLMs in recommender systems, the effective information can be boosted remarkably.
Firstly, the semantic content information (e.g., item descriptions) can be well understood and the associated open-world knowledge beyond the training data can also be injected.
Besides, with the blessing of planning and reasoning ability, users' latent intentions and preferences can be inferred accurately under semantic space.
Moreover, the traditional recommendation paradigm based on collaborative signals may also be upgraded to semantic signal dominance.
Based on the roles of LLMs in recommender systems, we divide these LLM-empowered recommender systems into two categories: \textit{LLMs for feature engineering} and \textit{LLMs for ranking}, shown in Fig.~\ref{fig:llm-rs}

\subsection{LLMs for feature engineering}
In this category, the role of LLMs is subservient, which is used to excavate beneficial semantic features and representations, and the traditional DL-based recommendation models still dominate the final ranking decision process.
Based on the subordinate adaptation of LLMs during the recommendation pipeline, the existing works can be mainly divided into two kinds: 1) feature generation, and 2) feature encoder. 
\subsubsection{LLMs for feature generation}
Feature generation is one of the most important stages for the industrial recommendation, especially before the era of deep learning, which aims to excavate beneficial features for better depicting the user-item relevance.
Benefiting from the excellent content generation capabilities and open-world knowledge, LLMs can be involved into the feature engineering stage for generating auxiliary textual features. 
By designing appropriate prompting strategies with several original features (e.g., item descriptions, user profiles, and behaviors), LLMs can produce informative textual features, thus enriching semantic information and alleviating the problem of feature scarcity in cold-start scenarios.
According to the type of augmented features, these works can be further classified into two groups: 1) user/item-level feature augmentation, and 2) instance-level feature augmentation. 

\textbf{User/item-level feature augmentation.}
Benefiting from the external open-world knowledge, LLMs can involve rich and even up-to-date information in the closed recommendation space, thus contributing to the feature augmentation, especially for the item features.
Besides, the powerful summarizing and reasoning ability of LLMs makes user preference modeling~\cite{disco} and item content understanding~\cite{xi2023towards} more accurate and efficient.
Served as a feature augmentation plug-in module for recommendation models, KAR~\cite{xi2023towards} leverages LLMs to generate user-level preference knowledge and item-level factual knowledge. 
Besides, SAGCN~\cite{liu2023understanding} designs a chain-based prompting approach to extract semantic aspect-aware user-item interaction features in a fine-grained manner, which are fed into the GCN for further modeling users’ preferences.
In contrast to employing a frozen and general base LLM for feature generation, LLaMA-E\cite{shi2023llama} and EcomGPT~\cite{li2023ecomgpt} engage in targeted fine-tuning of the base LLMs with intra-field knowledge before feature augmentation.

\textbf{Instance-level feature augmentation.}
Besides enhancing individual user/item-level features, LLMs are also employed to create synthetic samples thus contributing to a semantic-richer training dataset, which belongs to the instance-level feature augmentation.
GENRE framework~\cite{liu2023first} employs custom-designed prompts to extract not only user/news-level features via LLM-generated user profiles and news titles, but also instance-level features with LLM-generated synthetic news pieces.
Consequently, this LLM-augmented approach contributes to improving the prediction of news recommendations.
To efficiently compress and condensate the training data for recommendation, TF-DCon~\cite{wu2023leveraging} leverages LLMs to empower the generation of compact users and content compression, reducing the dataset by 95\% with similar performance impressively. 
Besides, LLMs play a crucial role in enhancing textual inputs for LLM-based recommenders~\cite{liu2023recprompt,sun2023large} and rewriting user queries~\cite{peng2023large,li2023agent4ranking}.

\subsubsection{LLMs for feature encoder}
For conventional ID-based recommender systems~\cite{wang2021dcn, zhou2018deep}, the structured features are first converted into one-hot features via one-hot encodings with semantic information loss, and following, a feature embedding layer parameterized as an embedding look-up table is deployed to map into dense embeddings, which acts as the default feature encoder in ID-based recommendation.
With the development of multi-modal content understanding technologies, the textual and visual features are no longer simply transformed into discrete one-hot features, but are further encoded by various models such as LLMs to excavate semantic information thus enriching the item/user representations.
Furthermore, benefiting from the bridge role of natural language, the adoption of LLMs as feature encoders also facilitates transfer learning and cross-domain recommendation~\cite{tian2023ufin,gong2023unified}.

For item representation enhancement, news recommendation~\cite{liu2023first,li2023pbnr,li2023exploring,runfeng2023lkpnr} is the most widely used scenario due to abundant textual features (e.g., news titles, entities, keywords, and abstract).
\citet{li2023exploring} harnessed LLMs as news encoders within traditional recommendation models to generate news representations for personalized news recommendations.
Besides, LKPNR~\cite{runfeng2023lkpnr} presents a novel framework that combines LLMs and Knowledge Graphs (KG) into news semantic representations.
In addition to news recommendations, LLMs are also served as item encoder applied to software purchase~\cite{john2024llmrs}, social networking~\cite{jiang2023social}, tour itinerary recommendation~\cite{ho2023utilizing}, and other general recommendation scenarios~\cite{chen2023tbin,harte2023leveraging}.

For user representation enhancement, the strength of LLMs lies in understanding the diverse interests and dynamic evolving preferences of users. 
To better assist aspect-based recommendations, LLM4ARec~\cite{li2023prompt} utilizes GPT2 for extracting personalized aspect terms and user representations from user profiles and reviews. 
In recommendation, users' interests can be adequately reflected by the historical interactive item IDs.
To overcome the semantic ID index problem, TIGER~\cite{rajput2023recommender} and LMIndexer~\cite{jin2023language} propose vector quantization (VQ) based methods by compressing each item into a tuple of discrete semantic tokens, thus facilitating the sequential modeling of LLMs.

\subsection{LLMs for ranking}
In this category, the role of the leader gradually shifts from traditional DL-based models to LLMs, and the recommendation paradigm shifts from collaborative signals to semantic signals.
To further improve the recommendation performance, plenty of works focus on injecting collaborative knowledge from traditional ID-based recommendation models into LLMs~\cite{zhang2023collm,li2023e4srec,zhu2023collaborative,kang2023llms}.
Based on the different tasks of LLM, the existing works can be divided into two groups: 1) scoring-based LLMs, and 2) generation-based LLMs. 

\subsubsection{Scoring-based LLMs}
For item-scoring tasks, the LLMs act as a pointwise function to estimate the prediction relevance between user and candidate items.
In this task, LLM-empowered recommendation models mainly perform Click-through Rate (CTR) prediction, which are widely-used in the \textbf{ranking stage} for large-scale industrial recommender systems.
To obtain the relevance score from LLMs, several approaches are proposed.

The first one is single-tower paradigm~\cite{clickprompt,zhang2023collm,li2023e4srec,zhu2023collaborative,kang2023llms} that discards the original language modeling decoder head and deploys an extra projection layer to calculate the score.
CoLLM\cite{zhang2023collm} and  E4SRec~\cite{li2023e4srec} integrate collaborative information pre-learned from ID-based recommendation models into LLMs and deploy LoRA for effective parameter fine-tuning.
Kang et al.~\cite{kang2023llms} fine-tuned the LLMs with different sizes for rating prediction and performed detailed experiments under zero-shot and few-show settings.

The second one is two-tower paradigm~\cite{tang2023one,torbati2023recommendations,yang2023collaborative,li2023text} that also discards the original decoder head and outputs the user and item representations from two heads. Based on this, the relevance score can be obtained via distance metric between the two representations such as cosine similarity.
Recformer~\cite{li2023text} organizes the raw features into text and leverages the language model for extracting item and user sequence representations.
Besides, LLM-Rec~\cite{tang2023one} further explores the multi-domain behavior modeling for LLMs.

The last one is preserving the decoder head and obtaining the score over limited tokens (e.g., ``Yes.'' and ``No.'') via Softmax function~\cite{bao2023tallrec,lin2023rella,luo2024integrating}.
To align the LLMs with recommendations, TALLRec~\cite{bao2023tallrec} designs a framework to build LLM-based recommendation models, which enables the effective and efficient finetuning of LLMs with lightweight LoRA. In these settings, LLMs are required to provide feedback of a user to the new item with ``Yes./No.''.
Moreover, ReLLa~\cite{lin2023rella} further extends to user long behavior modeling and mitigates the incomprehension problem of LLMs with a retrieval-enhanced approach.

\subsubsection{Generation-based LLMs}
Different from the item scoring tasks that estimate the relevance score between user and candidate items, item generation tasks require the LLMs to directly generate the final ranked list of items.
According to the difference in whether items generated by LLMs are restricted, generation-based LLMs can be further categorized into two groups: 1) open-set generation, and 2) closed-set generation.

\textbf{Open-set generation.}
In this category, LLMs directly generate a ranked item list without relying on the predefined candidate item set due to the massive scale of items, which is similar to the current \textbf{recall stage} in multi-stage industrial recommender systems. However, due to the smaller number of items generated by LLMs compared to the recall stage, these approaches belong to novel single-stage generative recommendation~\cite{ji2023text,bao2023bi,jiang2023reformulating}.
GenRec~\cite{ji2023text} prompts the LLMs to generate the target item directly. To make sure that the generated items are in the item pool, several post-processing operations are proposed, including L2 distance mapping in BIGRec~\cite{bao2023bi} and cosine similarity in LANCER~\cite{jiang2023reformulating}.

\textbf{Closed-set generation.}
As for closed-set item generation tasks, a pre-filtered candidate item list (up to 20 usually) is provided by DL-based recommendation models, and the LLMs are expected to select several items that users are most interested in~\cite{touvron2023llama,wang2023drdt,gao2024llm}.
This paradigm is similar to the \textbf{reranking stage} in industrial recommender systems that LLMs rerank the candidate items provided by the previous ranking models.
LlamaRec~\cite{yue2023llamarec} proposes a two-stage recommendation framework and deploys the LLaMA~\cite{touvron2023llama} for candidate ranking.
DRDT~\cite{wang2023drdt} also uses the retriever-reranker two-stage framework and employs iterative multi-round reflection to refine the final ranked list with given candidates gradually.

\subsection{Discussion}
LLM-empowered recommendation breaks down the recommendation internal knowledge barriers and empowers the semantic understanding of the entire recommendation pipeline from the feature engineering to model ranking, thus raising effective information remarkably.
Existing LLM-empowered recommendation models attempt different methods to involve in-domain collaborative knowledge with semantic information, such as tuning LLMs~\cite{bao2023tallrec,lin2023rella,li2023text} or inferring with collaborative models~\cite{li2023e4srec,xi2023towards,li2023e4srec}, thus achieving better performance compared with the traditional ID-based list-wise recommendation models. The rich world knowledge and excellent reasoning ability take understanding of users and items to new levels~\cite{xi2023towards,liu2023first}.

Although the incorporation of LLMs in recommender systems has brought about better ways of understanding content,  the interaction interface of LLM-empowered recommenders remains unchanged relying on implicit feedback like clicks, where users' current intentions or real needs are difficult to accurately detect and the degree of satisfaction with recommendation results are also difficult to reflect.
To this end, the interactive LLMs present a promising alternative, by offering a more active and adaptive form of user interaction in a conversational style.
Besides, this conversation-styled recommendation could engage in real-time interactions with users to obtain more accurate feedback about the recommendation results. 

\section{Conversational Recommender Systems Before LLM Era}\label{sec:pre_llm_crs}

In this section, we are pivoting our focus from the previous list-wise recommendation architectures to conversational recommender systems (CRS).
We will review the early-day developments of CRS before the LLM era \cite{gao2021advances}.

\begin{figure}[h]
    \centering
    \includegraphics[width=0.9\textwidth]{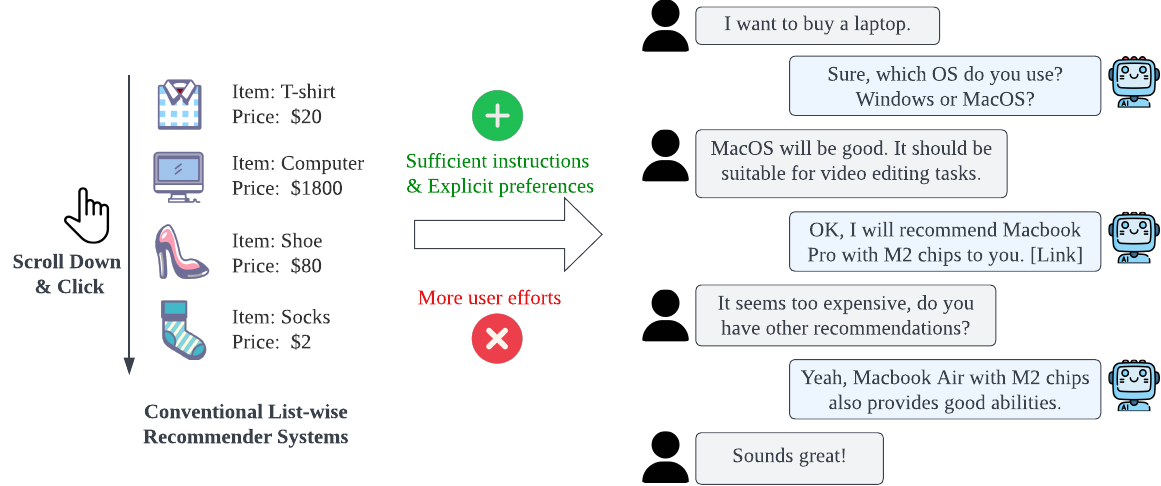}
    \caption{The illustration of conventional list-wise recommendation and conversational recommender systems (CRS).}
    \label{fig:list2crs}
\end{figure}

Compared with CRS, the list-wise recommendation platforms have one major shortcoming: the ambiguity in user understanding.
The user satisfaction and intentions are ambiguously expressed in traditional list-wise recommenders. 
The user feedback is usually implicit, such as clicking items or simply thumb-scrolling, as shown in Fig.~\ref{fig:list2crs}.
The implicit feedback makes it difficult for the system to determine what the user really likes or the exact intention of the user.
The user might not really like the item she clicked (wrong decision/big thumb) or even not really know what she likes before the system shows her all the options.

To overcome the above-mentioned shortcoming of the static recommendation systems, conversational recommender systems (CRS) are proposed \cite{gao2021advances}.
The definition of CRS could be a recommender system that has \textit{multi-turn interactions} with the users, and the recommendation results could be adjusted in real-time according to user's interactions.
Natural language is the major form of multi-turn interactions, users could express their delicate intentions and preferences by using natural language.
Furthermore, the system could provide more accurate and appropriate responses thus increasing the effective information obtained by the users.
An example of the recommendation process of CRS is illustrated in Fig.~\ref{fig:list2crs}.

In the following sections, we will introduce the two key aspects of designing a CRS: \textit{question asking} and \textit{maintaining conversations}.
We also review the methods of evaluating such systems and discuss the applications and limitations of CRS before the LLM era.
It should be noticed that the CRS models based on the early-day language models \cite{devlin2018bert,radford2019language} such as Bert \cite{xu2020user} and GPT-2 \cite{zhou2020towards} are also included in this section.

\subsection{Questions Asking}
For a CRS, the most important task is asking the user questions and trying to narrow down the candidate items according to the user's answers.
The different asking mechanisms could be categorized into item-level and attribute-level.

\subsubsection{Item-level}
The early-day CRS models ask directly if the user likes the item/recommended list or not \cite{gao2021advances,wang2018online,zhao2013interactive,christakopoulou2016towards}.
The choice-based models give users options and let them choose which one they like best.
The options could be two items \cite{sepliarskaia2018preference}, a list of items \cite{jiang2014choice,graus2015improving}, or two different lists \cite{loepp2014choice}.
The recommendation model will be updated according to the user's choice. 

Interactive recommendation is another research branch of item-level CRS. 
It is mainly based on reinforcement learning such as multi-arm bandits (MAB) \cite{zhao2013interactive,wang2018online} (balancing exploration\&exploitation) and deep RL methods \cite{zhao2018recommendations,chen2019large,xian2019reinforcement,chen2019top,zou2020pseudo} which could model the dynamic preference and long-term utility.
Other methods such as Bayesian models \cite{chajewska1998utility,vendrov2020gradient} are also proposed for item-level question asking in CRS.

Although many research works are proposed for item-level question asking, they are not very applicable in real systems because users could be bored of being asked many questions about specific items.
It is more suitable to ask coarse-grained questions about the attributes or concepts that the user likes. 
Thus, attribute-level question asking methods are proposed.

\subsubsection{Attribute-level}
Attribute-level question asking is more efficient than item-level because an attribute corresponds to a set of items which could help the system reduce the candidate set significantly.
The mainstream approaches include fitting historical interactions, critiquing-based methods, RL-based methods, and graph-based methods.

Historical interaction learning methods deem conversation records to be a sequence. 
And the training objective is to predict the next attribute to ask/item to recommend. 
These methods are usually based on sequential models like GRU or LSTM \cite{christakopoulou2018q,zhang2018towards}.
However, these methods do not consider how to respond to user's rejecting feedback.

The critiquing-based method means that users give critical opinions on certain attributes, such as "don't like green," and the system adjusts the recommended results accordingly.
The system usually removes all the candidate items with rejected attributes using static and heuristic rules \cite{chen2012critiquing,smyth2004compound,viappiani2007conversational,smyth2003analysis}.
Apart from the rule-based models, the vector-based methods embed the critic signals into the latent representation of items \cite{wu2019deep,luo2020deep}.

RL-based models usually use policy networks that select suitable attributes to ask the users \cite{lei2020estimation,lei2020interactive}. 
Graph-based models include path-walking methods according to the conversations \cite{lei2020interactive} and GCN-enhanced methods \cite{chen2019towards,liao2020topic}.

\subsection{Maintain Conversations} \label{sec:good_conv}
The question-asking methods focus on "what to ask." in this section, we discuss the methods of maintaining a good conversation with users ("when to ask").
A good conversation instead of continuing interrogating is much better for user experience.
We will describe the timing of question asking and how to lead the conversations.

\subsubsection{When to ask \& recommend}
The strategy for determining when to ask and recommend could be rule-based (making recommendations every $k$ turn of asking \cite{zhang2020conversational}) or random policy \cite{christakopoulou2018q}.
A more sophisticated strategy is implemented by utilizing reinforcement learning.
The RL-based models always utilize a deep policy network with multiple actions. The actions include facets to ask and yielding a recommendation \cite{sun2018conversational,lei2020estimation}.
The policy-based models are trained using a policy gradient.

\subsubsection{Leading the conversation}
Users may not be so sure about what they really like.
Thus the CRS should not only interrogate users but lead the conversation and make recommendations naturally without affecting users' minds.

There are some research works about multi-topic learning.
The system could proactively lead a conversation and naturally switch from a non-recommendation dialog to a recommendation dialog \cite{liu2020towards}.
These models define the recommendation task as the major goal and topic transitions as short-term goals \cite{zhou2020towards}.

A good CRS should lead the conversation with special abilities such as suggesting, negotiating, and persuading \cite{gao2021advances}.
With the development of large language models (LLMs), it could be easier for CRS to deal with the conversations more naturally.

\subsection{Evaluation of CRS}
The evaluation of CRS is not a trivial task.
The lack of good benchmark datasets is an important issue. 
The existing datasets are usually generated under specific constraints or rules.
The dataset scale is not large enough compared to real-world scenarios, resulting in insufficient evaluation \cite{gao2021advances}.

There are two levels of CRS evaluation: \textit{turn-level} and \textit{conversation-level}.
The turn-level evaluation focuses on a single turn of conversation.
It evaluates conversations from both language generation and recommendation accuracy.
For language generation evaluation, CRS mainly uses metrics such as BLEU \cite{papineni2002bleu} and Rouge \cite{lin2004rouge}.
For recommendation accuracy, rating-based (RMSE) or ranking-based (NDCG, MAP, MRR, etc) metrics are utilized.

For conversation-level evaluation, the most accurate method is to conduct an online test.
The metrics include average turns (AT) \cite{lei2020interactive,li2021seamlessly}, recommendation success rate (SR), and cumulative performance of each turn. 
Other conversation-level evaluation methods that do not require online tests include counterfactual evaluations \cite{jagerman2019people,mcinerney2020counterfactual} and user simulation \cite{zhang2020evaluating,christakopoulou2016towards}. 

\subsection{Discussion}
Although conversational recommender systems offer users personalized item recommendations through natural language interfaces, they have not emerged as the dominant product type.
One of the core limitations of CRS could be the extra user efforts of using natural language to interact with the system.
The early-day CRS models tend to ask the users a lot of questions, which may be wordy and rigid \cite{gao2021advances}.
It is not convenient for users to type all the words to tell the system what they want in the recommendation scenario.
Because the recommendation scenario corresponds to the "hanging out" behaviors on the platforms instead of explicit information needs.
The potential improvements of the recommendation accuracy by CRS may be not enough to cover the extra cost of user efforts introduced by the conversations.
Apart from the accuracy of a CRS, we should also be aware of the extra cost of user efforts, which will largely influence user experiences, which is less discussed in early-day CRS research.

\section{Conversational Recommender Systems in LLM Era}\label{sec:llm-crs}
The LLM-enhanced recommenders (detailed in Section~\ref{sec:llm_enhanced_rec}), akin to traditional recommenders, are a single-turn interaction process that faces challenges in precisely identifying users' current interests or actual needs and accurately reflecting satisfaction levels with recommendation outcomes. In contrast, CRS has the natural advantage of real-time understanding of user intents and the ability to adapt recommendations by interacting with users in a timely manner. To enhance comprehension of user preferences and needs, the dialogue module of CRS necessitates advanced language understanding and reasoning abilities. Yet, pre-LLM era CRS approaches (discussed in Section~\ref{sec:pre_llm_crs}) predominantly utilized pre-trained language models like GPT-2~\cite{radford2019language}, BERT~\cite{devlin2018bert}, BART~\cite{lewis2020bart}, and T5~\cite{raffel2020t5}. Owing to their limited training corpora and model capacities, these models demonstrate insufficient abilities for effective conversational recommendation.

\begin{figure*}[t]
    \centering
    \subfigure[Prompting-based LLM-CRS]
    {
        \label{fig:p_llm_crs}    
        \includegraphics[trim={0.0cm 0.0cm 0.0cm 0cm},clip,width=0.48\textwidth]{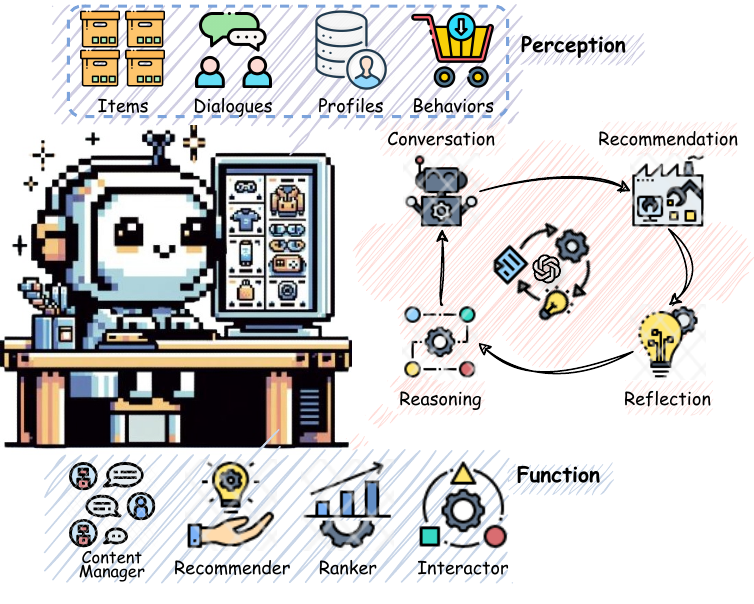}
    }
    \hfill
    \subfigure[Fine-tuning-based LLM-CRS]
    {
        \label{fig:ft_llm_crs}  
       \includegraphics[trim={0cm 0.0cm 0.0cm 0.0cm},clip,width=0.48\textwidth]{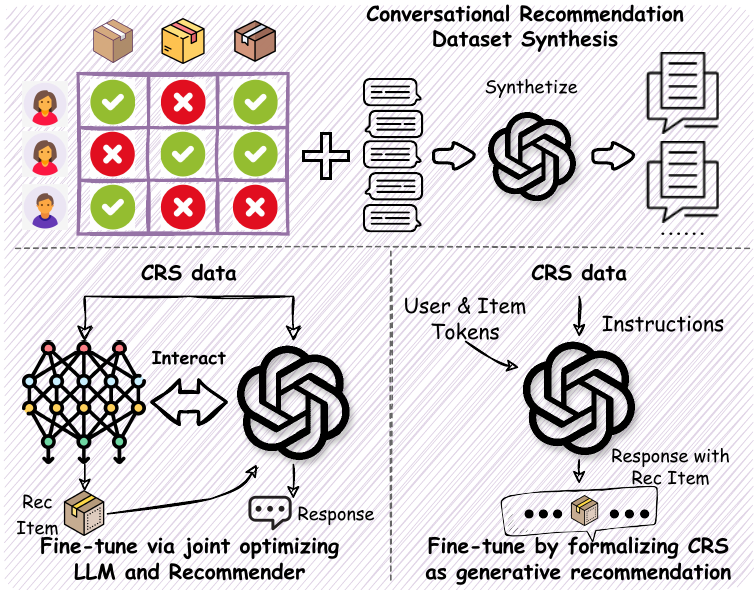}
    }
    \vspace{-0.3cm}
    \caption{A overview of the LLM-based conversational recommendation system (CRS): (a) The prompting-based LLM-CRS showcases an organized composition of interconnected modules, each serving a distinct purpose. The LLM acts as a controller, interacting with off-the-shelf recommenders to manage information about items, user profiles, dialogues, user behaviors and etc., through prompts, as well as enhancing performance via reasoning and reflection. (b) The fine-tuning-based LLM-CRS can be divided into two structural approaches: one aims to jointly optimize LLM and recommender, while the other formulates CRS as a generative recommendation task, tackled primarily through the LLM. Besides, the LLM is also utilized to generate CRS data, addressing data scarcity issues.}
    \label{fig:llm_crs}
\end{figure*}

Through pre-training on extensive world knowledge and alignment with high-quality instruction data, LLMs like GPT-4~\cite{achiam2023gpt4}, PaLM~\cite{chowdhery2023palm}, LLaMA~\cite{touvron2023llama} and ChatGLM~\cite{du2022glm,zeng2022glm} excel in knowledge retention, language understanding, content generation, and instruction following. These advancements offer innovative avenues for developing sophisticated conversational recommender systems. However, LLMs are generally trained by publicly available sources on the internet, which lack visibility into the data that resides within private platforms, resulting in the sub-optimal understanding capabilities of such data. Thus, LLM-based CRS focuses on integrating the LLMs with off-the-shelf recommenders to equip them with domain-specific knowledge. Based on the methods of applying LLM, LLM-based CRS can be roughly categorized into prompt engineering (prompting-based LLM-CRS, illustrated in Fig.~\ref{fig:p_llm_crs}) and fine-tuning (Fine-tuning-based LLM-CRS, depicted in Fig.~\ref{fig:ft_llm_crs}) approaches.

\subsection{Prompt Engineering Specialized for CRS}
Recent studies show that prompt serves as a critical interface for human-language model interactions in various tasks. A well crafted prompt strategy, such as CoT~\cite{wei2022chain}, ToT~\cite{yao2023tree}, ReAct~\cite{yao2023react}, can efficiently harness the advanced capabilities of LLMs. Thus, some work proposes to integrate recommendations and LLM through prompt engineering, enabling LLM to seamlessly interact with recommendation systems and leverage LLM's linguistic capabilities to achieve conversational recommendations~\cite{xu2024prompting}.

\subsubsection{Prompting-based LLM-CRS Architecture}
When prompting LLMs as recommender systems, despite variations in specific prompt designs across studies~\cite{murakhovska2023salespeople,gao2023chat,wang2023generativert,lin2023sparks,wang2023recmind,he2023large,spurlock2024chatgpt}, the format remains substantially similar with only slight differences. Since existing LLMs lack comprehensive conversational recommendation capabilities, prompting solutions dissect CRS into interconnected sub-modules, each aligned with a specific CRS function, including intent recognition, dialogue management, state tracking, response generation, recommendation calls, etc. Then, specific prompts are crafted for each sub-module to activate the reasoning abilities of LLMs. The conversational recommendation is facilitated through the collaborative interaction among sub-modules. Chat-REC~\cite{gao2023chat} augments LLMs by converting user proﬁles and historical interactions into prompts, and connecting users and products through in-context learning. Through linking traditional recommender systems with LLMs, Chat-REC achieves efficient and effective outcomes, even in cold-start scenarios. GeneRec~\cite{wang2023generativert} focuses on personalized AIGC, which adopts an AI generator to personalize content generation and leverages user instructions to acquire users’ information needs for recommendation. RecMind~\cite{wang2023recmind} introduces the planning mechanism, termed self-inspiring, to build an LLM-powered autonomous recommender agent, which can dynamically explore historical states to improve the recommendation. He \textit{et. al.}~\cite{he2023large} directly treat LLMs as zero-shot CRS and use hand-crafted task description template, format requirement and conversational context to prompt LLM. Spurlock \textit{et. al.}~\cite{spurlock2024chatgpt} further build a full pipeline, from to prompt creation and content analysis to final recommendation and feedback, around LLMs to simulate how a user might realistically probe the model for recommendations.

\subsubsection{Prompting-based LLM-CRS Evaluation}
Liu \textit{et. al.}~\cite{lin2023sparks} investigate the feasibility of developing artificial general recommender (AGR) via LLMs. Specifically, they propose ten fundamental principles, such as inconsistency detection, behavioral analysis, etc., that AGR should adhere to. Then they proceed to assess whether LLM can comply with the proposed principles by engaging in recommendation-oriented dialogues with the model while observing its behavior. They demonstrate the potential for LLM to serve as AGR, but with several limitations to be addressed. iEvaLM~\cite{wang2023rethinking} reveals the inadequacy of the existing CRS evaluation protocol, which overemphasizes the matching with ground-truth items annotated by humans while neglecting the interactive nature of CRS. iEvaLM utilizes LLM to build user simulators, which can simulate various system-user interaction scenarios and provide an in-depth evaluation of CRS.

\subsection{Fine-tuning Specialized for CRS}
LLM-based CRS via prompt engineering has shown improvements in effectiveness, informativeness, and user experience. However, a notable disparity exists between textual information and user-item collaborative signals, posing challenges for LLMs to capture key information accurately in recommendation scenarios. To address this, approaches focus on finer problem decomposition and advanced prompt techniques to ensure effectiveness. Nonetheless, this leads to complex execution processes and low efficiency. Some endeavors~\cite{friedman2023leveraging,liu2023conversational,mysore2023large,zhu2023collaborative,feng2023all,ravaut2024parameter} involve fine-tuning LLM to enhance CRS effectiveness by injecting recommendation capabilities.

\subsubsection{Data Acquisition}
Fine-tuning involves tuning LLM with massive high-quality domain-specific conversational recommendation data. However, existing datasets for conversational recommendation are typically limited in scale and quality. Meanwhile, the data in most recommendation scenarios primarily consists of explicit or implicit user-item interactions, but lacks conversation context. To alleviate this issue, recent work, such as RecLLM~\cite{friedman2023leveraging} and iEvaLM~\cite{wang2023rethinking}, proposes to construct a user simulator using LLM for conversational data generation. Specifically, RecLLM~\cite{friedman2023leveraging} builds a controllable LLM-based user simulator to generate synthetic conversations through fine-grained controls, while iEvaLM~\cite{wang2023rethinking} simulates various system-user interaction scenarios via carefully designed task instructions. Liu \textit{et. al.}~\cite{liu2023conversational} directly collects a real-world E-commerce pre-sales dialogue dataset. MINT~\cite{mysore2023large} uses LLM to re-purpose the publicly available user-item interaction datasets to form the synthetic narrative-driven recommendation data. CLLM4Rec~\cite{zhu2023collaborative} extends the vocabulary of LLM to cover the user/item IDs and then fuse them into LLM, thereby existing datasets can be easily adapted for model training.

\subsubsection{Model Architecture}
In consideration of the domain-specific nature of recommendation data and models, it is imperative for Large Language Models (LLMs) to effectively harness these resources to generate and convey recommendation outcomes to users. Liu \textit{et. al.}~\cite{liu2023conversational} have implemented a collaborative framework wherein the CRS model and LLMs are synergistically integrated to augment each other's functionality. Additional research~\cite{friedman2023leveraging,feng2023all} has segmented CRS into various sub-components or tasks, assigning roles such as dialogue management, execution, explanation, and ranking to LLMs, thereby facilitating interaction with users and recommendation systems. MINT~\cite{mysore2023large} conceptualizes CRS as a retrieval problem, subsequently refining LLMs for use as narrative-based retrieval agents. Similarly, CLLM4Rec~\cite{zhu2023collaborative} formalizes CRS as a generative recommendation, proposing the resolution of CRS through the direct application of fine-tuned LLMs.

\subsubsection{Fine-tuning Strategy}
The fine-tuning technique is critical in determining the ultimate quality of LLM-based CRS, with a well-designed and effective strategy significantly enhancing model performance and capabilities. Liu \textit{et. al.}~\cite{liu2023conversational} propose a sequential tuning paradigm for the collaboration between LLM and CRS in two ways, \textit{i.e.}, \textit{LLM assisting CRS} and \textit{CRS assisting LLM}. CLLM4Rec~\cite{zhu2023collaborative} regards CRS as a generative recommendation task and also adopts the sequential tuning strategy, where a pre-training stage to capture user/item collaborative information via language modeling objective and a fine-tuning stage to enhance recommendation generation. RecLLM~\cite{friedman2023leveraging} devises an integrated architecture with multiple modules powered by LLMs, where each module is optimized individually by the synthetic data from the user simulator. MINT~\cite{mysore2023large} formalizes CRS as a retrieval task and directly tunes the LLM via constructed instruction data. LLMCRS~\cite{feng2023all} relies on the ability of LLM to handle sub-tasks, and it optimizes the overall performance by jointly adapting the model through reinforcement learning for CRS performance feedback.

\subsection{Discussion}
In general, LLM-based CRS via prompt engineering facilitates the conversational recommendation through refined task decomposition and eliciting the reasoning abilities of LLMs with in-context learning, without the need for training and fine-tuning the LLMs. This allows for rapid adjustment and deployment across various conversational recommendation scenarios. Moreover, compared to conventional CRS, LLM possesses stronger content understanding and response generation abilities. Thus, LLM-based CRS could achieve better user interest capture and guidance, thereby elevating interaction efficiency and user experience. Meanwhile, Fine-tuning LLMs on conversational recommendation data enhances their ability to comprehend user-specific needs and preferences, thereby improving the accuracy and personalization of recommendations through better domain-specific language understanding. Moreover, the fine-tuned LLMs exhibit improved conversational fluency and naturalness. They demonstrate better intent understanding and response capabilities, streamlining prompt engineering processes and boosting efficiency. Additionally, they can effectively leverage historical interaction data for refined personalization, ensuring higher-quality interactions.

\subsubsection{Advantages}
Thanks to the extensive world knowledge, LLMs demonstrate enhanced proficiency in discerning user intent, comprehending content, and engaging in dialogue. Compared to conventional CRS, LLM-based CRS exhibits superior accuracy in intention recognition and swiftly identifies users' interests, thereby diminishing interaction rounds and associated costs. Moreover, leveraging LLMs' robust expressive capabilities, these methods markedly advance in response informativeness, offering elaborate item descriptions and credible explanations, while effectively navigating users through dialogues.

\subsubsection{Limitations}
Despite the promising performance achieved, existing LLM-based CRS methods still rely on the traditional conversational recommendation paradigms, primarily incorporating and aligning LLMs with conventional CRS methodologies via either prompt engineering or fine-tuning techniques. For instance, prompt engineering solutions focus on task decomposition for specific CRS scenarios, employing LLM as controller and sub-task handlers to deal with them separately, such as item ranking, dialogue management, response generation, and reflection, and interacting with recommendation engines to enable conversational recommendation. While fine-tuning solutions involve instruction tuning the LLM with the domain dataset to enable it to recommend and interact with CRS tasks. In summary, existing LLM-based CRSs are deficient in systemic innovation. Firstly, task decomposition and fine-tuning strategies based on specific needs cannot be generalized across more scenarios, resulting in inferior transferability of the existing solutions. Furthermore, these solutions lack advanced adaptive decision-making and process control capabilities, unable to adapt to users' rapidly changing and diverse needs.

\section{LLM-powered Recommendation Agent}~\label{sec:recagent}

The LLM-enhanced list-wise recommender systems (as shown in Section~\ref{sec:llm_enhanced_rec}) are facing the challenge of precisely identifying the users’ current needs by modeling with users' noisy interaction history (such as clicks, skips, ratings, and reviews).
Meanwhile, the LLM-based conversational recommender systems (as shown in Section~\ref{sec:llm-crs}) are deficient in adapting to users’ rapidly changing and diverse needs in various information-seeking scenarios by integrating LLMs with off-the-shelf recommenders to equip them with domain-specific knowledge.
In addition, due to the existence of scaling laws for LLMs~\cite{kaplan2020scaling}, we can anticipate further enhancement of LLM-based recommender systems in feature engineering, ranking performance, and conversational skills.
Driven by the powerful perception and reasoning abilities provided by LLMs, the previously mentioned two paths of LLM-based recommender systems (i.e.\ list-wise recommendation and conversational recommendation) converge at a point, called LLM-powered recommendation agents, that can maximally elicit the potential of LLMs under the scaling law.

An LLM-powered agent generally refers to an advanced AI system capable of perceiving its environment and taking autonomous actions based on the perceived information to achieve specific goals~\cite{franklin1996agent}.
Such an agent usually consists of several key components: perception, planning, memory, tool use, and actions~\cite{weng2023prompt,xi2023rise,wang2023survey}.
To this end, an LLM-powered recommendation agent is capable of perceiving and identifying the users' rapidly changing needs, autonomously planning the information-seeking tasks by decomposing and reflecting the tasks, retrieving the users' profiles and preferences from memory, invoking tools for searching and recommending, and taking actions by providing a ranking list of items, giving replies, or asking clarification questions.
An LLM-powered recommendation agent can avoid issues of inaccurate intent recognition by leveraging the LLM-enhanced list-wise recommender systems as a part of tools or memory. At the same time, it can also address the limitations of the LLM-based conversational recommender systems in generalizability across different scenarios and adaptability to diverse users' information needs.

In this Section, we first illustrate the architecture of an LLM-powered recommendation agent.
Then, we summarise the recent advances in LLM-based agents for recommendation.
We also define the future developmental stages of LLM-powered recommendation agents and analyze their capabilities at each stage.
We finally discuss some possible directions for LLM-powered recommendation agents.

\subsection{Architecture of LLM-Powered Recommendation Agent}

LLM-powered recommendation agents have been recently adapted from generic LLM-powered autonomous agents that frame an LLM as a powerful general problem solver, complemented by several key components, such as perception, planning, memory, and tool use, and actions~\cite{weng2023prompt,xi2023rise,wang2023survey}.
There exist several proof-of-concept projects to construct LLM-powered agents, such as AutoGPT\footnote{\url{https://github.com/Significant-Gravitas/AutoGPT}}, BabyAGI\footnote{\url{https://github.com/yoheinakajima/babyagi}}, MetaGPT\footnote{\url{https://github.com/geekan/MetaGPT}}~\cite{hong2023metagpt}, XAgent\footnote{\url{https://github.com/OpenBMB/XAgent}}, and GPTs\footnote{\url{https://openai.com/blog/introducing-gpts}}.
These LLM-powered agents have been proven to be powerful in achieving various real-world tasks, such as shopping online~\cite{yao2022webshop}, developing software~\cite{hong2023metagpt}, and video creation\footnote{\url{https://github.com/RayVentura/ShortGPT}}.
LLM-powered agents provide a generic solution for addressing users' complex information needs regarding recommendations.

\begin{figure}[t]
  \centering
  \includegraphics[width=0.86\linewidth]{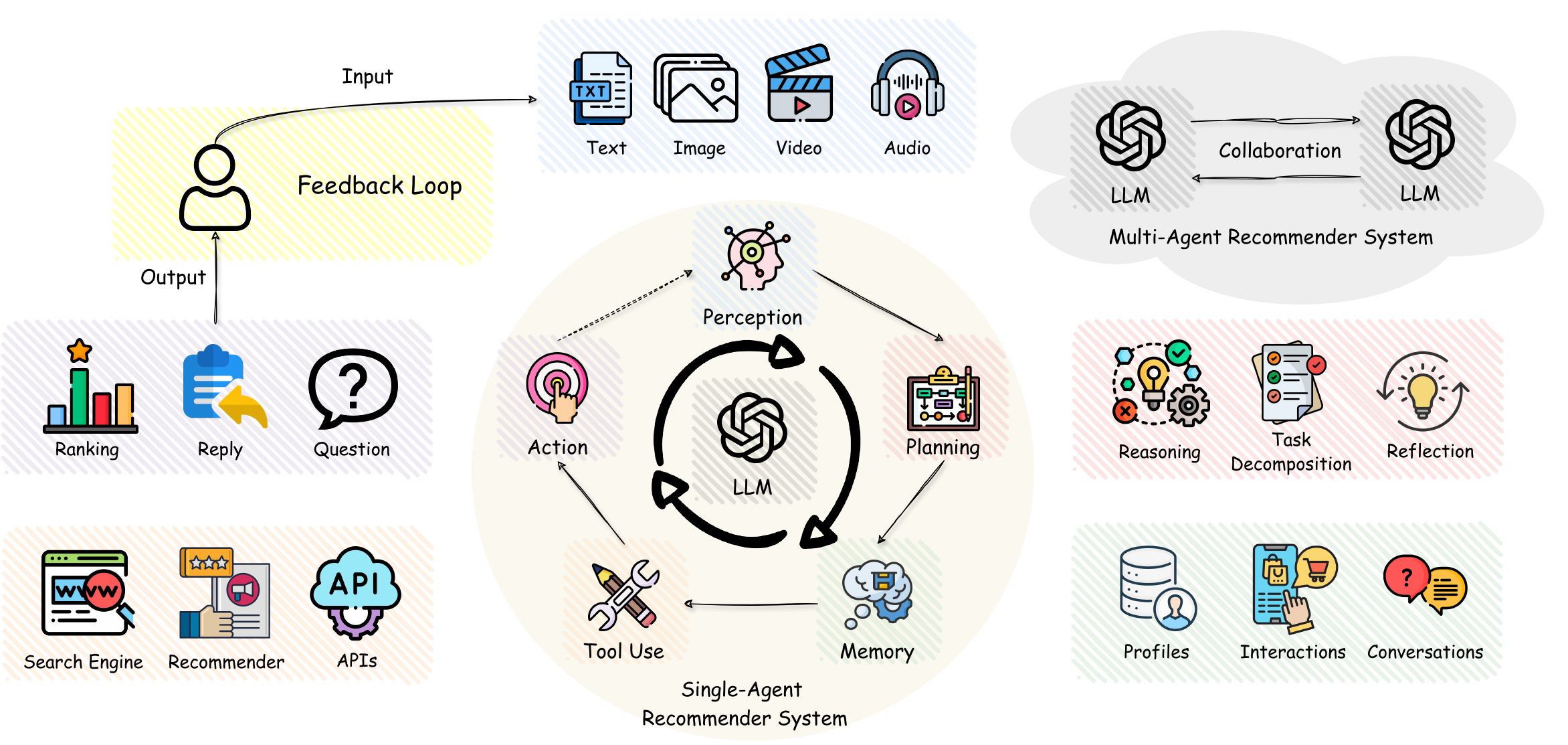}
  \caption{An overview of the LLM-powered recommendation agent.}
  \label{fig:generic-agent-framework}
\end{figure}

Fig.~\ref{fig:generic-agent-framework} depicts an overview of the LLM-powered recommendation agents.
In particular, an LLM-powered recommendation agent can engage in natural-language or multi-modal interactions, decompose the information-seeking task, retrieve knowledge, recall from memory, make informed decisions, and adapt to unfamiliar scenarios with its inherent generalization and adaptability~\cite{xi2023rise,wang2023survey}. 
In addition, LLM-powered agents can collaborate and divide tasks among each other, forming a multi-agent system~\cite{wooldridge2009introduction} to accomplish tasks jointly.
A single-agent recommender system leverages a single LLM as a controller to control the recommendation process, while a multi-agent recommender system leverages multiple LLMs as various roles (such as planning, criticizing, and reflection) in the recommendation process.
To this end, we define ``\textit{an LLM-powered recommendation agent as an advanced information-seeking system that is capable of perceiving its environment (including users and context),  autonomously planning the information-seeking tasks (such as identifying users' intention, decomposing the tasks, and reflection), retrieving memory (including profiles, interactions, and conversations), invoking tool use (such as search engines and recommenders), and taking actions (such as providing a ranking list of items, giving replies, or asking clarification questions) to effectively satisfy users' information needs with minimal user effort}''.

\subsection{Advances in LLM-powered Recommendation Agents}

Recent research on LLM-powered recommendation agents mainly focuses on the design of agent architectures.
In particular, LLM-powered recommendation agents can be roughly categorized into single-agent and multi-agents for recommendations.
Note that we mainly concentrate on recommender systems with LLM-powered agents in this survey, rather than LLM-powered user agents (such as RecAgent~\cite{wang2023large} and Agent4Rec~\cite{zhang2023agent4rec}) for user simulation.

\paragraph{Single-Agent for Recommendation}
Single-agent recommender systems resemble the LLM-based conversational recommender systems (detailed in Section~\ref{sec:llm-crs}) that consider off-the-shelf recommenders as a source of domain-specific knowledge.
Different from the LLM-based conversational recommender systems that heavily rely on prompt engineering with well-crafted prompt strategies, such as CoT~\cite{wei2022chain}, ToT~\cite{yao2023tree}, ReAct~\cite{yao2023react}, single-agent recommender systems emphasize the rational coordination of various components, such as perception, planning, memory, tools, and actions.
For instance, InteRecAgent~\cite{huang2023recommender} (\textbf{Inte}ractive \textbf{Rec}ommender \textbf{Agent}) proposed an efficient workflow for information-seeking task execution, incorporating key components such as a candidate memory bus, dynamic demonstration-augmented task planning, and reflection.
In particular, InteRecAgent introduced an efficient framework that employs LLMs as the brain and contains a minimal set of essential tools (an information query module to address the users' inquiries, an item retrieval tool to propose a list of item candidates and an item ranking tool to tailor personalized content for users) required to transform LLMs into agents.

\paragraph{Multi-Agents for Recommendation}

Multi-agent recommender systems are keen on designing different agent roles and facilitating their collaboration to accomplish a complex information-seeking task together.
Multi-agent systems have proven to be effective in reducing the difficulty of the task.
For instance, RAH~\cite{shu2023rah}  (Recommender system, Assistant, and Human) designed a framework with several LLM-powered agents, such as Perceive, Learn, Act, Critic, and Reflect, utilizing the so-called Learn-Act-Critic loop and a reflection mechanism for improving the alignment with user personalities.
Within this framework, the assistant facilitates two key workflows: (1)
\textbf{RecSys→Assistant→Human}:
This workflow focuses on the assistant filtering personalized recommendations for the end user.
(2) \textbf{Human→Assistant→RecSys}: This workflow enables the assistant to learn from user feedback and accordingly tune recommender systems.

\subsection{Development Levels}

The LLM-powered recommendation agent is a burgeoning new field that is still in rapid development.
Despite the above-mentioned PoC (Proof of Concept) attempts in LLM-powered recommendation agents, there are still many shortcomings.
For instance, perception modalities are singular, manually designing planning processes is inefficient, memory functions are underutilized, tools are not abundant, and actions are decoupled.
To better promote and guide the development of LLM-powered recommendation agents, Table~\ref{tab:levels} summarizes the four development levels of LLM-powered recommendation agents and the corresponding capabilities. 
Current agents for recommendation mostly fall in L1 and L2, only a few studies cover the capabilities of L3 or L4.
This highlights a need for further research to explore related directions in this domain.

\begin{table}[tb]
\centering
\caption{Different levels of intelligence for LLM-powered recommendation agents with CAPS.}
\label{tab:levels}
\resizebox{0.98\textwidth}{!}{
\begin{tabular}{l|p{4cm}|p{9cm}} 
\Xhline{1pt}
Level & Key Characteristics & Capabilities\\
\Xhline{0.5pt}
L1-Compliance & LLM-powered recommendation agents complete the recommendation task by following the instructions pre-defined by the users or the developers. &  \noindent$\bullet$ Uni-modal perception, primarily text-based. \newline \noindent$\bullet$ Crafted planning procedure. \newline \noindent$\bullet$ Short-term memory only within current conversational session. \newline \noindent$\bullet$ Limited tools, using only recommendation models and search engines. \newline \noindent$\bullet$ Decoupled actions, either ranking lists or textual responses. \\
\Xhline{0.5pt}
L2-Autonomy & LLM-powered recommendation agents autonomously plan the recommendation task using memory \& various tools, and iterate the plan based on both internal and external feedback until completion. & \noindent$\bullet$ Uni-modal perception, primarily text-based. \newline \noindent$\bullet$ Autonomous planning process with task decomposition and reflection. \newline \noindent$\bullet$ Short/long-term memory across conversational sessions. Extracted personalized memory from raw conversation history. \newline \noindent$\bullet$ A set of tools for a certain domain, including recommendation models, search engines, and online APIs. \newline \noindent$\bullet$ Explainable actions, such as ranking lists with explanations or informative textual responses. \\
\Xhline{0.5pt}
L3-Personification & LLM-powered recommendation agents are equipped with professional knowledge and skills (such as salespersons, tourist guides), and can proactively provide personalized recommendation services at appropriate times in various domains. & \noindent$\bullet$ Multi-modal perception, including text, image, video, and audio. \newline \noindent$\bullet$ Adaptive autonomous planning procedure leveraging techniques such as graphs. \newline \noindent$\bullet$ Comprehensive memory, including profiles, interactions, conversations, and extracted personalized memory. \newline \noindent$\bullet$ Various professional tools for different domains. \newline \noindent$\bullet$ Reactive \& proactive actions, including ranking lists with informative explanations, informative textual responses, and clarification questions. \\
\Xhline{0.5pt}
L4-Self-Evolution & LLM-powered recommendation agents are able to autonomously improve, adapt, or transform themselves over time, driven by internal feedback mechanisms or external stimuli. & \noindent$\bullet$ Mixture of multi-modal perception, including text, image, video, and audio. \newline \noindent$\bullet$ Adaptive autonomous planning procedure with task graph optimization. \newline \noindent$\bullet$ Comprehensive memory, including profiles, interactions, conversations, and extracted personalized memory. \newline \noindent$\bullet$ All available tools. \newline \noindent$\bullet$ Mixture of reactive \& proactive actions. \\
\Xhline{1pt}
\end{tabular}}
\end{table}

\subsection{Discussion}

Despite the above-mentioned initial attempts in the recommendation domain, there is still significant exploration needed to fully develop recommendation systems utilizing LLM-based agents.
This includes (1) establishing reasonable evaluation mechanisms for recommendation agents, (2) devising rational optimization mechanisms for recommendation agents, such as prompt engineering, optimizing agent structure/workflow, and fine-tuning language models (SFT, RLHF/RLAIF) to achieve parameter optimization, and (3) modeling more complex user behaviors within recommendation agents, including dialogue behaviors and click behaviors.

\section{Open problems and Future directions}

The remarkable progress made by the recommendation community over the past decades has facilitated the success of various great products.
Currently, we are on a brand new path that leads to the next generation of recommender systems.
Based on the lessons learned from history, we raise several problems and directions that may have a critical impact on the technical evolution along this path. 

\begin{itemize}[leftmargin=*]

\item \textbf{From Browsing to Experience}. Current recommender systems have primarily focused on presenting users with items they might like, while future recommender systems can transform this passive browsing into an immersive journey. 
Instead of an insipid information stream, an organic synthesis of information like a table d'hôte is much more charming and attractive.
Future research could explore how to collect, exploit, and aggregate information from diverse resources and modalities to facilitate dynamic, context-aware, and personalized experiences that anticipate and adapt to user needs in both real-time and long term.
This shall involve innovative interfaces and interaction designs that provide not just recommendations but engagement opportunities that resonate with users' evolving preferences and situational context.

\item \textbf{From Virtual to Reality} Recommender systems are predominantly operated in the digital domain, yet there lies immense potential in bridging this virtual interaction with the physical world. 
Future developments could enhance real-world experiences through augmented reality (AR) and virtual reality (VR), offering contextual recommendations that seamlessly blend digital content with physical environments. 
Moreover, the connection between recommender systems and physical assets through the Internet of Things (IoT) or smart environments would also pose great opportunities for embodied recommendation, which brings completely different user experience.
While this convergence promises significant advancements, it also presents challenges in ensuring seamless synchronization between virtual recommendations and real-world dynamics, which future research needs to address.

\item \textbf{From Perception to Cognition} Current recommender systems operate largely on perceptual data, recognizing patterns and correlations among user behaviors and preferences. 
The next leap involves transitioning towards cognitive models that understand and predict user motivations, desires, and cognitive states. 
With the capacity of LLMs for nuanced comprehension and generation, we can develop more sophisticated cognitive frameworks within recommender systems. 
These systems would not only recommend based on surface-level behavior but also infer deeper, more abstract user intentions and emotions by understanding and reasoning on long-term and even life-long user data.
Research in this direction may also integrate cognitive science principles and user psychology into the design and functionality of recommender systems, moving towards a more profound understanding of user cognition and experience.

\item \textbf{From Tool to Assistant} The role of recommender systems is gradually evolving from being static tools that suggest options to becoming lively assistants that build friendship and trust with users.
Leveraging the powerful audio and video generation abilities of multimodal large foundation models, future recommender systems can be designed to engage with users like intimates rather than machines. 
Researchers may focus on unifying recommender systems with digital assistants, which can provide users not just reactive but proactive and personalized digital companions.

\item \textbf{From Usability to Responsibility}. As recommender systems become deeply integrated into users' lives, ensuring responsible and ethical usage becomes indispensable.
Future research must address usability not only from a functional perspective but also from ethical, privacy, and bias mitigation standpoints. 
The broad applications of large foundation models have posed great risks related to data privacy, algorithmic bias, transparency, and accountability. 
Researchers need to develop frameworks that ensure fair and unbiased recommendations, protect user privacy, and provide transparency in how recommendation results are generated, exploited, and disseminated.
In this framework, AI-generated content should be verified and moderated, and proper watermarking techniques are necessary.
The misuse of future recommender systems may lead to inconspicuous distortion of public perception, which should be carefully prevented and traced.

\item \textbf{From Product to Ecosystem}. A future destination of recommender systems lies in their transition from standalone products to integral components of a broader digital ecosystem. 
In leveraging the general intelligence of foundation model capabilities, recommender agents may coordinate an interconnected network that shares insights and collaborates across different domains and platforms.
This involves not only interoperable standards and protocols that facilitate efficient and secure data exchange but also incentive mechanisms that ensure fair and inclusive profit sharing. 
This trend may ultimately evolve recommender systems into a unified digital ecosystem, which relies on responsible computing mechanisms that link numerous platforms and individuals in various channels.

\end{itemize}

\section{Conclusion}
Recommender systems play a crucial role as a bridge between humans and massive information, helping users obtain more useful information with less effort. As LLMs revolutionize the way humans engage with the online world, they also bring about new opportunities for exploring and improving recommender systems. At this pivotal point, this paper systematically reviews recommender systems' two main development paths: traditional list-wise recommendation and conversational recommendation (depicted in Fig.~\ref{fig:trend}), each representing a different approach to better human-computer interaction. After thorough analysis, we believe that ``all roads lead to Rome'', that is LLM-powered recommendation agents with more effective information and less interactive cost.
Finally, we point out several potential opportunities across the LLM era for future directions.

\bibliographystyle{ACM-Reference-Format}
\bibliography{sample-manuscript}


\begin{thebibliography}{189}


\ifx \showCODEN    \undefined \def \showCODEN     #1{\unskip}     \fi
\ifx \showDOI      \undefined \def \showDOI       #1{#1}\fi
\ifx \showISBNx    \undefined \def \showISBNx     #1{\unskip}     \fi
\ifx \showISBNxiii \undefined \def \showISBNxiii  #1{\unskip}     \fi
\ifx \showISSN     \undefined \def \showISSN      #1{\unskip}     \fi
\ifx \showLCCN     \undefined \def \showLCCN      #1{\unskip}     \fi
\ifx \shownote     \undefined \def \shownote      #1{#1}          \fi
\ifx \showarticletitle \undefined \def \showarticletitle #1{#1}   \fi
\ifx \showURL      \undefined \def \showURL       {\relax}        \fi
\providecommand\bibfield[2]{#2}
\providecommand\bibinfo[2]{#2}
\providecommand\natexlab[1]{#1}
\providecommand\showeprint[2][]{arXiv:#2}

\bibitem[Abdool et~al\mbox{.}(2020)]%
        {abdool2020managing}
\bibfield{author}{\bibinfo{person}{Mustafa Abdool}, \bibinfo{person}{Malay Haldar}, \bibinfo{person}{Prashant Ramanathan}, \bibinfo{person}{Tyler Sax}, \bibinfo{person}{Lanbo Zhang}, \bibinfo{person}{Aamir Manaswala}, \bibinfo{person}{Lynn Yang}, \bibinfo{person}{Bradley Turnbull}, \bibinfo{person}{Qing Zhang}, {and} \bibinfo{person}{Thomas Legrand}.} \bibinfo{year}{2020}\natexlab{}.
\newblock \showarticletitle{Managing diversity in airbnb search}. In \bibinfo{booktitle}{\emph{Proceedings of the 26th ACM SIGKDD International Conference on Knowledge Discovery \& Data Mining}}. \bibinfo{pages}{2952--2960}.
\newblock


\bibitem[Ai et~al\mbox{.}(2018)]%
        {ai2018learning}
\bibfield{author}{\bibinfo{person}{Qingyao Ai}, \bibinfo{person}{Keping Bi}, \bibinfo{person}{Jiafeng Guo}, {and} \bibinfo{person}{W~Bruce Croft}.} \bibinfo{year}{2018}\natexlab{}.
\newblock \showarticletitle{Learning a deep listwise context model for ranking refinement}. In \bibinfo{booktitle}{\emph{The 41st international ACM SIGIR conference on research \& development in information retrieval}}. \bibinfo{pages}{135--144}.
\newblock


\bibitem[Bao et~al\mbox{.}(2023a)]%
        {bao2023bi}
\bibfield{author}{\bibinfo{person}{Keqin Bao}, \bibinfo{person}{Jizhi Zhang}, \bibinfo{person}{Wenjie Wang}, \bibinfo{person}{Yang Zhang}, \bibinfo{person}{Zhengyi Yang}, \bibinfo{person}{Yancheng Luo}, \bibinfo{person}{Fuli Feng}, \bibinfo{person}{Xiangnaan He}, {and} \bibinfo{person}{Qi Tian}.} \bibinfo{year}{2023}\natexlab{a}.
\newblock \showarticletitle{A bi-step grounding paradigm for large language models in recommendation systems}.
\newblock \bibinfo{journal}{\emph{arXiv preprint arXiv:2308.08434}} (\bibinfo{year}{2023}).
\newblock


\bibitem[Bao et~al\mbox{.}(2023b)]%
        {bao2023tallrec}
\bibfield{author}{\bibinfo{person}{Keqin Bao}, \bibinfo{person}{Jizhi Zhang}, \bibinfo{person}{Yang Zhang}, \bibinfo{person}{Wenjie Wang}, \bibinfo{person}{Fuli Feng}, {and} \bibinfo{person}{Xiangnan He}.} \bibinfo{year}{2023}\natexlab{b}.
\newblock \showarticletitle{TALLRec: An Effective and Efficient Tuning Framework to Align Large Language Model with Recommendation}.
\newblock \bibinfo{journal}{\emph{arXiv preprint arXiv:2305.00447}} (\bibinfo{year}{2023}).
\newblock


\bibitem[Bawden and Robinson(2020)]%
        {bawden2020information}
\bibfield{author}{\bibinfo{person}{David Bawden} {and} \bibinfo{person}{Lyn Robinson}.} \bibinfo{year}{2020}\natexlab{}.
\newblock \showarticletitle{Information overload: An overview}.
\newblock  (\bibinfo{year}{2020}).
\newblock


\bibitem[Bobadilla et~al\mbox{.}(2013)]%
        {bobadilla2013recommender}
\bibfield{author}{\bibinfo{person}{Jes{\'u}s Bobadilla}, \bibinfo{person}{Fernando Ortega}, \bibinfo{person}{Antonio Hernando}, {and} \bibinfo{person}{Abraham Guti{\'e}rrez}.} \bibinfo{year}{2013}\natexlab{}.
\newblock \showarticletitle{Recommender systems survey}.
\newblock \bibinfo{journal}{\emph{Knowledge-based systems}}  \bibinfo{volume}{46} (\bibinfo{year}{2013}), \bibinfo{pages}{109--132}.
\newblock


\bibitem[Brown et~al\mbox{.}(2020)]%
        {brown2020language}
\bibfield{author}{\bibinfo{person}{Tom Brown}, \bibinfo{person}{Benjamin Mann}, \bibinfo{person}{Nick Ryder}, \bibinfo{person}{Melanie Subbiah}, \bibinfo{person}{Jared~D Kaplan}, \bibinfo{person}{Prafulla Dhariwal}, \bibinfo{person}{Arvind Neelakantan}, \bibinfo{person}{Pranav Shyam}, \bibinfo{person}{Girish Sastry}, \bibinfo{person}{Amanda Askell}, {et~al\mbox{.}}} \bibinfo{year}{2020}\natexlab{}.
\newblock \showarticletitle{Language models are few-shot learners}.
\newblock \bibinfo{journal}{\emph{Advances in neural information processing systems}}  \bibinfo{volume}{33} (\bibinfo{year}{2020}), \bibinfo{pages}{1877--1901}.
\newblock


\bibitem[Bubeck et~al\mbox{.}(2023)]%
        {bubeck2023sparks}
\bibfield{author}{\bibinfo{person}{S{\'e}bastien Bubeck}, \bibinfo{person}{Varun Chandrasekaran}, \bibinfo{person}{Ronen Eldan}, \bibinfo{person}{Johannes Gehrke}, \bibinfo{person}{Eric Horvitz}, \bibinfo{person}{Ece Kamar}, \bibinfo{person}{Peter Lee}, \bibinfo{person}{Yin~Tat Lee}, \bibinfo{person}{Yuanzhi Li}, \bibinfo{person}{Scott Lundberg}, {et~al\mbox{.}}} \bibinfo{year}{2023}\natexlab{}.
\newblock \showarticletitle{Sparks of artificial general intelligence: Early experiments with gpt-4}.
\newblock \bibinfo{journal}{\emph{arXiv preprint arXiv:2303.12712}} (\bibinfo{year}{2023}).
\newblock


\bibitem[Caruana(1997)]%
        {caruana1997multitask}
\bibfield{author}{\bibinfo{person}{Rich Caruana}.} \bibinfo{year}{1997}\natexlab{}.
\newblock \showarticletitle{Multitask learning}.
\newblock \bibinfo{journal}{\emph{Machine learning}}  \bibinfo{volume}{28} (\bibinfo{year}{1997}), \bibinfo{pages}{41--75}.
\newblock


\bibitem[Chajewska et~al\mbox{.}(1998)]%
        {chajewska1998utility}
\bibfield{author}{\bibinfo{person}{Urszula Chajewska}, \bibinfo{person}{Lise Getoor}, \bibinfo{person}{Joseph Norman}, {and} \bibinfo{person}{Yuval Shahar}.} \bibinfo{year}{1998}\natexlab{}.
\newblock \showarticletitle{Utility Elicitation as a Classification Problem.}. In \bibinfo{booktitle}{\emph{UAI}}. \bibinfo{pages}{79--88}.
\newblock


\bibitem[Chen et~al\mbox{.}(2019b)]%
        {chen2019large}
\bibfield{author}{\bibinfo{person}{Haokun Chen}, \bibinfo{person}{Xinyi Dai}, \bibinfo{person}{Han Cai}, \bibinfo{person}{Weinan Zhang}, \bibinfo{person}{Xuejian Wang}, \bibinfo{person}{Ruiming Tang}, \bibinfo{person}{Yuzhou Zhang}, {and} \bibinfo{person}{Yong Yu}.} \bibinfo{year}{2019}\natexlab{b}.
\newblock \showarticletitle{Large-scale interactive recommendation with tree-structured policy gradient}. In \bibinfo{booktitle}{\emph{Proceedings of the AAAI conference on artificial intelligence}}, Vol.~\bibinfo{volume}{33}. \bibinfo{pages}{3312--3320}.
\newblock


\bibitem[Chen and Pu(2012)]%
        {chen2012critiquing}
\bibfield{author}{\bibinfo{person}{Li Chen} {and} \bibinfo{person}{Pearl Pu}.} \bibinfo{year}{2012}\natexlab{}.
\newblock \showarticletitle{Critiquing-based recommenders: survey and emerging trends}.
\newblock \bibinfo{journal}{\emph{User Modeling and User-Adapted Interaction}}  \bibinfo{volume}{22} (\bibinfo{year}{2012}), \bibinfo{pages}{125--150}.
\newblock


\bibitem[Chen et~al\mbox{.}(2019a)]%
        {chen2019top}
\bibfield{author}{\bibinfo{person}{Minmin Chen}, \bibinfo{person}{Alex Beutel}, \bibinfo{person}{Paul Covington}, \bibinfo{person}{Sagar Jain}, \bibinfo{person}{Francois Belletti}, {and} \bibinfo{person}{Ed~H Chi}.} \bibinfo{year}{2019}\natexlab{a}.
\newblock \showarticletitle{Top-k off-policy correction for a REINFORCE recommender system}. In \bibinfo{booktitle}{\emph{Proceedings of the Twelfth ACM International Conference on Web Search and Data Mining}}. \bibinfo{pages}{456--464}.
\newblock


\bibitem[Chen et~al\mbox{.}(2019c)]%
        {chen2019towards}
\bibfield{author}{\bibinfo{person}{Qibin Chen}, \bibinfo{person}{Junyang Lin}, \bibinfo{person}{Yichang Zhang}, \bibinfo{person}{Ming Ding}, \bibinfo{person}{Yukuo Cen}, \bibinfo{person}{Hongxia Yang}, {and} \bibinfo{person}{Jie Tang}.} \bibinfo{year}{2019}\natexlab{c}.
\newblock \showarticletitle{Towards knowledge-based recommender dialog system}.
\newblock \bibinfo{journal}{\emph{arXiv preprint arXiv:1908.05391}} (\bibinfo{year}{2019}).
\newblock


\bibitem[Chen et~al\mbox{.}(2021)]%
        {chen2021end}
\bibfield{author}{\bibinfo{person}{Qiwei Chen}, \bibinfo{person}{Changhua Pei}, \bibinfo{person}{Shanshan Lv}, \bibinfo{person}{Chao Li}, \bibinfo{person}{Junfeng Ge}, {and} \bibinfo{person}{Wenwu Ou}.} \bibinfo{year}{2021}\natexlab{}.
\newblock \showarticletitle{End-to-end user behavior retrieval in click-through rateprediction model}.
\newblock \bibinfo{journal}{\emph{arXiv preprint arXiv:2108.04468}} (\bibinfo{year}{2021}).
\newblock


\bibitem[Chen et~al\mbox{.}(2023)]%
        {chen2023tbin}
\bibfield{author}{\bibinfo{person}{Shuwei Chen}, \bibinfo{person}{Xiang Li}, \bibinfo{person}{Jian Dong}, \bibinfo{person}{Jin Zhang}, \bibinfo{person}{Yongkang Wang}, {and} \bibinfo{person}{Xingxing Wang}.} \bibinfo{year}{2023}\natexlab{}.
\newblock \showarticletitle{TBIN: Modeling Long Textual Behavior Data for CTR Prediction}.
\newblock \bibinfo{journal}{\emph{arXiv preprint arXiv:2308.08483}} (\bibinfo{year}{2023}).
\newblock


\bibitem[Chowdhery et~al\mbox{.}(2023)]%
        {chowdhery2023palm}
\bibfield{author}{\bibinfo{person}{Aakanksha Chowdhery}, \bibinfo{person}{Sharan Narang}, \bibinfo{person}{Jacob Devlin}, \bibinfo{person}{Maarten Bosma}, \bibinfo{person}{Gaurav Mishra}, \bibinfo{person}{Adam Roberts}, \bibinfo{person}{Paul Barham}, \bibinfo{person}{Hyung~Won Chung}, \bibinfo{person}{Charles Sutton}, \bibinfo{person}{Sebastian Gehrmann}, {et~al\mbox{.}}} \bibinfo{year}{2023}\natexlab{}.
\newblock \showarticletitle{Palm: Scaling language modeling with pathways}.
\newblock \bibinfo{journal}{\emph{Journal of Machine Learning Research}} \bibinfo{volume}{24}, \bibinfo{number}{240} (\bibinfo{year}{2023}), \bibinfo{pages}{1--113}.
\newblock


\bibitem[Christakopoulou et~al\mbox{.}(2018)]%
        {christakopoulou2018q}
\bibfield{author}{\bibinfo{person}{Konstantina Christakopoulou}, \bibinfo{person}{Alex Beutel}, \bibinfo{person}{Rui Li}, \bibinfo{person}{Sagar Jain}, {and} \bibinfo{person}{Ed~H Chi}.} \bibinfo{year}{2018}\natexlab{}.
\newblock \showarticletitle{Q\&R: A two-stage approach toward interactive recommendation}. In \bibinfo{booktitle}{\emph{Proceedings of the 24th ACM SIGKDD International Conference on Knowledge Discovery \& Data Mining}}. \bibinfo{pages}{139--148}.
\newblock


\bibitem[Christakopoulou et~al\mbox{.}(2016)]%
        {christakopoulou2016towards}
\bibfield{author}{\bibinfo{person}{Konstantina Christakopoulou}, \bibinfo{person}{Filip Radlinski}, {and} \bibinfo{person}{Katja Hofmann}.} \bibinfo{year}{2016}\natexlab{}.
\newblock \showarticletitle{Towards conversational recommender systems}. In \bibinfo{booktitle}{\emph{Proceedings of the 22nd ACM SIGKDD international conference on knowledge discovery and data mining}}. \bibinfo{pages}{815--824}.
\newblock


\bibitem[Covington et~al\mbox{.}(2016)]%
        {covington2016deep}
\bibfield{author}{\bibinfo{person}{Paul Covington}, \bibinfo{person}{Jay Adams}, {and} \bibinfo{person}{Emre Sargin}.} \bibinfo{year}{2016}\natexlab{}.
\newblock \showarticletitle{Deep neural networks for youtube recommendations}. In \bibinfo{booktitle}{\emph{Proceedings of the 10th ACM conference on recommender systems}}. \bibinfo{pages}{191--198}.
\newblock


\bibitem[Cybenko(1989)]%
        {cybenko1989approximation}
\bibfield{author}{\bibinfo{person}{George Cybenko}.} \bibinfo{year}{1989}\natexlab{}.
\newblock \showarticletitle{Approximation by superpositions of a sigmoidal function}.
\newblock \bibinfo{journal}{\emph{Mathematics of control, signals and systems}} \bibinfo{volume}{2}, \bibinfo{number}{4} (\bibinfo{year}{1989}), \bibinfo{pages}{303--314}.
\newblock


\bibitem[Devlin et~al\mbox{.}(2018)]%
        {devlin2018bert}
\bibfield{author}{\bibinfo{person}{Jacob Devlin}, \bibinfo{person}{Ming-Wei Chang}, \bibinfo{person}{Kenton Lee}, {and} \bibinfo{person}{Kristina Toutanova}.} \bibinfo{year}{2018}\natexlab{}.
\newblock \showarticletitle{Bert: Pre-training of deep bidirectional transformers for language understanding}.
\newblock \bibinfo{journal}{\emph{arXiv preprint arXiv:1810.04805}} (\bibinfo{year}{2018}).
\newblock


\bibitem[Du et~al\mbox{.}(2024)]%
        {disco}
\bibfield{author}{\bibinfo{person}{Kounianhua Du}, \bibinfo{person}{Jizheng Chen}, \bibinfo{person}{Jianghao Lin}, \bibinfo{person}{Yunjia Xi}, \bibinfo{person}{Hangyu Wang}, \bibinfo{person}{Xinyi Dai}, \bibinfo{person}{Bo Chen}, \bibinfo{person}{Ruiming Tang}, {and} \bibinfo{person}{Weinan Zhang}.} \bibinfo{year}{2024}\natexlab{}.
\newblock \showarticletitle{DisCo: Towards Harmonious Disentanglement and Collaboration between Tabular and Semantic Space for Recommendation}.
\newblock \bibinfo{journal}{\emph{arXiv preprint arXiv:2406.00011}} (\bibinfo{year}{2024}).
\newblock


\bibitem[Du et~al\mbox{.}(2022)]%
        {du2022glm}
\bibfield{author}{\bibinfo{person}{Zhengxiao Du}, \bibinfo{person}{Yujie Qian}, \bibinfo{person}{Xiao Liu}, \bibinfo{person}{Ming Ding}, \bibinfo{person}{Jiezhong Qiu}, \bibinfo{person}{Zhilin Yang}, {and} \bibinfo{person}{Jie Tang}.} \bibinfo{year}{2022}\natexlab{}.
\newblock \showarticletitle{GLM: General Language Model Pretraining with Autoregressive Blank Infilling}. In \bibinfo{booktitle}{\emph{Proceedings of the 60th Annual Meeting of the Association for Computational Linguistics (Volume 1: Long Papers)}}. \bibinfo{pages}{320--335}.
\newblock


\bibitem[Fan et~al\mbox{.}(2023)]%
        {fan2023recommender}
\bibfield{author}{\bibinfo{person}{Wenqi Fan}, \bibinfo{person}{Zihuai Zhao}, \bibinfo{person}{Jiatong Li}, \bibinfo{person}{Yunqing Liu}, \bibinfo{person}{Xiaowei Mei}, \bibinfo{person}{Yiqi Wang}, \bibinfo{person}{Zhen Wen}, \bibinfo{person}{Fei Wang}, \bibinfo{person}{Xiangyu Zhao}, \bibinfo{person}{Jiliang Tang}, {and} \bibinfo{person}{Qing Li}.} \bibinfo{year}{2023}\natexlab{}.
\newblock \bibinfo{title}{Recommender Systems in the Era of Large Language Models (LLMs)}.
\newblock
\newblock
\showeprint[arxiv]{2307.02046}~[cs.IR]


\bibitem[Feng et~al\mbox{.}(2023)]%
        {feng2023all}
\bibfield{author}{\bibinfo{person}{Yue Feng}, \bibinfo{person}{Shuchang Liu}, \bibinfo{person}{Zhenghai Xue}, \bibinfo{person}{Qingpeng Cai}, \bibinfo{person}{Lantao Hu}, \bibinfo{person}{Peng Jiang}, \bibinfo{person}{Kun Gai}, {and} \bibinfo{person}{Fei Sun}.} \bibinfo{year}{2023}\natexlab{}.
\newblock \showarticletitle{A Large Language Model Enhanced Conversational Recommender System}.
\newblock \bibinfo{journal}{\emph{ArXiv}}  \bibinfo{volume}{abs/2308.06212} (\bibinfo{year}{2023}).
\newblock


\bibitem[Feng et~al\mbox{.}(2018)]%
        {feng2018greedy}
\bibfield{author}{\bibinfo{person}{Yue Feng}, \bibinfo{person}{Jun Xu}, \bibinfo{person}{Yanyan Lan}, \bibinfo{person}{Jiafeng Guo}, \bibinfo{person}{Wei Zeng}, {and} \bibinfo{person}{Xueqi Cheng}.} \bibinfo{year}{2018}\natexlab{}.
\newblock \showarticletitle{From greedy selection to exploratory decision-making: Diverse ranking with policy-value networks}. In \bibinfo{booktitle}{\emph{The 41st international ACM SIGIR conference on research \& development in information retrieval}}. \bibinfo{pages}{125--134}.
\newblock


\bibitem[Franklin and Graesser(1996)]%
        {franklin1996agent}
\bibfield{author}{\bibinfo{person}{Stan Franklin} {and} \bibinfo{person}{Art Graesser}.} \bibinfo{year}{1996}\natexlab{}.
\newblock \showarticletitle{Is it an Agent, or just a Program?: A Taxonomy for Autonomous Agents}. In \bibinfo{booktitle}{\emph{International workshop on agent theories, architectures, and languages}}. Springer, \bibinfo{pages}{21--35}.
\newblock


\bibitem[Friedman et~al\mbox{.}(2023)]%
        {friedman2023leveraging}
\bibfield{author}{\bibinfo{person}{Luke Friedman}, \bibinfo{person}{Sameer Ahuja}, \bibinfo{person}{David Allen}, \bibinfo{person}{Zhenning Tan}, \bibinfo{person}{Hakim Sidahmed}, \bibinfo{person}{Changbo Long}, \bibinfo{person}{Jun Xie}, \bibinfo{person}{Gabriel Schubiner}, \bibinfo{person}{Ajay Patel}, \bibinfo{person}{Harsh Lara}, \bibinfo{person}{Brian Chu}, \bibinfo{person}{Zexiang Chen}, {and} \bibinfo{person}{Manoj Tiwari}.} \bibinfo{year}{2023}\natexlab{}.
\newblock \showarticletitle{Leveraging Large Language Models in Conversational Recommender Systems}.
\newblock \bibinfo{journal}{\emph{ArXiv}}  \bibinfo{volume}{abs/2305.07961} (\bibinfo{year}{2023}).
\newblock


\bibitem[Gao et~al\mbox{.}(2021)]%
        {gao2021advances}
\bibfield{author}{\bibinfo{person}{Chongming Gao}, \bibinfo{person}{Wenqiang Lei}, \bibinfo{person}{Xiangnan He}, \bibinfo{person}{Maarten de Rijke}, {and} \bibinfo{person}{Tat-Seng Chua}.} \bibinfo{year}{2021}\natexlab{}.
\newblock \showarticletitle{Advances and challenges in conversational recommender systems: A survey}.
\newblock \bibinfo{journal}{\emph{AI Open}}  \bibinfo{volume}{2} (\bibinfo{year}{2021}), \bibinfo{pages}{100--126}.
\newblock


\bibitem[Gao et~al\mbox{.}(2024)]%
        {gao2024llm}
\bibfield{author}{\bibinfo{person}{Jingtong Gao}, \bibinfo{person}{Bo Chen}, \bibinfo{person}{Xiangyu Zhao}, \bibinfo{person}{Weiwen Liu}, \bibinfo{person}{Xiangyang Li}, \bibinfo{person}{Yichao Wang}, \bibinfo{person}{Zijian Zhang}, \bibinfo{person}{Wanyu Wang}, \bibinfo{person}{Yuyang Ye}, \bibinfo{person}{Shanru Lin}, {et~al\mbox{.}}} \bibinfo{year}{2024}\natexlab{}.
\newblock \showarticletitle{LLM-enhanced Reranking in Recommender Systems}.
\newblock \bibinfo{journal}{\emph{arXiv preprint arXiv:2406.12433}} (\bibinfo{year}{2024}).
\newblock


\bibitem[Gao et~al\mbox{.}(2023)]%
        {gao2023chat}
\bibfield{author}{\bibinfo{person}{Yunfan Gao}, \bibinfo{person}{Tao Sheng}, \bibinfo{person}{Youlin Xiang}, \bibinfo{person}{Yun Xiong}, \bibinfo{person}{Haofen Wang}, {and} \bibinfo{person}{Jiawei Zhang}.} \bibinfo{year}{2023}\natexlab{}.
\newblock \showarticletitle{Chat-REC: Towards Interactive and Explainable LLMs-Augmented Recommender System}.
\newblock \bibinfo{journal}{\emph{ArXiv}}  \bibinfo{volume}{abs/2303.14524} (\bibinfo{year}{2023}).
\newblock


\bibitem[Gong et~al\mbox{.}(2023)]%
        {gong2023unified}
\bibfield{author}{\bibinfo{person}{Yuqi Gong}, \bibinfo{person}{Xichen Ding}, \bibinfo{person}{Yehui Su}, \bibinfo{person}{Kaiming Shen}, \bibinfo{person}{Zhongyi Liu}, {and} \bibinfo{person}{Guannan Zhang}.} \bibinfo{year}{2023}\natexlab{}.
\newblock \showarticletitle{An Unified Search and Recommendation Foundation Model for Cold-Start Scenario}. In \bibinfo{booktitle}{\emph{Proceedings of the 32nd ACM International Conference on Information and Knowledge Management}}. \bibinfo{pages}{4595--4601}.
\newblock


\bibitem[Gong et~al\mbox{.}(2020)]%
        {gong2020edgerec}
\bibfield{author}{\bibinfo{person}{Yu Gong}, \bibinfo{person}{Ziwen Jiang}, \bibinfo{person}{Yufei Feng}, \bibinfo{person}{Binbin Hu}, \bibinfo{person}{Kaiqi Zhao}, \bibinfo{person}{Qingwen Liu}, {and} \bibinfo{person}{Wenwu Ou}.} \bibinfo{year}{2020}\natexlab{}.
\newblock \showarticletitle{EdgeRec: recommender system on edge in Mobile Taobao}. In \bibinfo{booktitle}{\emph{Proceedings of the 29th ACM International Conference on Information \& Knowledge Management}}. \bibinfo{pages}{2477--2484}.
\newblock


\bibitem[Graus and Willemsen(2015)]%
        {graus2015improving}
\bibfield{author}{\bibinfo{person}{Mark~P Graus} {and} \bibinfo{person}{Martijn~C Willemsen}.} \bibinfo{year}{2015}\natexlab{}.
\newblock \showarticletitle{Improving the user experience during cold start through choice-based preference elicitation}. In \bibinfo{booktitle}{\emph{Proceedings of the 9th ACM Conference on Recommender Systems}}. \bibinfo{pages}{273--276}.
\newblock


\bibitem[Guo et~al\mbox{.}(2017)]%
        {guo2017deepfm}
\bibfield{author}{\bibinfo{person}{Huifeng Guo}, \bibinfo{person}{Ruiming Tang}, \bibinfo{person}{Yunming Ye}, \bibinfo{person}{Zhenguo Li}, {and} \bibinfo{person}{Xiuqiang He}.} \bibinfo{year}{2017}\natexlab{}.
\newblock \showarticletitle{DeepFM: a factorization-machine based neural network for CTR prediction}.
\newblock \bibinfo{journal}{\emph{arXiv preprint arXiv:1703.04247}} (\bibinfo{year}{2017}).
\newblock


\bibitem[Harte et~al\mbox{.}(2023)]%
        {harte2023leveraging}
\bibfield{author}{\bibinfo{person}{Jesse Harte}, \bibinfo{person}{Wouter Zorgdrager}, \bibinfo{person}{Panos Louridas}, \bibinfo{person}{Asterios Katsifodimos}, \bibinfo{person}{Dietmar Jannach}, {and} \bibinfo{person}{Marios Fragkoulis}.} \bibinfo{year}{2023}\natexlab{}.
\newblock \showarticletitle{Leveraging large language models for sequential recommendation}. In \bibinfo{booktitle}{\emph{Proceedings of the 17th ACM Conference on Recommender Systems}}. \bibinfo{pages}{1096--1102}.
\newblock


\bibitem[He et~al\mbox{.}(2016)]%
        {he2016interactive}
\bibfield{author}{\bibinfo{person}{Chen He}, \bibinfo{person}{Denis Parra}, {and} \bibinfo{person}{Katrien Verbert}.} \bibinfo{year}{2016}\natexlab{}.
\newblock \showarticletitle{Interactive recommender systems: A survey of the state of the art and future research challenges and opportunities}.
\newblock \bibinfo{journal}{\emph{Expert Systems with Applications}}  \bibinfo{volume}{56} (\bibinfo{year}{2016}), \bibinfo{pages}{9--27}.
\newblock


\bibitem[He et~al\mbox{.}(2023a)]%
        {he2023survey}
\bibfield{author}{\bibinfo{person}{Zhicheng He}, \bibinfo{person}{Weiwen Liu}, \bibinfo{person}{Wei Guo}, \bibinfo{person}{Jiarui Qin}, \bibinfo{person}{Yingxue Zhang}, \bibinfo{person}{Yaochen Hu}, {and} \bibinfo{person}{Ruiming Tang}.} \bibinfo{year}{2023}\natexlab{a}.
\newblock \showarticletitle{A Survey on User Behavior Modeling in Recommender Systems}.
\newblock \bibinfo{journal}{\emph{arXiv preprint arXiv:2302.11087}} (\bibinfo{year}{2023}).
\newblock


\bibitem[He et~al\mbox{.}(2023b)]%
        {he2023large}
\bibfield{author}{\bibinfo{person}{Zhankui He}, \bibinfo{person}{Zhouhang Xie}, \bibinfo{person}{Rahul Jha}, \bibinfo{person}{Harald Steck}, \bibinfo{person}{Dawen Liang}, \bibinfo{person}{Yesu Feng}, \bibinfo{person}{Bodhisattwa~Prasad Majumder}, \bibinfo{person}{Nathan Kallus}, {and} \bibinfo{person}{Julian McAuley}.} \bibinfo{year}{2023}\natexlab{b}.
\newblock \showarticletitle{Large language models as zero-shot conversational recommenders}. In \bibinfo{booktitle}{\emph{Proceedings of the 32nd ACM international conference on information and knowledge management}}. \bibinfo{pages}{720--730}.
\newblock


\bibitem[Ho and Lim(2023)]%
        {ho2023utilizing}
\bibfield{author}{\bibinfo{person}{Ngai~Lam Ho} {and} \bibinfo{person}{Kwan~Hui Lim}.} \bibinfo{year}{2023}\natexlab{}.
\newblock \showarticletitle{Utilizing Language Models for Tour Itinerary Recommendation}.
\newblock \bibinfo{journal}{\emph{arXiv preprint arXiv:2311.12355}} (\bibinfo{year}{2023}).
\newblock


\bibitem[Hong et~al\mbox{.}(2023)]%
        {hong2023metagpt}
\bibfield{author}{\bibinfo{person}{Sirui Hong}, \bibinfo{person}{Xiawu Zheng}, \bibinfo{person}{Jonathan Chen}, \bibinfo{person}{Yuheng Cheng}, \bibinfo{person}{Jinlin Wang}, \bibinfo{person}{Ceyao Zhang}, \bibinfo{person}{Zili Wang}, \bibinfo{person}{Steven Ka~Shing Yau}, \bibinfo{person}{Zijuan Lin}, \bibinfo{person}{Liyang Zhou}, {et~al\mbox{.}}} \bibinfo{year}{2023}\natexlab{}.
\newblock \showarticletitle{Metagpt: Meta programming for multi-agent collaborative framework}.
\newblock \bibinfo{journal}{\emph{arXiv preprint arXiv:2308.00352}} (\bibinfo{year}{2023}).
\newblock


\bibitem[Hu et~al\mbox{.}(2018)]%
        {hu2018squeeze}
\bibfield{author}{\bibinfo{person}{Jie Hu}, \bibinfo{person}{Li Shen}, {and} \bibinfo{person}{Gang Sun}.} \bibinfo{year}{2018}\natexlab{}.
\newblock \showarticletitle{Squeeze-and-excitation networks}. In \bibinfo{booktitle}{\emph{Proceedings of the IEEE conference on computer vision and pattern recognition}}. \bibinfo{pages}{7132--7141}.
\newblock


\bibitem[Huang et~al\mbox{.}(2020)]%
        {huang2020personalized}
\bibfield{author}{\bibinfo{person}{Jinhong Huang}, \bibinfo{person}{Yang Li}, \bibinfo{person}{Shan Sun}, \bibinfo{person}{Bufeng Zhang}, {and} \bibinfo{person}{Jin Huang}.} \bibinfo{year}{2020}\natexlab{}.
\newblock \showarticletitle{Personalized flight itinerary ranking at fliggy}. In \bibinfo{booktitle}{\emph{Proceedings of the 29th ACM International Conference on Information \& Knowledge Management}}. \bibinfo{pages}{2541--2548}.
\newblock


\bibitem[Huang et~al\mbox{.}(2019)]%
        {huang2019fibinet}
\bibfield{author}{\bibinfo{person}{Tongwen Huang}, \bibinfo{person}{Zhiqi Zhang}, {and} \bibinfo{person}{Junlin Zhang}.} \bibinfo{year}{2019}\natexlab{}.
\newblock \showarticletitle{FiBiNET: combining feature importance and bilinear feature interaction for click-through rate prediction}. In \bibinfo{booktitle}{\emph{Proceedings of the 13th ACM conference on recommender systems}}. \bibinfo{pages}{169--177}.
\newblock


\bibitem[Huang et~al\mbox{.}(2023)]%
        {huang2023recommender}
\bibfield{author}{\bibinfo{person}{Xu Huang}, \bibinfo{person}{Jianxun Lian}, \bibinfo{person}{Yuxuan Lei}, \bibinfo{person}{Jing Yao}, \bibinfo{person}{Defu Lian}, {and} \bibinfo{person}{Xing Xie}.} \bibinfo{year}{2023}\natexlab{}.
\newblock \showarticletitle{Recommender AI Agent: Integrating Large Language Models for Interactive Recommendations}.
\newblock \bibinfo{journal}{\emph{ArXiv}}  \bibinfo{volume}{abs/2308.16505} (\bibinfo{year}{2023}).
\newblock


\bibitem[Huzhang et~al\mbox{.}(2021)]%
        {huzhang2021aliexpress}
\bibfield{author}{\bibinfo{person}{Guangda Huzhang}, \bibinfo{person}{Zhen-Jia Pang}, \bibinfo{person}{Yongqing Gao}, \bibinfo{person}{Yawen Liu}, \bibinfo{person}{Weijie Shen}, \bibinfo{person}{Wen-Ji Zhou}, \bibinfo{person}{Qianying Lin}, \bibinfo{person}{Qing Da}, \bibinfo{person}{An-Xiang Zeng}, \bibinfo{person}{Han Yu}, {et~al\mbox{.}}} \bibinfo{year}{2021}\natexlab{}.
\newblock \showarticletitle{AliExpress Learning-To-Rank: Maximizing online model performance without going online}.
\newblock \bibinfo{journal}{\emph{IEEE Transactions on Knowledge and Data Engineering}} \bibinfo{volume}{35}, \bibinfo{number}{2} (\bibinfo{year}{2021}), \bibinfo{pages}{1214--1226}.
\newblock


\bibitem[Jagerman et~al\mbox{.}(2019)]%
        {jagerman2019people}
\bibfield{author}{\bibinfo{person}{Rolf Jagerman}, \bibinfo{person}{Ilya Markov}, {and} \bibinfo{person}{Maarten de Rijke}.} \bibinfo{year}{2019}\natexlab{}.
\newblock \showarticletitle{When people change their mind: Off-policy evaluation in non-stationary recommendation environments}. In \bibinfo{booktitle}{\emph{Proceedings of the Twelfth ACM International Conference on Web Search and Data Mining}}. \bibinfo{pages}{447--455}.
\newblock


\bibitem[Jannach et~al\mbox{.}(2021)]%
        {jannach2021survey}
\bibfield{author}{\bibinfo{person}{Dietmar Jannach}, \bibinfo{person}{Ahtsham Manzoor}, \bibinfo{person}{Wanling Cai}, {and} \bibinfo{person}{Li Chen}.} \bibinfo{year}{2021}\natexlab{}.
\newblock \showarticletitle{A survey on conversational recommender systems}.
\newblock \bibinfo{journal}{\emph{ACM Computing Surveys (CSUR)}} \bibinfo{volume}{54}, \bibinfo{number}{5} (\bibinfo{year}{2021}), \bibinfo{pages}{1--36}.
\newblock


\bibitem[Ji et~al\mbox{.}(2023)]%
        {ji2023text}
\bibfield{author}{\bibinfo{person}{Jianchao Ji}, \bibinfo{person}{Zelong Li}, \bibinfo{person}{Shuyuan Xu}, \bibinfo{person}{Wenyue Hua}, \bibinfo{person}{Yingqiang Ge}, \bibinfo{person}{Juntao Tan}, {and} \bibinfo{person}{Yongfeng Zhang}.} \bibinfo{year}{2023}\natexlab{}.
\newblock \showarticletitle{Text based Large Language Model for Recommendation}.
\newblock \bibinfo{journal}{\emph{arXiv preprint arXiv:2307.00457}} (\bibinfo{year}{2023}).
\newblock


\bibitem[Jiang et~al\mbox{.}(2014)]%
        {jiang2014choice}
\bibfield{author}{\bibinfo{person}{Hai Jiang}, \bibinfo{person}{Xin Qi}, {and} \bibinfo{person}{He Sun}.} \bibinfo{year}{2014}\natexlab{}.
\newblock \showarticletitle{Choice-based recommender systems: a unified approach to achieving relevancy and diversity}.
\newblock \bibinfo{journal}{\emph{Operations Research}} \bibinfo{volume}{62}, \bibinfo{number}{5} (\bibinfo{year}{2014}), \bibinfo{pages}{973--993}.
\newblock


\bibitem[Jiang and Ferrara(2023)]%
        {jiang2023social}
\bibfield{author}{\bibinfo{person}{Julie Jiang} {and} \bibinfo{person}{Emilio Ferrara}.} \bibinfo{year}{2023}\natexlab{}.
\newblock \showarticletitle{Social-LLM: Modeling User Behavior at Scale using Language Models and Social Network Data}.
\newblock \bibinfo{journal}{\emph{arXiv preprint arXiv:2401.00893}} (\bibinfo{year}{2023}).
\newblock


\bibitem[Jiang et~al\mbox{.}(2023)]%
        {jiang2023reformulating}
\bibfield{author}{\bibinfo{person}{Junzhe Jiang}, \bibinfo{person}{Shang Qu}, \bibinfo{person}{Mingyue Cheng}, {and} \bibinfo{person}{Qi Liu}.} \bibinfo{year}{2023}\natexlab{}.
\newblock \showarticletitle{Reformulating Sequential Recommendation: Learning Dynamic User Interest with Content-enriched Language Modeling}.
\newblock \bibinfo{journal}{\emph{arXiv preprint arXiv:2309.10435}} (\bibinfo{year}{2023}).
\newblock


\bibitem[Jiang et~al\mbox{.}(2018)]%
        {jiang2018beyond}
\bibfield{author}{\bibinfo{person}{Ray Jiang}, \bibinfo{person}{Sven Gowal}, \bibinfo{person}{Timothy~A Mann}, {and} \bibinfo{person}{Danilo~J Rezende}.} \bibinfo{year}{2018}\natexlab{}.
\newblock \showarticletitle{Beyond greedy ranking: Slate optimization via list-CVAE}.
\newblock \bibinfo{journal}{\emph{arXiv preprint arXiv:1803.01682}} (\bibinfo{year}{2018}).
\newblock


\bibitem[Jin et~al\mbox{.}(2023)]%
        {jin2023language}
\bibfield{author}{\bibinfo{person}{Bowen Jin}, \bibinfo{person}{Hansi Zeng}, \bibinfo{person}{Guoyin Wang}, \bibinfo{person}{Xiusi Chen}, \bibinfo{person}{Tianxin Wei}, \bibinfo{person}{Ruirui Li}, \bibinfo{person}{Zhengyang Wang}, \bibinfo{person}{Zheng Li}, \bibinfo{person}{Yang Li}, \bibinfo{person}{Hanqing Lu}, {et~al\mbox{.}}} \bibinfo{year}{2023}\natexlab{}.
\newblock \showarticletitle{Language Models As Semantic Indexers}.
\newblock \bibinfo{journal}{\emph{arXiv preprint arXiv:2310.07815}} (\bibinfo{year}{2023}).
\newblock


\bibitem[John et~al\mbox{.}(2024)]%
        {john2024llmrs}
\bibfield{author}{\bibinfo{person}{Angela John}, \bibinfo{person}{Theophilus Aidoo}, \bibinfo{person}{Hamayoon Behmanush}, \bibinfo{person}{Irem~B Gunduz}, \bibinfo{person}{Hewan Shrestha}, \bibinfo{person}{Maxx~Richard Rahman}, {and} \bibinfo{person}{Wolfgang Maa{\ss}}.} \bibinfo{year}{2024}\natexlab{}.
\newblock \showarticletitle{LLMRS: Unlocking Potentials of LLM-Based Recommender Systems for Software Purchase}.
\newblock \bibinfo{journal}{\emph{arXiv preprint arXiv:2401.06676}} (\bibinfo{year}{2024}).
\newblock


\bibitem[Kang et~al\mbox{.}(2023)]%
        {kang2023llms}
\bibfield{author}{\bibinfo{person}{Wang-Cheng Kang}, \bibinfo{person}{Jianmo Ni}, \bibinfo{person}{Nikhil Mehta}, \bibinfo{person}{Maheswaran Sathiamoorthy}, \bibinfo{person}{Lichan Hong}, \bibinfo{person}{Ed Chi}, {and} \bibinfo{person}{Derek~Zhiyuan Cheng}.} \bibinfo{year}{2023}\natexlab{}.
\newblock \showarticletitle{Do LLMs Understand User Preferences? Evaluating LLMs On User Rating Prediction}.
\newblock \bibinfo{journal}{\emph{arXiv preprint arXiv:2305.06474}} (\bibinfo{year}{2023}).
\newblock


\bibitem[Kaplan et~al\mbox{.}(2020)]%
        {kaplan2020scaling}
\bibfield{author}{\bibinfo{person}{Jared Kaplan}, \bibinfo{person}{Sam McCandlish}, \bibinfo{person}{Tom Henighan}, \bibinfo{person}{Tom~B Brown}, \bibinfo{person}{Benjamin Chess}, \bibinfo{person}{Rewon Child}, \bibinfo{person}{Scott Gray}, \bibinfo{person}{Alec Radford}, \bibinfo{person}{Jeffrey Wu}, {and} \bibinfo{person}{Dario Amodei}.} \bibinfo{year}{2020}\natexlab{}.
\newblock \showarticletitle{Scaling laws for neural language models}.
\newblock \bibinfo{journal}{\emph{arXiv preprint arXiv:2001.08361}} (\bibinfo{year}{2020}).
\newblock


\bibitem[Ko et~al\mbox{.}(2022)]%
        {ko2022survey}
\bibfield{author}{\bibinfo{person}{Hyeyoung Ko}, \bibinfo{person}{Suyeon Lee}, \bibinfo{person}{Yoonseo Park}, {and} \bibinfo{person}{Anna Choi}.} \bibinfo{year}{2022}\natexlab{}.
\newblock \showarticletitle{A survey of recommendation systems: recommendation models, techniques, and application fields}.
\newblock \bibinfo{journal}{\emph{Electronics}} \bibinfo{volume}{11}, \bibinfo{number}{1} (\bibinfo{year}{2022}), \bibinfo{pages}{141}.
\newblock


\bibitem[Koren et~al\mbox{.}(2009)]%
        {koren2009matrix}
\bibfield{author}{\bibinfo{person}{Yehuda Koren}, \bibinfo{person}{Robert Bell}, {and} \bibinfo{person}{Chris Volinsky}.} \bibinfo{year}{2009}\natexlab{}.
\newblock \showarticletitle{Matrix factorization techniques for recommender systems}.
\newblock \bibinfo{journal}{\emph{Computer}} \bibinfo{volume}{42}, \bibinfo{number}{8} (\bibinfo{year}{2009}), \bibinfo{pages}{30--37}.
\newblock


\bibitem[Lei et~al\mbox{.}(2020a)]%
        {lei2020estimation}
\bibfield{author}{\bibinfo{person}{Wenqiang Lei}, \bibinfo{person}{Xiangnan He}, \bibinfo{person}{Yisong Miao}, \bibinfo{person}{Qingyun Wu}, \bibinfo{person}{Richang Hong}, \bibinfo{person}{Min-Yen Kan}, {and} \bibinfo{person}{Tat-Seng Chua}.} \bibinfo{year}{2020}\natexlab{a}.
\newblock \showarticletitle{Estimation-action-reflection: Towards deep interaction between conversational and recommender systems}. In \bibinfo{booktitle}{\emph{Proceedings of the 13th International Conference on Web Search and Data Mining}}. \bibinfo{pages}{304--312}.
\newblock


\bibitem[Lei et~al\mbox{.}(2020b)]%
        {lei2020interactive}
\bibfield{author}{\bibinfo{person}{Wenqiang Lei}, \bibinfo{person}{Gangyi Zhang}, \bibinfo{person}{Xiangnan He}, \bibinfo{person}{Yisong Miao}, \bibinfo{person}{Xiang Wang}, \bibinfo{person}{Liang Chen}, {and} \bibinfo{person}{Tat-Seng Chua}.} \bibinfo{year}{2020}\natexlab{b}.
\newblock \showarticletitle{Interactive path reasoning on graph for conversational recommendation}. In \bibinfo{booktitle}{\emph{Proceedings of the 26th ACM SIGKDD international conference on knowledge discovery \& data mining}}. \bibinfo{pages}{2073--2083}.
\newblock


\bibitem[Lewis et~al\mbox{.}(2020)]%
        {lewis2020bart}
\bibfield{author}{\bibinfo{person}{Mike Lewis}, \bibinfo{person}{Yinhan Liu}, \bibinfo{person}{Naman Goyal}, \bibinfo{person}{Marjan Ghazvininejad}, \bibinfo{person}{Abdelrahman Mohamed}, \bibinfo{person}{Omer Levy}, \bibinfo{person}{Veselin Stoyanov}, {and} \bibinfo{person}{Luke Zettlemoyer}.} \bibinfo{year}{2020}\natexlab{}.
\newblock \showarticletitle{{BART}: Denoising Sequence-to-Sequence Pre-training for Natural Language Generation, Translation, and Comprehension}. In \bibinfo{booktitle}{\emph{Proceedings of the 58th Annual Meeting of the Association for Computational Linguistics}}. \bibinfo{publisher}{Association for Computational Linguistics}, \bibinfo{pages}{7871--7880}.
\newblock


\bibitem[Li et~al\mbox{.}(2023c)]%
        {li2023taggpt}
\bibfield{author}{\bibinfo{person}{Chen Li}, \bibinfo{person}{Yixiao Ge}, \bibinfo{person}{Jiayong Mao}, \bibinfo{person}{Dian Li}, {and} \bibinfo{person}{Ying Shan}.} \bibinfo{year}{2023}\natexlab{c}.
\newblock \showarticletitle{TagGPT: Large Language Models are Zero-shot Multimodal Taggers}.
\newblock \bibinfo{journal}{\emph{arXiv preprint arXiv:2304.03022}} (\bibinfo{year}{2023}).
\newblock


\bibitem[Li et~al\mbox{.}(2023g)]%
        {li2023text}
\bibfield{author}{\bibinfo{person}{Jiacheng Li}, \bibinfo{person}{Ming Wang}, \bibinfo{person}{Jin Li}, \bibinfo{person}{Jinmiao Fu}, \bibinfo{person}{Xin Shen}, \bibinfo{person}{Jingbo Shang}, {and} \bibinfo{person}{Julian McAuley}.} \bibinfo{year}{2023}\natexlab{g}.
\newblock \showarticletitle{Text Is All You Need: Learning Language Representations for Sequential Recommendation}.
\newblock \bibinfo{journal}{\emph{arXiv preprint arXiv:2305.13731}} (\bibinfo{year}{2023}).
\newblock


\bibitem[Li et~al\mbox{.}(2023f)]%
        {li2023prompt}
\bibfield{author}{\bibinfo{person}{Pan Li}, \bibinfo{person}{Yuyan Wang}, \bibinfo{person}{Ed~H Chi}, {and} \bibinfo{person}{Minmin Chen}.} \bibinfo{year}{2023}\natexlab{f}.
\newblock \showarticletitle{Prompt Tuning Large Language Models on Personalized Aspect Extraction for Recommendations}.
\newblock \bibinfo{journal}{\emph{arXiv preprint arXiv:2306.01475}} (\bibinfo{year}{2023}).
\newblock


\bibitem[Li et~al\mbox{.}(2021)]%
        {li2021seamlessly}
\bibfield{author}{\bibinfo{person}{Shijun Li}, \bibinfo{person}{Wenqiang Lei}, \bibinfo{person}{Qingyun Wu}, \bibinfo{person}{Xiangnan He}, \bibinfo{person}{Peng Jiang}, {and} \bibinfo{person}{Tat-Seng Chua}.} \bibinfo{year}{2021}\natexlab{}.
\newblock \showarticletitle{Seamlessly unifying attributes and items: Conversational recommendation for cold-start users}.
\newblock \bibinfo{journal}{\emph{ACM Transactions on Information Systems (TOIS)}} \bibinfo{volume}{39}, \bibinfo{number}{4} (\bibinfo{year}{2021}), \bibinfo{pages}{1--29}.
\newblock


\bibitem[Li et~al\mbox{.}(2023a)]%
        {li2023ctrl}
\bibfield{author}{\bibinfo{person}{Xiangyang Li}, \bibinfo{person}{Bo Chen}, \bibinfo{person}{Lu Hou}, {and} \bibinfo{person}{Ruiming Tang}.} \bibinfo{year}{2023}\natexlab{a}.
\newblock \showarticletitle{CTRL: Connect Tabular and Language Model for CTR Prediction}.
\newblock \bibinfo{journal}{\emph{arXiv preprint arXiv:2306.02841}} (\bibinfo{year}{2023}).
\newblock


\bibitem[Li et~al\mbox{.}(2023b)]%
        {li2023e4srec}
\bibfield{author}{\bibinfo{person}{Xinhang Li}, \bibinfo{person}{Chong Chen}, \bibinfo{person}{Xiangyu Zhao}, \bibinfo{person}{Yong Zhang}, {and} \bibinfo{person}{Chunxiao Xing}.} \bibinfo{year}{2023}\natexlab{b}.
\newblock \showarticletitle{E4SRec: An Elegant Effective Efficient Extensible Solution of Large Language Models for Sequential Recommendation}.
\newblock \bibinfo{journal}{\emph{arXiv preprint arXiv:2312.02443}} (\bibinfo{year}{2023}).
\newblock


\bibitem[Li et~al\mbox{.}(2023e)]%
        {li2023agent4ranking}
\bibfield{author}{\bibinfo{person}{Xiaopeng Li}, \bibinfo{person}{Lixin Su}, \bibinfo{person}{Pengyue Jia}, \bibinfo{person}{Xiangyu Zhao}, \bibinfo{person}{Suqi Cheng}, \bibinfo{person}{Junfeng Wang}, {and} \bibinfo{person}{Dawei Yin}.} \bibinfo{year}{2023}\natexlab{e}.
\newblock \showarticletitle{Agent4Ranking: Semantic Robust Ranking via Personalized Query Rewriting Using Multi-agent LLM}.
\newblock \bibinfo{journal}{\emph{arXiv preprint arXiv:2312.15450}} (\bibinfo{year}{2023}).
\newblock


\bibitem[Li et~al\mbox{.}(2023h)]%
        {li2023exploring}
\bibfield{author}{\bibinfo{person}{Xinyi Li}, \bibinfo{person}{Yongfeng Zhang}, {and} \bibinfo{person}{Edward~C Malthouse}.} \bibinfo{year}{2023}\natexlab{h}.
\newblock \showarticletitle{Exploring Fine-tuning ChatGPT for News Recommendation}.
\newblock \bibinfo{journal}{\emph{arXiv preprint arXiv:2311.05850}} (\bibinfo{year}{2023}).
\newblock


\bibitem[Li et~al\mbox{.}(2023i)]%
        {li2023pbnr}
\bibfield{author}{\bibinfo{person}{Xinyi Li}, \bibinfo{person}{Yongfeng Zhang}, {and} \bibinfo{person}{Edward~C Malthouse}.} \bibinfo{year}{2023}\natexlab{i}.
\newblock \showarticletitle{PBNR: Prompt-based News Recommender System}.
\newblock \bibinfo{journal}{\emph{arXiv preprint arXiv:2304.07862}} (\bibinfo{year}{2023}).
\newblock


\bibitem[Li et~al\mbox{.}(2023d)]%
        {li2023ecomgpt}
\bibfield{author}{\bibinfo{person}{Yangning Li}, \bibinfo{person}{Shirong Ma}, \bibinfo{person}{Xiaobin Wang}, \bibinfo{person}{Shen Huang}, \bibinfo{person}{Chengyue Jiang}, \bibinfo{person}{Hai-Tao Zheng}, \bibinfo{person}{Pengjun Xie}, \bibinfo{person}{Fei Huang}, {and} \bibinfo{person}{Yong Jiang}.} \bibinfo{year}{2023}\natexlab{d}.
\newblock \showarticletitle{EcomGPT: Instruction-tuning Large Language Model with Chain-of-Task Tasks for E-commerce}.
\newblock \bibinfo{journal}{\emph{arXiv preprint arXiv:2308.06966}} (\bibinfo{year}{2023}).
\newblock


\bibitem[Li et~al\mbox{.}(2019)]%
        {li2019fi}
\bibfield{author}{\bibinfo{person}{Zekun Li}, \bibinfo{person}{Zeyu Cui}, \bibinfo{person}{Shu Wu}, \bibinfo{person}{Xiaoyu Zhang}, {and} \bibinfo{person}{Liang Wang}.} \bibinfo{year}{2019}\natexlab{}.
\newblock \showarticletitle{Fi-gnn: Modeling feature interactions via graph neural networks for ctr prediction}. In \bibinfo{booktitle}{\emph{Proceedings of the 28th ACM international conference on information and knowledge management}}. \bibinfo{pages}{539--548}.
\newblock


\bibitem[Liao et~al\mbox{.}(2020)]%
        {liao2020topic}
\bibfield{author}{\bibinfo{person}{Lizi Liao}, \bibinfo{person}{Ryuichi Takanobu}, \bibinfo{person}{Yunshan Ma}, \bibinfo{person}{Xun Yang}, \bibinfo{person}{Minlie Huang}, {and} \bibinfo{person}{Tat-Seng Chua}.} \bibinfo{year}{2020}\natexlab{}.
\newblock \showarticletitle{Topic-guided relational conversational recommender in multi-domain}.
\newblock \bibinfo{journal}{\emph{IEEE Transactions on Knowledge and Data Engineering}} (\bibinfo{year}{2020}).
\newblock


\bibitem[Lin(2004)]%
        {lin2004rouge}
\bibfield{author}{\bibinfo{person}{Chin-Yew Lin}.} \bibinfo{year}{2004}\natexlab{}.
\newblock \showarticletitle{Rouge: A package for automatic evaluation of summaries}. In \bibinfo{booktitle}{\emph{Text summarization branches out}}. \bibinfo{pages}{74--81}.
\newblock


\bibitem[Lin and Zhang(2023)]%
        {lin2023sparks}
\bibfield{author}{\bibinfo{person}{Guo Lin} {and} \bibinfo{person}{Yongfeng Zhang}.} \bibinfo{year}{2023}\natexlab{}.
\newblock \showarticletitle{Sparks of Artificial General Recommender (AGR): Experiments with ChatGPT}.
\newblock \bibinfo{journal}{\emph{Algorithms}}  \bibinfo{volume}{16} (\bibinfo{year}{2023}), \bibinfo{pages}{432}.
\newblock


\bibitem[Lin et~al\mbox{.}(2024)]%
        {clickprompt}
\bibfield{author}{\bibinfo{person}{Jianghao Lin}, \bibinfo{person}{Bo Chen}, \bibinfo{person}{Hangyu Wang}, \bibinfo{person}{Yunjia Xi}, \bibinfo{person}{Yanru Qu}, \bibinfo{person}{Xinyi Dai}, \bibinfo{person}{Kangning Zhang}, \bibinfo{person}{Ruiming Tang}, \bibinfo{person}{Yong Yu}, {and} \bibinfo{person}{Weinan Zhang}.} \bibinfo{year}{2024}\natexlab{}.
\newblock \showarticletitle{ClickPrompt: CTR Models are Strong Prompt Generators for Adapting Language Models to CTR Prediction}. In \bibinfo{booktitle}{\emph{Proceedings of the ACM on Web Conference 2024}}. \bibinfo{pages}{3319--3330}.
\newblock


\bibitem[Lin et~al\mbox{.}(2023a)]%
        {lin2023can}
\bibfield{author}{\bibinfo{person}{Jianghao Lin}, \bibinfo{person}{Xinyi Dai}, \bibinfo{person}{Yunjia Xi}, \bibinfo{person}{Weiwen Liu}, \bibinfo{person}{Bo Chen}, \bibinfo{person}{Xiangyang Li}, \bibinfo{person}{Chenxu Zhu}, \bibinfo{person}{Huifeng Guo}, \bibinfo{person}{Yong Yu}, \bibinfo{person}{Ruiming Tang}, {et~al\mbox{.}}} \bibinfo{year}{2023}\natexlab{a}.
\newblock \showarticletitle{How Can Recommender Systems Benefit from Large Language Models: A Survey}.
\newblock \bibinfo{journal}{\emph{arXiv preprint arXiv:2306.05817}} (\bibinfo{year}{2023}).
\newblock


\bibitem[Lin et~al\mbox{.}(2023b)]%
        {lin2023rella}
\bibfield{author}{\bibinfo{person}{Jianghao Lin}, \bibinfo{person}{Rong Shan}, \bibinfo{person}{Chenxu Zhu}, \bibinfo{person}{Kounianhua Du}, \bibinfo{person}{Bo Chen}, \bibinfo{person}{Shigang Quan}, \bibinfo{person}{Ruiming Tang}, \bibinfo{person}{Yong Yu}, {and} \bibinfo{person}{Weinan Zhang}.} \bibinfo{year}{2023}\natexlab{b}.
\newblock \showarticletitle{ReLLa: Retrieval-enhanced Large Language Models for Lifelong Sequential Behavior Comprehension in Recommendation}.
\newblock \bibinfo{journal}{\emph{arXiv preprint arXiv:2308.11131}} (\bibinfo{year}{2023}).
\newblock


\bibitem[Liu et~al\mbox{.}(2019)]%
        {liu2019feature}
\bibfield{author}{\bibinfo{person}{Bin Liu}, \bibinfo{person}{Ruiming Tang}, \bibinfo{person}{Yingzhi Chen}, \bibinfo{person}{Jinkai Yu}, \bibinfo{person}{Huifeng Guo}, {and} \bibinfo{person}{Yuzhou Zhang}.} \bibinfo{year}{2019}\natexlab{}.
\newblock \showarticletitle{Feature generation by convolutional neural network for click-through rate prediction}. In \bibinfo{booktitle}{\emph{The World Wide Web Conference}}. \bibinfo{pages}{1119--1129}.
\newblock


\bibitem[Liu et~al\mbox{.}(2023e)]%
        {liu2023recprompt}
\bibfield{author}{\bibinfo{person}{Dairui Liu}, \bibinfo{person}{Boming Yang}, \bibinfo{person}{Honghui Du}, \bibinfo{person}{Derek Greene}, \bibinfo{person}{Aonghus Lawlor}, \bibinfo{person}{Ruihai Dong}, {and} \bibinfo{person}{Irene Li}.} \bibinfo{year}{2023}\natexlab{e}.
\newblock \showarticletitle{RecPrompt: A Prompt Tuning Framework for News Recommendation Using Large Language Models}.
\newblock \bibinfo{journal}{\emph{arXiv preprint arXiv:2312.10463}} (\bibinfo{year}{2023}).
\newblock


\bibitem[Liu et~al\mbox{.}(2023c)]%
        {liu2023understanding}
\bibfield{author}{\bibinfo{person}{Fan Liu}, \bibinfo{person}{Yaqi Liu}, \bibinfo{person}{Zhiyong Cheng}, \bibinfo{person}{Liqiang Nie}, {and} \bibinfo{person}{Mohan Kankanhalli}.} \bibinfo{year}{2023}\natexlab{c}.
\newblock \showarticletitle{Understanding Before Recommendation: Semantic Aspect-Aware Review Exploitation via Large Language Models}.
\newblock \bibinfo{journal}{\emph{arXiv preprint arXiv:2312.16275}} (\bibinfo{year}{2023}).
\newblock


\bibitem[Liu et~al\mbox{.}(2023f)]%
        {liu2023pre}
\bibfield{author}{\bibinfo{person}{Peng Liu}, \bibinfo{person}{Lemei Zhang}, {and} \bibinfo{person}{Jon~Atle Gulla}.} \bibinfo{year}{2023}\natexlab{f}.
\newblock \showarticletitle{Pre-train, prompt and recommendation: A comprehensive survey of language modelling paradigm adaptations in recommender systems}.
\newblock \bibinfo{journal}{\emph{arXiv preprint arXiv:2302.03735}} (\bibinfo{year}{2023}).
\newblock


\bibitem[Liu et~al\mbox{.}(2023b)]%
        {liu2023first}
\bibfield{author}{\bibinfo{person}{Qijiong Liu}, \bibinfo{person}{Nuo Chen}, \bibinfo{person}{Tetsuya Sakai}, {and} \bibinfo{person}{Xiao-Ming Wu}.} \bibinfo{year}{2023}\natexlab{b}.
\newblock \showarticletitle{A First Look at LLM-Powered Generative News Recommendation}.
\newblock \bibinfo{journal}{\emph{arXiv preprint arXiv:2305.06566}} (\bibinfo{year}{2023}).
\newblock


\bibitem[Liu et~al\mbox{.}(2021)]%
        {liu2021variation}
\bibfield{author}{\bibinfo{person}{Shuchang Liu}, \bibinfo{person}{Fei Sun}, \bibinfo{person}{Yingqiang Ge}, \bibinfo{person}{Changhua Pei}, {and} \bibinfo{person}{Yongfeng Zhang}.} \bibinfo{year}{2021}\natexlab{}.
\newblock \showarticletitle{Variation control and evaluation for generative slate recommendations}. In \bibinfo{booktitle}{\emph{Proceedings of the Web Conference 2021}}. \bibinfo{pages}{436--448}.
\newblock


\bibitem[Liu et~al\mbox{.}(2020a)]%
        {liu2020personalized}
\bibfield{author}{\bibinfo{person}{Weiwen Liu}, \bibinfo{person}{Qing Liu}, \bibinfo{person}{Ruiming Tang}, \bibinfo{person}{Junyang Chen}, \bibinfo{person}{Xiuqiang He}, {and} \bibinfo{person}{Pheng~Ann Heng}.} \bibinfo{year}{2020}\natexlab{a}.
\newblock \showarticletitle{Personalized Re-ranking with Item Relationships for E-commerce}. In \bibinfo{booktitle}{\emph{Proceedings of the 29th ACM International Conference on Information \& Knowledge Management}}. \bibinfo{pages}{925--934}.
\newblock


\bibitem[Liu et~al\mbox{.}(2023d)]%
        {liu2023personalized}
\bibfield{author}{\bibinfo{person}{Weiwen Liu}, \bibinfo{person}{Yunjia Xi}, \bibinfo{person}{Jiarui Qin}, \bibinfo{person}{Xinyi Dai}, \bibinfo{person}{Ruiming Tang}, \bibinfo{person}{Shuai Li}, \bibinfo{person}{Weinan Zhang}, {and} \bibinfo{person}{Rui Zhang}.} \bibinfo{year}{2023}\natexlab{d}.
\newblock \showarticletitle{Personalized Diversification for Neural Re-ranking in Recommendation}. In \bibinfo{booktitle}{\emph{2023 IEEE 39th International Conference on Data Engineering (ICDE)}}. IEEE, \bibinfo{pages}{802--815}.
\newblock


\bibitem[Liu et~al\mbox{.}(2022)]%
        {liu2022neural}
\bibfield{author}{\bibinfo{person}{Weiwen Liu}, \bibinfo{person}{Yunjia Xi}, \bibinfo{person}{Jiarui Qin}, \bibinfo{person}{Fei Sun}, \bibinfo{person}{Bo Chen}, \bibinfo{person}{Weinan Zhang}, \bibinfo{person}{Rui Zhang}, {and} \bibinfo{person}{Ruiming Tang}.} \bibinfo{year}{2022}\natexlab{}.
\newblock \showarticletitle{Neural re-ranking in multi-stage recommender systems: A review}.
\newblock \bibinfo{journal}{\emph{arXiv preprint arXiv:2202.06602}} (\bibinfo{year}{2022}).
\newblock


\bibitem[Liu et~al\mbox{.}(2023a)]%
        {liu2023conversational}
\bibfield{author}{\bibinfo{person}{Yuanxing Liu}, \bibinfo{person}{Weinan  }, \bibinfo{person}{Yifan Chen}, \bibinfo{person}{Yuchi Zhang}, \bibinfo{person}{Haopeng Bai}, \bibinfo{person}{Fan Feng}, \bibinfo{person}{Hengbin Cui}, \bibinfo{person}{Yongbin Li}, {and} \bibinfo{person}{Wanxiang Che}.} \bibinfo{year}{2023}\natexlab{a}.
\newblock \showarticletitle{Conversational Recommender System and Large Language Model Are Made for Each Other in {E}-commerce Pre-sales Dialogue}. In \bibinfo{booktitle}{\emph{Findings of the Association for Computational Linguistics: EMNLP 2023}}. \bibinfo{pages}{9587--9605}.
\newblock


\bibitem[Liu et~al\mbox{.}(2020b)]%
        {liu2020towards}
\bibfield{author}{\bibinfo{person}{Zeming Liu}, \bibinfo{person}{Haifeng Wang}, \bibinfo{person}{Zheng-Yu Niu}, \bibinfo{person}{Hua Wu}, \bibinfo{person}{Wanxiang Che}, {and} \bibinfo{person}{Ting Liu}.} \bibinfo{year}{2020}\natexlab{b}.
\newblock \showarticletitle{Towards conversational recommendation over multi-type dialogs}.
\newblock \bibinfo{journal}{\emph{arXiv preprint arXiv:2005.03954}} (\bibinfo{year}{2020}).
\newblock


\bibitem[Loepp et~al\mbox{.}(2014)]%
        {loepp2014choice}
\bibfield{author}{\bibinfo{person}{Benedikt Loepp}, \bibinfo{person}{Tim Hussein}, {and} \bibinfo{person}{J{\"u}ergen Ziegler}.} \bibinfo{year}{2014}\natexlab{}.
\newblock \showarticletitle{Choice-based preference elicitation for collaborative filtering recommender systems}. In \bibinfo{booktitle}{\emph{Proceedings of the SIGCHI Conference on Human Factors in Computing Systems}}. \bibinfo{pages}{3085--3094}.
\newblock


\bibitem[Luo et~al\mbox{.}(2020)]%
        {luo2020deep}
\bibfield{author}{\bibinfo{person}{Kai Luo}, \bibinfo{person}{Hojin Yang}, \bibinfo{person}{Ga Wu}, {and} \bibinfo{person}{Scott Sanner}.} \bibinfo{year}{2020}\natexlab{}.
\newblock \showarticletitle{Deep critiquing for VAE-based recommender systems}. In \bibinfo{booktitle}{\emph{Proceedings of the 43rd International ACM SIGIR conference on research and development in Information Retrieval}}. \bibinfo{pages}{1269--1278}.
\newblock


\bibitem[Luo et~al\mbox{.}(2024)]%
        {luo2024integrating}
\bibfield{author}{\bibinfo{person}{Sichun Luo}, \bibinfo{person}{Yuxuan Yao}, \bibinfo{person}{Bowei He}, \bibinfo{person}{Yinya Huang}, \bibinfo{person}{Aojun Zhou}, \bibinfo{person}{Xinyi Zhang}, \bibinfo{person}{Yuanzhang Xiao}, \bibinfo{person}{Mingjie Zhan}, {and} \bibinfo{person}{Linqi Song}.} \bibinfo{year}{2024}\natexlab{}.
\newblock \bibinfo{title}{Integrating Large Language Models into Recommendation via Mutual Augmentation and Adaptive Aggregation}.
\newblock
\newblock
\showeprint[arxiv]{2401.13870}~[cs.IR]


\bibitem[Ma et~al\mbox{.}(2018b)]%
        {mmoe}
\bibfield{author}{\bibinfo{person}{Jiaqi Ma}, \bibinfo{person}{Zhe Zhao}, \bibinfo{person}{Xinyang Yi}, \bibinfo{person}{Jilin Chen}, \bibinfo{person}{Lichan Hong}, {and} \bibinfo{person}{Ed~H Chi}.} \bibinfo{year}{2018}\natexlab{b}.
\newblock \showarticletitle{Modeling task relationships in multi-task learning with multi-gate mixture-of-experts}. In \bibinfo{booktitle}{\emph{SIGKDD}}. \bibinfo{pages}{1930--1939}.
\newblock


\bibitem[Ma et~al\mbox{.}(2018a)]%
        {ma2018entire}
\bibfield{author}{\bibinfo{person}{Xiao Ma}, \bibinfo{person}{Liqin Zhao}, \bibinfo{person}{Guan Huang}, \bibinfo{person}{Zhi Wang}, \bibinfo{person}{Zelin Hu}, \bibinfo{person}{Xiaoqiang Zhu}, {and} \bibinfo{person}{Kun Gai}.} \bibinfo{year}{2018}\natexlab{a}.
\newblock \showarticletitle{Entire space multi-task model: An effective approach for estimating post-click conversion rate}. In \bibinfo{booktitle}{\emph{The 41st International ACM SIGIR Conference on Research \& Development in Information Retrieval}}. \bibinfo{pages}{1137--1140}.
\newblock


\bibitem[McInerney et~al\mbox{.}(2020)]%
        {mcinerney2020counterfactual}
\bibfield{author}{\bibinfo{person}{James McInerney}, \bibinfo{person}{Brian Brost}, \bibinfo{person}{Praveen Chandar}, \bibinfo{person}{Rishabh Mehrotra}, {and} \bibinfo{person}{Benjamin Carterette}.} \bibinfo{year}{2020}\natexlab{}.
\newblock \showarticletitle{Counterfactual evaluation of slate recommendations with sequential reward interactions}. In \bibinfo{booktitle}{\emph{Proceedings of the 26th ACM SIGKDD International Conference on Knowledge Discovery \& Data Mining}}. \bibinfo{pages}{1779--1788}.
\newblock


\bibitem[Murakhovs{'}ka et~al\mbox{.}(2023)]%
        {murakhovska2023salespeople}
\bibfield{author}{\bibinfo{person}{Lidiya Murakhovs{'}ka}, \bibinfo{person}{Philippe  }, \bibinfo{person}{Tian Xie}, \bibinfo{person}{Caiming  }, {and} \bibinfo{person}{Chien-Sheng Wu}.} \bibinfo{year}{2023}\natexlab{}.
\newblock \showarticletitle{Salespeople vs {S}ales{B}ot: Exploring the Role of Educational Value in Conversational Recommender Systems}. In \bibinfo{booktitle}{\emph{Findings of the Association for Computational Linguistics: EMNLP 2023}}. \bibinfo{pages}{9823--9838}.
\newblock


\bibitem[Mysore et~al\mbox{.}(2023)]%
        {mysore2023large}
\bibfield{author}{\bibinfo{person}{Sheshera Mysore}, \bibinfo{person}{Andrew Mccallum}, {and} \bibinfo{person}{Hamed Zamani}.} \bibinfo{year}{2023}\natexlab{}.
\newblock \showarticletitle{Large Language Model Augmented Narrative Driven Recommendations}. In \bibinfo{booktitle}{\emph{Proceedings of the 17th ACM Conference on Recommender Systems}} \emph{(\bibinfo{series}{RecSys '23})}. \bibinfo{pages}{777–783}.
\newblock


\bibitem[OpenAI(2023)]%
        {achiam2023gpt4}
\bibfield{author}{\bibinfo{person}{OpenAI}.} \bibinfo{year}{2023}\natexlab{}.
\newblock \showarticletitle{GPT-4 Technical Report}.
\newblock \bibinfo{journal}{\emph{ArXiv}}  \bibinfo{volume}{abs/2303.08774} (\bibinfo{year}{2023}).
\newblock


\bibitem[Papineni et~al\mbox{.}(2002)]%
        {papineni2002bleu}
\bibfield{author}{\bibinfo{person}{Kishore Papineni}, \bibinfo{person}{Salim Roukos}, \bibinfo{person}{Todd Ward}, {and} \bibinfo{person}{Wei-Jing Zhu}.} \bibinfo{year}{2002}\natexlab{}.
\newblock \showarticletitle{Bleu: a method for automatic evaluation of machine translation}. In \bibinfo{booktitle}{\emph{Proceedings of the 40th annual meeting of the Association for Computational Linguistics}}. \bibinfo{pages}{311--318}.
\newblock


\bibitem[Pei et~al\mbox{.}(2019)]%
        {pei2019personalized}
\bibfield{author}{\bibinfo{person}{Changhua Pei}, \bibinfo{person}{Yi Zhang}, \bibinfo{person}{Yongfeng Zhang}, \bibinfo{person}{Fei Sun}, \bibinfo{person}{Xiao Lin}, \bibinfo{person}{Hanxiao Sun}, \bibinfo{person}{Jian Wu}, \bibinfo{person}{Peng Jiang}, \bibinfo{person}{Junfeng Ge}, \bibinfo{person}{Wenwu Ou}, {et~al\mbox{.}}} \bibinfo{year}{2019}\natexlab{}.
\newblock \showarticletitle{Personalized re-ranking for recommendation}. In \bibinfo{booktitle}{\emph{Proceedings of the 13th ACM conference on recommender systems}}. \bibinfo{pages}{3--11}.
\newblock


\bibitem[Peng et~al\mbox{.}(2023)]%
        {peng2023large}
\bibfield{author}{\bibinfo{person}{Wenjun Peng}, \bibinfo{person}{Guiyang Li}, \bibinfo{person}{Yue Jiang}, \bibinfo{person}{Zilong Wang}, \bibinfo{person}{Dan Ou}, \bibinfo{person}{Xiaoyi Zeng}, \bibinfo{person}{Enhong Chen}, {et~al\mbox{.}}} \bibinfo{year}{2023}\natexlab{}.
\newblock \showarticletitle{Large Language Model based Long-tail Query Rewriting in Taobao Search}.
\newblock \bibinfo{journal}{\emph{arXiv preprint arXiv:2311.03758}} (\bibinfo{year}{2023}).
\newblock


\bibitem[Pi et~al\mbox{.}(2020)]%
        {pi2020search}
\bibfield{author}{\bibinfo{person}{Qi Pi}, \bibinfo{person}{Guorui Zhou}, \bibinfo{person}{Yujing Zhang}, \bibinfo{person}{Zhe Wang}, \bibinfo{person}{Lejian Ren}, \bibinfo{person}{Ying Fan}, \bibinfo{person}{Xiaoqiang Zhu}, {and} \bibinfo{person}{Kun Gai}.} \bibinfo{year}{2020}\natexlab{}.
\newblock \showarticletitle{Search-based user interest modeling with lifelong sequential behavior data for click-through rate prediction}. In \bibinfo{booktitle}{\emph{Proceedings of the 29th ACM International Conference on Information \& Knowledge Management}}. \bibinfo{pages}{2685--2692}.
\newblock


\bibitem[Qin et~al\mbox{.}(2022)]%
        {rankflow}
\bibfield{author}{\bibinfo{person}{Jiarui Qin}, \bibinfo{person}{Jiachen Zhu}, \bibinfo{person}{Bo Chen}, \bibinfo{person}{Zhirong Liu}, \bibinfo{person}{Weiwen Liu}, \bibinfo{person}{Ruiming Tang}, \bibinfo{person}{Rui Zhang}, \bibinfo{person}{Yong Yu}, {and} \bibinfo{person}{Weinan Zhang}.} \bibinfo{year}{2022}\natexlab{}.
\newblock \showarticletitle{Rankflow: Joint optimization of multi-stage cascade ranking systems as flows}. In \bibinfo{booktitle}{\emph{Proceedings of the 45th International ACM SIGIR Conference on Research and Development in Information Retrieval}}. \bibinfo{pages}{814--824}.
\newblock


\bibitem[Qin et~al\mbox{.}(2020)]%
        {qin2020diversifying}
\bibfield{author}{\bibinfo{person}{Xubo Qin}, \bibinfo{person}{Zhicheng Dou}, {and} \bibinfo{person}{Ji-Rong Wen}.} \bibinfo{year}{2020}\natexlab{}.
\newblock \showarticletitle{Diversifying search results using self-attention network}. In \bibinfo{booktitle}{\emph{Proceedings of the 29th ACM International Conference on Information \& Knowledge Management}}. \bibinfo{pages}{1265--1274}.
\newblock


\bibitem[Qu et~al\mbox{.}(2016)]%
        {qu2016product}
\bibfield{author}{\bibinfo{person}{Yanru Qu}, \bibinfo{person}{Han Cai}, \bibinfo{person}{Kan Ren}, \bibinfo{person}{Weinan Zhang}, \bibinfo{person}{Yong Yu}, \bibinfo{person}{Ying Wen}, {and} \bibinfo{person}{Jun Wang}.} \bibinfo{year}{2016}\natexlab{}.
\newblock \showarticletitle{Product-based neural networks for user response prediction}. In \bibinfo{booktitle}{\emph{2016 IEEE 16th international conference on data mining (ICDM)}}. IEEE, \bibinfo{pages}{1149--1154}.
\newblock


\bibitem[Radford et~al\mbox{.}(2019)]%
        {radford2019language}
\bibfield{author}{\bibinfo{person}{Alec Radford}, \bibinfo{person}{Jeffrey Wu}, \bibinfo{person}{Rewon Child}, \bibinfo{person}{David Luan}, \bibinfo{person}{Dario Amodei}, \bibinfo{person}{Ilya Sutskever}, {et~al\mbox{.}}} \bibinfo{year}{2019}\natexlab{}.
\newblock \showarticletitle{Language models are unsupervised multitask learners}.
\newblock \bibinfo{journal}{\emph{OpenAI blog}} \bibinfo{volume}{1}, \bibinfo{number}{8} (\bibinfo{year}{2019}), \bibinfo{pages}{9}.
\newblock


\bibitem[Raffel et~al\mbox{.}(2020)]%
        {raffel2020t5}
\bibfield{author}{\bibinfo{person}{Colin Raffel}, \bibinfo{person}{Noam Shazeer}, \bibinfo{person}{Adam Roberts}, \bibinfo{person}{Katherine Lee}, \bibinfo{person}{Sharan Narang}, \bibinfo{person}{Michael Matena}, \bibinfo{person}{Yanqi Zhou}, \bibinfo{person}{Wei Li}, {and} \bibinfo{person}{Peter~J. Liu}.} \bibinfo{year}{2020}\natexlab{}.
\newblock \showarticletitle{Exploring the Limits of Transfer Learning with a Unified Text-to-Text Transformer}.
\newblock \bibinfo{journal}{\emph{J. Mach. Learn. Res.}}  \bibinfo{volume}{21} (\bibinfo{date}{jan} \bibinfo{year}{2020}), \bibinfo{numpages}{67}~pages.
\newblock


\bibitem[Rajput et~al\mbox{.}(2023)]%
        {rajput2023recommender}
\bibfield{author}{\bibinfo{person}{Shashank Rajput}, \bibinfo{person}{Nikhil Mehta}, \bibinfo{person}{Anima Singh}, \bibinfo{person}{Raghunandan~H Keshavan}, \bibinfo{person}{Trung Vu}, \bibinfo{person}{Lukasz Heldt}, \bibinfo{person}{Lichan Hong}, \bibinfo{person}{Yi Tay}, \bibinfo{person}{Vinh~Q Tran}, \bibinfo{person}{Jonah Samost}, {et~al\mbox{.}}} \bibinfo{year}{2023}\natexlab{}.
\newblock \showarticletitle{Recommender Systems with Generative Retrieval}.
\newblock \bibinfo{journal}{\emph{arXiv preprint arXiv:2305.05065}} (\bibinfo{year}{2023}).
\newblock


\bibitem[Ravaut et~al\mbox{.}(2024)]%
        {ravaut2024parameter}
\bibfield{author}{\bibinfo{person}{Mathieu Ravaut}, \bibinfo{person}{Hao Zhang}, \bibinfo{person}{Lu Xu}, \bibinfo{person}{Aixin Sun}, {and} \bibinfo{person}{Yong Liu}.} \bibinfo{year}{2024}\natexlab{}.
\newblock \showarticletitle{Parameter-Efficient Conversational Recommender System as a Language Processing Task}. In \bibinfo{booktitle}{\emph{Proceedings of the 18th Conference of the European Chapter of the Association for Computational Linguistics (Volume 1: Long Papers)}}. \bibinfo{publisher}{Association for Computational Linguistics}, \bibinfo{address}{St. Julian{'}s, Malta}, \bibinfo{pages}{152--165}.
\newblock


\bibitem[Ren et~al\mbox{.}(2023)]%
        {ren2023kecr}
\bibfield{author}{\bibinfo{person}{Xuhui Ren}, \bibinfo{person}{Tong Chen}, \bibinfo{person}{Quoc Viet~Hung Nguyen}, \bibinfo{person}{Li zhen Cui}, \bibinfo{person}{Zi-Liang Huang}, {and} \bibinfo{person}{Hongzhi Yin}.} \bibinfo{year}{2023}\natexlab{}.
\newblock \showarticletitle{Explicit Knowledge Graph Reasoning for Conversational Recommendation}.
\newblock \bibinfo{journal}{\emph{ArXiv}}  \bibinfo{volume}{abs/2305.00783} (\bibinfo{year}{2023}).
\newblock


\bibitem[Runfeng et~al\mbox{.}(2023)]%
        {runfeng2023lkpnr}
\bibfield{author}{\bibinfo{person}{Xie Runfeng}, \bibinfo{person}{Cui Xiangyang}, \bibinfo{person}{Yan Zhou}, \bibinfo{person}{Wang Xin}, \bibinfo{person}{Xuan Zhanwei}, \bibinfo{person}{Zhang Kai}, {et~al\mbox{.}}} \bibinfo{year}{2023}\natexlab{}.
\newblock \showarticletitle{Lkpnr: Llm and kg for personalized news recommendation framework}.
\newblock \bibinfo{journal}{\emph{arXiv preprint arXiv:2308.12028}} (\bibinfo{year}{2023}).
\newblock


\bibitem[Sepliarskaia et~al\mbox{.}(2018)]%
        {sepliarskaia2018preference}
\bibfield{author}{\bibinfo{person}{Anna Sepliarskaia}, \bibinfo{person}{Julia Kiseleva}, \bibinfo{person}{Filip Radlinski}, {and} \bibinfo{person}{Maarten de Rijke}.} \bibinfo{year}{2018}\natexlab{}.
\newblock \showarticletitle{Preference elicitation as an optimization problem}. In \bibinfo{booktitle}{\emph{Proceedings of the 12th ACM Conference on Recommender Systems}}. \bibinfo{pages}{172--180}.
\newblock


\bibitem[Shapira et~al\mbox{.}(2013)]%
        {shapira2013facebook}
\bibfield{author}{\bibinfo{person}{Bracha Shapira}, \bibinfo{person}{Lior Rokach}, {and} \bibinfo{person}{Shirley Freilikhman}.} \bibinfo{year}{2013}\natexlab{}.
\newblock \showarticletitle{Facebook single and cross domain data for recommendation systems}.
\newblock \bibinfo{journal}{\emph{User Modeling and User-Adapted Interaction}}  \bibinfo{volume}{23} (\bibinfo{year}{2013}), \bibinfo{pages}{211--247}.
\newblock


\bibitem[Sheng et~al\mbox{.}(2021)]%
        {sheng2021one}
\bibfield{author}{\bibinfo{person}{Xiang-Rong Sheng}, \bibinfo{person}{Liqin Zhao}, \bibinfo{person}{Guorui Zhou}, \bibinfo{person}{Xinyao Ding}, \bibinfo{person}{Binding Dai}, \bibinfo{person}{Qiang Luo}, \bibinfo{person}{Siran Yang}, \bibinfo{person}{Jingshan Lv}, \bibinfo{person}{Chi Zhang}, \bibinfo{person}{Hongbo Deng}, {et~al\mbox{.}}} \bibinfo{year}{2021}\natexlab{}.
\newblock \showarticletitle{One model to serve all: Star topology adaptive recommender for multi-domain ctr prediction}. In \bibinfo{booktitle}{\emph{Proceedings of the 30th ACM International Conference on Information \& Knowledge Management}}. \bibinfo{pages}{4104--4113}.
\newblock


\bibitem[Shi et~al\mbox{.}(2023)]%
        {shi2023llama}
\bibfield{author}{\bibinfo{person}{Kaize Shi}, \bibinfo{person}{Xueyao Sun}, \bibinfo{person}{Dingxian Wang}, \bibinfo{person}{Yinlin Fu}, \bibinfo{person}{Guandong Xu}, {and} \bibinfo{person}{Qing Li}.} \bibinfo{year}{2023}\natexlab{}.
\newblock \showarticletitle{LLaMA-E: Empowering E-commerce Authoring with Multi-Aspect Instruction Following}.
\newblock \bibinfo{journal}{\emph{arXiv preprint arXiv:2308.04913}} (\bibinfo{year}{2023}).
\newblock


\bibitem[Shu et~al\mbox{.}(2023)]%
        {shu2023rah}
\bibfield{author}{\bibinfo{person}{Yubo Shu}, \bibinfo{person}{Haonan Zhang}, \bibinfo{person}{Hansu Gu}, \bibinfo{person}{Peng Zhang}, \bibinfo{person}{Tun Lu}, \bibinfo{person}{Dongsheng Li}, {and} \bibinfo{person}{Ning Gu}.} \bibinfo{year}{2023}\natexlab{}.
\newblock \showarticletitle{RAH! RecSys-Assistant-Human: A Human-Centered Recommendation Framework with LLM Agents}.
\newblock \bibinfo{journal}{\emph{ArXiv}}  \bibinfo{volume}{abs/2308.09904} (\bibinfo{year}{2023}).
\newblock


\bibitem[Smyth and McGinty(2003)]%
        {smyth2003analysis}
\bibfield{author}{\bibinfo{person}{Barry Smyth} {and} \bibinfo{person}{Lorraine McGinty}.} \bibinfo{year}{2003}\natexlab{}.
\newblock \showarticletitle{An analysis of feedback strategies in conversational recommenders}. In \bibinfo{booktitle}{\emph{the Fourteenth Irish Artificial Intelligence and Cognitive Science Conference (AICS 2003)}}. Citeseer.
\newblock


\bibitem[Smyth et~al\mbox{.}(2004)]%
        {smyth2004compound}
\bibfield{author}{\bibinfo{person}{Barry Smyth}, \bibinfo{person}{Lorraine McGinty}, \bibinfo{person}{James Reilly}, {and} \bibinfo{person}{Kevin McCarthy}.} \bibinfo{year}{2004}\natexlab{}.
\newblock \showarticletitle{Compound critiques for conversational recommender systems}. In \bibinfo{booktitle}{\emph{IEEE/WIC/ACM International Conference on Web Intelligence (WI'04)}}. IEEE, \bibinfo{pages}{145--151}.
\newblock


\bibitem[Song et~al\mbox{.}(2020)]%
        {song2020co}
\bibfield{author}{\bibinfo{person}{Junshuai Song}, \bibinfo{person}{Zhao Li}, \bibinfo{person}{Chang Zhou}, \bibinfo{person}{Jinze Bai}, \bibinfo{person}{Zhenpeng Li}, \bibinfo{person}{Jian Li}, {and} \bibinfo{person}{Jun Gao}.} \bibinfo{year}{2020}\natexlab{}.
\newblock \showarticletitle{Co-displayed items aware list recommendation}.
\newblock \bibinfo{journal}{\emph{IEEE Access}}  \bibinfo{volume}{8} (\bibinfo{year}{2020}), \bibinfo{pages}{64591--64602}.
\newblock


\bibitem[Song et~al\mbox{.}(2019)]%
        {song2019autoint}
\bibfield{author}{\bibinfo{person}{Weiping Song}, \bibinfo{person}{Chence Shi}, \bibinfo{person}{Zhiping Xiao}, \bibinfo{person}{Zhijian Duan}, \bibinfo{person}{Yewen Xu}, \bibinfo{person}{Ming Zhang}, {and} \bibinfo{person}{Jian Tang}.} \bibinfo{year}{2019}\natexlab{}.
\newblock \showarticletitle{Autoint: Automatic feature interaction learning via self-attentive neural networks}. In \bibinfo{booktitle}{\emph{Proceedings of the 28th ACM international conference on information and knowledge management}}. \bibinfo{pages}{1161--1170}.
\newblock


\bibitem[Spurlock et~al\mbox{.}(2024)]%
        {spurlock2024chatgpt}
\bibfield{author}{\bibinfo{person}{Kyle~Dylan Spurlock}, \bibinfo{person}{Cagla Acun}, \bibinfo{person}{Esin Saka}, {and} \bibinfo{person}{Olfa Nasraoui}.} \bibinfo{year}{2024}\natexlab{}.
\newblock \showarticletitle{ChatGPT for Conversational Recommendation: Refining Recommendations by Reprompting with Feedback}.
\newblock \bibinfo{journal}{\emph{ArXiv}}  \bibinfo{volume}{abs/2401.03605} (\bibinfo{year}{2024}).
\newblock


\bibitem[Sun et~al\mbox{.}(2019)]%
        {sun2019bert4rec}
\bibfield{author}{\bibinfo{person}{Fei Sun}, \bibinfo{person}{Jun Liu}, \bibinfo{person}{Jian Wu}, \bibinfo{person}{Changhua Pei}, \bibinfo{person}{Xiao Lin}, \bibinfo{person}{Wenwu Ou}, {and} \bibinfo{person}{Peng Jiang}.} \bibinfo{year}{2019}\natexlab{}.
\newblock \showarticletitle{BERT4Rec: Sequential recommendation with bidirectional encoder representations from transformer}. In \bibinfo{booktitle}{\emph{Proceedings of the 28th ACM international conference on information and knowledge management}}. \bibinfo{pages}{1441--1450}.
\newblock


\bibitem[Sun and Zhang(2018)]%
        {sun2018conversational}
\bibfield{author}{\bibinfo{person}{Yueming Sun} {and} \bibinfo{person}{Yi Zhang}.} \bibinfo{year}{2018}\natexlab{}.
\newblock \showarticletitle{Conversational recommender system}. In \bibinfo{booktitle}{\emph{The 41st international acm sigir conference on research \& development in information retrieval}}. \bibinfo{pages}{235--244}.
\newblock


\bibitem[Sun et~al\mbox{.}(2023)]%
        {sun2023large}
\bibfield{author}{\bibinfo{person}{Zhu Sun}, \bibinfo{person}{Hongyang Liu}, \bibinfo{person}{Xinghua Qu}, \bibinfo{person}{Kaidong Feng}, \bibinfo{person}{Yan Wang}, {and} \bibinfo{person}{Yew-Soon Ong}.} \bibinfo{year}{2023}\natexlab{}.
\newblock \showarticletitle{Large Language Models for Intent-Driven Session Recommendations}.
\newblock \bibinfo{journal}{\emph{arXiv preprint arXiv:2312.07552}} (\bibinfo{year}{2023}).
\newblock


\bibitem[Tang et~al\mbox{.}(2020)]%
        {PLE}
\bibfield{author}{\bibinfo{person}{Hongyan Tang}, \bibinfo{person}{Junning Liu}, \bibinfo{person}{Ming Zhao}, {and} \bibinfo{person}{Xudong Gong}.} \bibinfo{year}{2020}\natexlab{}.
\newblock \showarticletitle{Progressive layered extraction (ple): A novel multi-task learning (mtl) model for personalized recommendations}. In \bibinfo{booktitle}{\emph{RecSys}}. \bibinfo{pages}{269--278}.
\newblock


\bibitem[Tang et~al\mbox{.}(2023)]%
        {tang2023one}
\bibfield{author}{\bibinfo{person}{Zuoli Tang}, \bibinfo{person}{Zhaoxin Huan}, \bibinfo{person}{Zihao Li}, \bibinfo{person}{Xiaolu Zhang}, \bibinfo{person}{Jun Hu}, \bibinfo{person}{Chilin Fu}, \bibinfo{person}{Jun Zhou}, {and} \bibinfo{person}{Chenliang Li}.} \bibinfo{year}{2023}\natexlab{}.
\newblock \showarticletitle{One Model for All: Large Language Models are Domain-Agnostic Recommendation Systems}.
\newblock \bibinfo{journal}{\emph{arXiv preprint arXiv:2310.14304}} (\bibinfo{year}{2023}).
\newblock


\bibitem[Tian et~al\mbox{.}(2023)]%
        {tian2023ufin}
\bibfield{author}{\bibinfo{person}{Zhen Tian}, \bibinfo{person}{Changwang Zhang}, \bibinfo{person}{Wayne~Xin Zhao}, \bibinfo{person}{Xin Zhao}, \bibinfo{person}{Ji-Rong Wen}, {and} \bibinfo{person}{Zhao Cao}.} \bibinfo{year}{2023}\natexlab{}.
\newblock \showarticletitle{UFIN: Universal Feature Interaction Network for Multi-Domain Click-Through Rate Prediction}.
\newblock \bibinfo{journal}{\emph{arXiv preprint arXiv:2311.15493}} (\bibinfo{year}{2023}).
\newblock


\bibitem[Torbati et~al\mbox{.}(2023)]%
        {torbati2023recommendations}
\bibfield{author}{\bibinfo{person}{Ghazaleh~Haratinezhad Torbati}, \bibinfo{person}{Anna Tigunova}, \bibinfo{person}{Andrew Yates}, {and} \bibinfo{person}{Gerhard Weikum}.} \bibinfo{year}{2023}\natexlab{}.
\newblock \showarticletitle{Recommendations by Concise User Profiles from Review Text}.
\newblock \bibinfo{journal}{\emph{arXiv preprint arXiv:2311.01314}} (\bibinfo{year}{2023}).
\newblock


\bibitem[Touvron et~al\mbox{.}(2023)]%
        {touvron2023llama}
\bibfield{author}{\bibinfo{person}{Hugo Touvron}, \bibinfo{person}{Thibaut Lavril}, \bibinfo{person}{Gautier Izacard}, \bibinfo{person}{Xavier Martinet}, \bibinfo{person}{Marie-Anne Lachaux}, \bibinfo{person}{Timoth{\'e}e Lacroix}, \bibinfo{person}{Baptiste Rozi{\`e}re}, \bibinfo{person}{Naman Goyal}, \bibinfo{person}{Eric Hambro}, \bibinfo{person}{Faisal Azhar}, {et~al\mbox{.}}} \bibinfo{year}{2023}\natexlab{}.
\newblock \showarticletitle{Llama: Open and efficient foundation language models}.
\newblock \bibinfo{journal}{\emph{arXiv preprint arXiv:2302.13971}} (\bibinfo{year}{2023}).
\newblock


\bibitem[Vendrov et~al\mbox{.}(2020)]%
        {vendrov2020gradient}
\bibfield{author}{\bibinfo{person}{Ivan Vendrov}, \bibinfo{person}{Tyler Lu}, \bibinfo{person}{Qingqing Huang}, {and} \bibinfo{person}{Craig Boutilier}.} \bibinfo{year}{2020}\natexlab{}.
\newblock \showarticletitle{Gradient-based optimization for bayesian preference elicitation}. In \bibinfo{booktitle}{\emph{Proceedings of the AAAI Conference on Artificial Intelligence}}, Vol.~\bibinfo{volume}{34}. \bibinfo{pages}{10292--10301}.
\newblock


\bibitem[Viappiani et~al\mbox{.}(2007)]%
        {viappiani2007conversational}
\bibfield{author}{\bibinfo{person}{Paolo Viappiani}, \bibinfo{person}{Pearl Pu}, {and} \bibinfo{person}{Boi Faltings}.} \bibinfo{year}{2007}\natexlab{}.
\newblock \showarticletitle{Conversational recommenders with adaptive suggestions}. In \bibinfo{booktitle}{\emph{Proceedings of the 2007 ACM conference on Recommender systems}}. \bibinfo{pages}{89--96}.
\newblock


\bibitem[Wang et~al\mbox{.}(2019)]%
        {wang2019sequential}
\bibfield{author}{\bibinfo{person}{Fan Wang}, \bibinfo{person}{Xiaomin Fang}, \bibinfo{person}{Lihang Liu}, \bibinfo{person}{Yaxue Chen}, \bibinfo{person}{Jiucheng Tao}, \bibinfo{person}{Zhiming Peng}, \bibinfo{person}{Cihang Jin}, {and} \bibinfo{person}{Hao Tian}.} \bibinfo{year}{2019}\natexlab{}.
\newblock \showarticletitle{Sequential evaluation and generation framework for combinatorial recommender system}.
\newblock \bibinfo{journal}{\emph{arXiv preprint arXiv:1902.00245}} (\bibinfo{year}{2019}).
\newblock


\bibitem[Wang et~al\mbox{.}(2018b)]%
        {wang2018ripplenet}
\bibfield{author}{\bibinfo{person}{Hongwei Wang}, \bibinfo{person}{Fuzheng Zhang}, \bibinfo{person}{Jialin Wang}, \bibinfo{person}{Miao Zhao}, \bibinfo{person}{Wenjie Li}, \bibinfo{person}{Xing Xie}, {and} \bibinfo{person}{Minyi Guo}.} \bibinfo{year}{2018}\natexlab{b}.
\newblock \showarticletitle{Ripplenet: Propagating user preferences on the knowledge graph for recommender systems}. In \bibinfo{booktitle}{\emph{Proceedings of the 27th ACM international conference on information and knowledge management}}. \bibinfo{pages}{417--426}.
\newblock


\bibitem[Wang et~al\mbox{.}(2023d)]%
        {wang2023survey}
\bibfield{author}{\bibinfo{person}{Lei Wang}, \bibinfo{person}{Chen Ma}, \bibinfo{person}{Xueyang Feng}, \bibinfo{person}{Zeyu Zhang}, \bibinfo{person}{Hao Yang}, \bibinfo{person}{Jingsen Zhang}, \bibinfo{person}{Zhiyuan Chen}, \bibinfo{person}{Jiakai Tang}, \bibinfo{person}{Xu Chen}, \bibinfo{person}{Yankai Lin}, {et~al\mbox{.}}} \bibinfo{year}{2023}\natexlab{d}.
\newblock \showarticletitle{A survey on large language model based autonomous agents}.
\newblock \bibinfo{journal}{\emph{arXiv preprint arXiv:2308.11432}} (\bibinfo{year}{2023}).
\newblock


\bibitem[Wang et~al\mbox{.}(2023f)]%
        {wang2023large}
\bibfield{author}{\bibinfo{person}{Lei Wang}, \bibinfo{person}{Jingsen Zhang}, \bibinfo{person}{Hao Yang}, \bibinfo{person}{Zhiyuan Chen}, \bibinfo{person}{Jiakai Tang}, \bibinfo{person}{Zeyu Zhang}, \bibinfo{person}{Xu Chen}, \bibinfo{person}{Yankai Lin}, \bibinfo{person}{Ruihua Song}, \bibinfo{person}{Wayne~Xin Zhao}, \bibinfo{person}{Jun Xu}, \bibinfo{person}{Zhicheng Dou}, \bibinfo{person}{Jun Wang}, {and} \bibinfo{person}{Ji-Rong Wen}.} \bibinfo{year}{2023}\natexlab{f}.
\newblock \showarticletitle{When Large Language Model based Agent Meets User Behavior Analysis: A Novel User Simulation Paradigm}.
\newblock  (\bibinfo{year}{2023}).
\newblock
\showeprint[arxiv]{2306.02552}~[cs.IR]


\bibitem[Wang(2022)]%
        {wang2022recommendation}
\bibfield{author}{\bibinfo{person}{Pengda Wang}.} \bibinfo{year}{2022}\natexlab{}.
\newblock \showarticletitle{Recommendation algorithm in TikTok: Strengths, dilemmas, and possible directions}.
\newblock \bibinfo{journal}{\emph{Int'l J. Soc. Sci. Stud.}}  \bibinfo{volume}{10} (\bibinfo{year}{2022}), \bibinfo{pages}{60}.
\newblock


\bibitem[Wang et~al\mbox{.}(2018a)]%
        {wang2018online}
\bibfield{author}{\bibinfo{person}{Qing Wang}, \bibinfo{person}{Chunqiu Zeng}, \bibinfo{person}{Wubai Zhou}, \bibinfo{person}{Tao Li}, \bibinfo{person}{S~Sitharama Iyengar}, \bibinfo{person}{Larisa Shwartz}, {and} \bibinfo{person}{Genady~Ya Grabarnik}.} \bibinfo{year}{2018}\natexlab{a}.
\newblock \showarticletitle{Online interactive collaborative filtering using multi-armed bandit with dependent arms}.
\newblock \bibinfo{journal}{\emph{IEEE Transactions on Knowledge and Data Engineering}} \bibinfo{volume}{31}, \bibinfo{number}{8} (\bibinfo{year}{2018}), \bibinfo{pages}{1569--1580}.
\newblock


\bibitem[Wang et~al\mbox{.}(2017)]%
        {wang2017deep}
\bibfield{author}{\bibinfo{person}{Ruoxi Wang}, \bibinfo{person}{Bin Fu}, \bibinfo{person}{Gang Fu}, {and} \bibinfo{person}{Mingliang Wang}.} \bibinfo{year}{2017}\natexlab{}.
\newblock \showarticletitle{Deep \& cross network for ad click predictions}.
\newblock In \bibinfo{booktitle}{\emph{Proceedings of the ADKDD'17}}. \bibinfo{pages}{1--7}.
\newblock


\bibitem[Wang et~al\mbox{.}(2021)]%
        {wang2021dcn}
\bibfield{author}{\bibinfo{person}{Ruoxi Wang}, \bibinfo{person}{Rakesh Shivanna}, \bibinfo{person}{Derek Cheng}, \bibinfo{person}{Sagar Jain}, \bibinfo{person}{Dong Lin}, \bibinfo{person}{Lichan Hong}, {and} \bibinfo{person}{Ed Chi}.} \bibinfo{year}{2021}\natexlab{}.
\newblock \showarticletitle{Dcn v2: Improved deepfm \& cross network and practical lessons for web-scale learning to rank systems}. In \bibinfo{booktitle}{\emph{Proceedings of the web conference 2021}}. \bibinfo{pages}{1785--1797}.
\newblock


\bibitem[Wang et~al\mbox{.}(2023b)]%
        {wang2023generativert}
\bibfield{author}{\bibinfo{person}{Wenjie Wang}, \bibinfo{person}{Xinyu Lin}, \bibinfo{person}{Fuli Feng}, \bibinfo{person}{Xiangnan He}, {and} \bibinfo{person}{Tat seng Chua}.} \bibinfo{year}{2023}\natexlab{b}.
\newblock \showarticletitle{Generative Recommendation: Towards Next-generation Recommender Paradigm}.
\newblock \bibinfo{journal}{\emph{ArXiv}}  \bibinfo{volume}{abs/2304.03516} (\bibinfo{year}{2023}).
\newblock


\bibitem[Wang et~al\mbox{.}(2023e)]%
        {wang2023rethinking}
\bibfield{author}{\bibinfo{person}{Xiaolei Wang}, \bibinfo{person}{Xinyu Tang}, \bibinfo{person}{Wayne~Xin Zhao}, \bibinfo{person}{Jingyuan Wang}, {and} \bibinfo{person}{Ji rong Wen}.} \bibinfo{year}{2023}\natexlab{e}.
\newblock \showarticletitle{Rethinking the Evaluation for Conversational Recommendation in the Era of Large Language Models}.
\newblock \bibinfo{journal}{\emph{ArXiv}}  \bibinfo{volume}{abs/2305.13112} (\bibinfo{year}{2023}).
\newblock


\bibitem[Wang et~al\mbox{.}(2023a)]%
        {wang2023recmind}
\bibfield{author}{\bibinfo{person}{Yancheng Wang}, \bibinfo{person}{Ziyan Jiang}, \bibinfo{person}{Zheng Chen}, \bibinfo{person}{Fan Yang}, \bibinfo{person}{Yingxue Zhou}, \bibinfo{person}{Eunah Cho}, \bibinfo{person}{Xing Fan}, \bibinfo{person}{Xiaojiang Huang}, \bibinfo{person}{Yanbin Lu}, {and} \bibinfo{person}{Yingzhen Yang}.} \bibinfo{year}{2023}\natexlab{a}.
\newblock \showarticletitle{RecMind: Large Language Model Powered Agent For Recommendation}.
\newblock \bibinfo{journal}{\emph{ArXiv}}  \bibinfo{volume}{abs/2308.14296} (\bibinfo{year}{2023}).
\newblock


\bibitem[Wang et~al\mbox{.}(2023c)]%
        {wang2023drdt}
\bibfield{author}{\bibinfo{person}{Yu Wang}, \bibinfo{person}{Zhiwei Liu}, \bibinfo{person}{Jianguo Zhang}, \bibinfo{person}{Weiran Yao}, \bibinfo{person}{Shelby Heinecke}, {and} \bibinfo{person}{Philip~S Yu}.} \bibinfo{year}{2023}\natexlab{c}.
\newblock \showarticletitle{DRDT: Dynamic Reflection with Divergent Thinking for LLM-based Sequential Recommendation}.
\newblock \bibinfo{journal}{\emph{arXiv preprint arXiv:2312.11336}} (\bibinfo{year}{2023}).
\newblock


\bibitem[Wei et~al\mbox{.}(2022)]%
        {wei2022chain}
\bibfield{author}{\bibinfo{person}{Jason Wei}, \bibinfo{person}{Xuezhi Wang}, \bibinfo{person}{Dale Schuurmans}, \bibinfo{person}{Maarten Bosma}, \bibinfo{person}{Fei Xia}, \bibinfo{person}{Ed Chi}, \bibinfo{person}{Quoc~V Le}, \bibinfo{person}{Denny Zhou}, {et~al\mbox{.}}} \bibinfo{year}{2022}\natexlab{}.
\newblock \showarticletitle{Chain-of-thought prompting elicits reasoning in large language models}.
\newblock \bibinfo{journal}{\emph{Advances in Neural Information Processing Systems}}  \bibinfo{volume}{35} (\bibinfo{year}{2022}), \bibinfo{pages}{24824--24837}.
\newblock


\bibitem[Weng(2023)]%
        {weng2023prompt}
\bibfield{author}{\bibinfo{person}{Lilian Weng}.} \bibinfo{year}{2023}\natexlab{}.
\newblock \showarticletitle{LLM-powered Autonomous Agents}.
\newblock \bibinfo{journal}{\emph{lilianweng.github.io}} (\bibinfo{date}{Jun} \bibinfo{year}{2023}).
\newblock
\urldef\tempurl%
\url{https://lilianweng.github.io/posts/2023-06-23-agent/}
\showURL{%
\tempurl}


\bibitem[Wooldridge(2009)]%
        {wooldridge2009introduction}
\bibfield{author}{\bibinfo{person}{Michael Wooldridge}.} \bibinfo{year}{2009}\natexlab{}.
\newblock \bibinfo{booktitle}{\emph{An introduction to multiagent systems}}.
\newblock \bibinfo{publisher}{John wiley \& sons}.
\newblock


\bibitem[Wu et~al\mbox{.}(2019)]%
        {wu2019deep}
\bibfield{author}{\bibinfo{person}{Ga Wu}, \bibinfo{person}{Kai Luo}, \bibinfo{person}{Scott Sanner}, {and} \bibinfo{person}{Harold Soh}.} \bibinfo{year}{2019}\natexlab{}.
\newblock \showarticletitle{Deep language-based critiquing for recommender systems}. In \bibinfo{booktitle}{\emph{Proceedings of the 13th ACM Conference on Recommender Systems}}. \bibinfo{pages}{137--145}.
\newblock


\bibitem[Wu et~al\mbox{.}(2023a)]%
        {wu2023leveraging}
\bibfield{author}{\bibinfo{person}{Jiahao Wu}, \bibinfo{person}{Qijiong Liu}, \bibinfo{person}{Hengchang Hu}, \bibinfo{person}{Wenqi Fan}, \bibinfo{person}{Shengcai Liu}, \bibinfo{person}{Qing Li}, \bibinfo{person}{Xiao-Ming Wu}, {and} \bibinfo{person}{Ke Tang}.} \bibinfo{year}{2023}\natexlab{a}.
\newblock \showarticletitle{Leveraging Large Language Models (LLMs) to Empower Training-Free Dataset Condensation for Content-Based Recommendation}.
\newblock \bibinfo{journal}{\emph{arXiv preprint arXiv:2310.09874}} (\bibinfo{year}{2023}).
\newblock


\bibitem[Wu et~al\mbox{.}(2023b)]%
        {wu2023survey}
\bibfield{author}{\bibinfo{person}{Likang Wu}, \bibinfo{person}{Zhi Zheng}, \bibinfo{person}{Zhaopeng Qiu}, \bibinfo{person}{Hao Wang}, \bibinfo{person}{Hongchao Gu}, \bibinfo{person}{Tingjia Shen}, \bibinfo{person}{Chuan Qin}, \bibinfo{person}{Chen Zhu}, \bibinfo{person}{Hengshu Zhu}, \bibinfo{person}{Qi Liu}, {et~al\mbox{.}}} \bibinfo{year}{2023}\natexlab{b}.
\newblock \showarticletitle{A Survey on Large Language Models for Recommendation}.
\newblock \bibinfo{journal}{\emph{arXiv preprint arXiv:2305.19860}} (\bibinfo{year}{2023}).
\newblock


\bibitem[Xi et~al\mbox{.}(2021)]%
        {xi2021context}
\bibfield{author}{\bibinfo{person}{Yunjia Xi}, \bibinfo{person}{Weiwen Liu}, \bibinfo{person}{Xinyi Dai}, \bibinfo{person}{Ruiming Tang}, \bibinfo{person}{Weinan Zhang}, \bibinfo{person}{Qing Liu}, \bibinfo{person}{Xiuqiang He}, {and} \bibinfo{person}{Yong Yu}.} \bibinfo{year}{2021}\natexlab{}.
\newblock \showarticletitle{Context-aware reranking with utility maximization for recommendation}.
\newblock \bibinfo{journal}{\emph{arXiv preprint arXiv:2110.09059}} (\bibinfo{year}{2021}).
\newblock


\bibitem[Xi et~al\mbox{.}(2023b)]%
        {xi2023towards}
\bibfield{author}{\bibinfo{person}{Yunjia Xi}, \bibinfo{person}{Weiwen Liu}, \bibinfo{person}{Jianghao Lin}, \bibinfo{person}{Jieming Zhu}, \bibinfo{person}{Bo Chen}, \bibinfo{person}{Ruiming Tang}, \bibinfo{person}{Weinan Zhang}, \bibinfo{person}{Rui Zhang}, {and} \bibinfo{person}{Yong Yu}.} \bibinfo{year}{2023}\natexlab{b}.
\newblock \showarticletitle{Towards Open-World Recommendation with Knowledge Augmentation from Large Language Models}.
\newblock \bibinfo{journal}{\emph{arXiv preprint arXiv:2306.10933}} (\bibinfo{year}{2023}).
\newblock


\bibitem[Xi et~al\mbox{.}(2023a)]%
        {xi2023rise}
\bibfield{author}{\bibinfo{person}{Zhiheng Xi}, \bibinfo{person}{Wenxiang Chen}, \bibinfo{person}{Xin Guo}, \bibinfo{person}{Wei He}, \bibinfo{person}{Yiwen Ding}, \bibinfo{person}{Boyang Hong}, \bibinfo{person}{Ming Zhang}, \bibinfo{person}{Junzhe Wang}, \bibinfo{person}{Senjie Jin}, \bibinfo{person}{Enyu Zhou}, {et~al\mbox{.}}} \bibinfo{year}{2023}\natexlab{a}.
\newblock \showarticletitle{The rise and potential of large language model based agents: A survey}.
\newblock \bibinfo{journal}{\emph{arXiv preprint arXiv:2309.07864}} (\bibinfo{year}{2023}).
\newblock


\bibitem[Xian et~al\mbox{.}(2019)]%
        {xian2019reinforcement}
\bibfield{author}{\bibinfo{person}{Yikun Xian}, \bibinfo{person}{Zuohui Fu}, \bibinfo{person}{Shan Muthukrishnan}, \bibinfo{person}{Gerard De~Melo}, {and} \bibinfo{person}{Yongfeng Zhang}.} \bibinfo{year}{2019}\natexlab{}.
\newblock \showarticletitle{Reinforcement knowledge graph reasoning for explainable recommendation}. In \bibinfo{booktitle}{\emph{Proceedings of the 42nd international ACM SIGIR conference on research and development in information retrieval}}. \bibinfo{pages}{285--294}.
\newblock


\bibitem[Xie et~al\mbox{.}(2021)]%
        {xie2021hierarchical}
\bibfield{author}{\bibinfo{person}{Ruobing Xie}, \bibinfo{person}{Shaoliang Zhang}, \bibinfo{person}{Rui Wang}, \bibinfo{person}{Feng Xia}, {and} \bibinfo{person}{Leyu Lin}.} \bibinfo{year}{2021}\natexlab{}.
\newblock \showarticletitle{Hierarchical reinforcement learning for integrated recommendation}. In \bibinfo{booktitle}{\emph{Proceedings of the AAAI Conference on Artificial Intelligence}}, Vol.~\bibinfo{volume}{35}. \bibinfo{pages}{4521--4528}.
\newblock


\bibitem[Xu et~al\mbox{.}(2020)]%
        {xu2020user}
\bibfield{author}{\bibinfo{person}{Hu Xu}, \bibinfo{person}{Seungwhan Moon}, \bibinfo{person}{Honglei Liu}, \bibinfo{person}{Bing Liu}, \bibinfo{person}{Pararth Shah}, {and} \bibinfo{person}{Philip~S Yu}.} \bibinfo{year}{2020}\natexlab{}.
\newblock \showarticletitle{User memory reasoning for conversational recommendation}.
\newblock \bibinfo{journal}{\emph{arXiv preprint arXiv:2006.00184}} (\bibinfo{year}{2020}).
\newblock


\bibitem[Xu et~al\mbox{.}(2024)]%
        {xu2024prompting}
\bibfield{author}{\bibinfo{person}{Lanling Xu}, \bibinfo{person}{Junjie Zhang}, \bibinfo{person}{Bingqian Li}, \bibinfo{person}{Jinpeng Wang}, \bibinfo{person}{Mingchen Cai}, \bibinfo{person}{Wayne~Xin Zhao}, {and} \bibinfo{person}{Ji-Rong Wen}.} \bibinfo{year}{2024}\natexlab{}.
\newblock \showarticletitle{Prompting Large Language Models for Recommender Systems: A Comprehensive Framework and Empirical Analysis}.
\newblock \bibinfo{journal}{\emph{ArXiv}}  \bibinfo{volume}{abs/2401.04997} (\bibinfo{year}{2024}).
\newblock


\bibitem[Yan et~al\mbox{.}(2022)]%
        {yan2022apg}
\bibfield{author}{\bibinfo{person}{Bencheng Yan}, \bibinfo{person}{Pengjie Wang}, \bibinfo{person}{Kai Zhang}, \bibinfo{person}{Feng Li}, \bibinfo{person}{Hongbo Deng}, \bibinfo{person}{Jian Xu}, {and} \bibinfo{person}{Bo Zheng}.} \bibinfo{year}{2022}\natexlab{}.
\newblock \showarticletitle{Apg: Adaptive parameter generation network for click-through rate prediction}.
\newblock \bibinfo{journal}{\emph{Advances in Neural Information Processing Systems}}  \bibinfo{volume}{35} (\bibinfo{year}{2022}), \bibinfo{pages}{24740--24752}.
\newblock


\bibitem[Yan et~al\mbox{.}(2021)]%
        {yan2021diversification}
\bibfield{author}{\bibinfo{person}{Le Yan}, \bibinfo{person}{Zhen Qin}, \bibinfo{person}{Rama~Kumar Pasumarthi}, \bibinfo{person}{Xuanhui Wang}, {and} \bibinfo{person}{Michael Bendersky}.} \bibinfo{year}{2021}\natexlab{}.
\newblock \showarticletitle{Diversification-aware learning to rank using distributed representation}. In \bibinfo{booktitle}{\emph{Proceedings of the Web Conference 2021}}. \bibinfo{pages}{127--136}.
\newblock


\bibitem[Yang et~al\mbox{.}(2023)]%
        {yang2023collaborative}
\bibfield{author}{\bibinfo{person}{Shenghao Yang}, \bibinfo{person}{Chenyang Wang}, \bibinfo{person}{Yankai Liu}, \bibinfo{person}{Kangping Xu}, \bibinfo{person}{Weizhi Ma}, \bibinfo{person}{Yiqun Liu}, \bibinfo{person}{Min Zhang}, \bibinfo{person}{Haitao Zeng}, \bibinfo{person}{Junlan Feng}, {and} \bibinfo{person}{Chao Deng}.} \bibinfo{year}{2023}\natexlab{}.
\newblock \showarticletitle{Collaborative Word-based Pre-trained Item Representation for Transferable Recommendation}.
\newblock \bibinfo{journal}{\emph{arXiv preprint arXiv:2311.10501}} (\bibinfo{year}{2023}).
\newblock


\bibitem[Yao et~al\mbox{.}(2022)]%
        {yao2022webshop}
\bibfield{author}{\bibinfo{person}{Shunyu Yao}, \bibinfo{person}{Howard Chen}, \bibinfo{person}{John Yang}, {and} \bibinfo{person}{Karthik Narasimhan}.} \bibinfo{year}{2022}\natexlab{}.
\newblock \showarticletitle{Webshop: Towards scalable real-world web interaction with grounded language agents}.
\newblock \bibinfo{journal}{\emph{Advances in Neural Information Processing Systems}}  \bibinfo{volume}{35} (\bibinfo{year}{2022}), \bibinfo{pages}{20744--20757}.
\newblock


\bibitem[Yao et~al\mbox{.}(2023a)]%
        {yao2023tree}
\bibfield{author}{\bibinfo{person}{Shunyu Yao}, \bibinfo{person}{Dian Yu}, \bibinfo{person}{Jeffrey Zhao}, \bibinfo{person}{Izhak Shafran}, \bibinfo{person}{Thomas~L. Griffiths}, \bibinfo{person}{Yuan Cao}, {and} \bibinfo{person}{Karthik~R Narasimhan}.} \bibinfo{year}{2023}\natexlab{a}.
\newblock \showarticletitle{Tree of Thoughts: Deliberate Problem Solving with Large Language Models}. In \bibinfo{booktitle}{\emph{Thirty-seventh Conference on Neural Information Processing Systems}}.
\newblock


\bibitem[Yao et~al\mbox{.}(2023b)]%
        {yao2023react}
\bibfield{author}{\bibinfo{person}{Shunyu Yao}, \bibinfo{person}{Jeffrey Zhao}, \bibinfo{person}{Dian Yu}, \bibinfo{person}{Nan Du}, \bibinfo{person}{Izhak Shafran}, \bibinfo{person}{Karthik~R Narasimhan}, {and} \bibinfo{person}{Yuan Cao}.} \bibinfo{year}{2023}\natexlab{b}.
\newblock \showarticletitle{ReAct: Synergizing Reasoning and Acting in Language Models}. In \bibinfo{booktitle}{\emph{The Eleventh International Conference on Learning Representations}}.
\newblock


\bibitem[Ying et~al\mbox{.}(2018)]%
        {ying2018graph}
\bibfield{author}{\bibinfo{person}{Rex Ying}, \bibinfo{person}{Ruining He}, \bibinfo{person}{Kaifeng Chen}, \bibinfo{person}{Pong Eksombatchai}, \bibinfo{person}{William~L Hamilton}, {and} \bibinfo{person}{Jure Leskovec}.} \bibinfo{year}{2018}\natexlab{}.
\newblock \showarticletitle{Graph convolutional neural networks for web-scale recommender systems}. In \bibinfo{booktitle}{\emph{Proceedings of the 24th ACM SIGKDD international conference on knowledge discovery \& data mining}}. \bibinfo{pages}{974--983}.
\newblock


\bibitem[Yuan et~al\mbox{.}(2022)]%
        {yuan2022multi}
\bibfield{author}{\bibinfo{person}{Enming Yuan}, \bibinfo{person}{Wei Guo}, \bibinfo{person}{Zhicheng He}, \bibinfo{person}{Huifeng Guo}, \bibinfo{person}{Chengkai Liu}, {and} \bibinfo{person}{Ruiming Tang}.} \bibinfo{year}{2022}\natexlab{}.
\newblock \showarticletitle{Multi-behavior sequential transformer recommender}. In \bibinfo{booktitle}{\emph{Proceedings of the 45th international ACM SIGIR conference on research and development in information retrieval}}. \bibinfo{pages}{1642--1652}.
\newblock


\bibitem[Yue et~al\mbox{.}(2023)]%
        {yue2023llamarec}
\bibfield{author}{\bibinfo{person}{Zhenrui Yue}, \bibinfo{person}{Sara Rabhi}, \bibinfo{person}{Gabriel de Souza~Pereira Moreira}, \bibinfo{person}{Dong Wang}, {and} \bibinfo{person}{Even Oldridge}.} \bibinfo{year}{2023}\natexlab{}.
\newblock \showarticletitle{LlamaRec: Two-Stage Recommendation using Large Language Models for Ranking}.
\newblock \bibinfo{journal}{\emph{arXiv preprint arXiv:2311.02089}} (\bibinfo{year}{2023}).
\newblock


\bibitem[Zeng et~al\mbox{.}(2022)]%
        {zeng2022glm}
\bibfield{author}{\bibinfo{person}{Aohan Zeng}, \bibinfo{person}{Xiao Liu}, \bibinfo{person}{Zhengxiao Du}, \bibinfo{person}{Zihan Wang}, \bibinfo{person}{Hanyu Lai}, \bibinfo{person}{Ming Ding}, \bibinfo{person}{Zhuoyi Yang}, \bibinfo{person}{Yifan Xu}, \bibinfo{person}{Wendi Zheng}, \bibinfo{person}{Xiao Xia}, {et~al\mbox{.}}} \bibinfo{year}{2022}\natexlab{}.
\newblock \showarticletitle{Glm-130b: An open bilingual pre-trained model}.
\newblock \bibinfo{journal}{\emph{arXiv preprint arXiv:2210.02414}} (\bibinfo{year}{2022}).
\newblock


\bibitem[Zhai et~al\mbox{.}(2024)]%
        {zhai2024actions}
\bibfield{author}{\bibinfo{person}{Jiaqi Zhai}, \bibinfo{person}{Lucy Liao}, \bibinfo{person}{Xing Liu}, \bibinfo{person}{Yueming Wang}, \bibinfo{person}{Rui Li}, \bibinfo{person}{Xuan Cao}, \bibinfo{person}{Leon Gao}, \bibinfo{person}{Zhaojie Gong}, \bibinfo{person}{Fangda Gu}, \bibinfo{person}{Michael He}, \bibinfo{person}{Yinghai Lu}, {and} \bibinfo{person}{Yu Shi}.} \bibinfo{year}{2024}\natexlab{}.
\newblock \bibinfo{title}{Actions Speak Louder than Words: Trillion-Parameter Sequential Transducers for Generative Recommendations}.
\newblock
\newblock
\showeprint[arxiv]{2402.17152}~[cs.LG]


\bibitem[Zhang et~al\mbox{.}(2023b)]%
        {zhang2023agent4rec}
\bibfield{author}{\bibinfo{person}{An Zhang}, \bibinfo{person}{Leheng Sheng}, \bibinfo{person}{Yuxin Chen}, \bibinfo{person}{Hao Li}, \bibinfo{person}{Yang Deng}, \bibinfo{person}{Xiang Wang}, {and} \bibinfo{person}{Tat-Seng Chua}.} \bibinfo{year}{2023}\natexlab{b}.
\newblock \showarticletitle{On Generative Agents in Recommendation}.
\newblock \bibinfo{journal}{\emph{arXiv preprint arXiv:2310.10108}} (\bibinfo{year}{2023}).
\newblock


\bibitem[Zhang et~al\mbox{.}(2022a)]%
        {zhang2022leaving}
\bibfield{author}{\bibinfo{person}{Qianqian Zhang}, \bibinfo{person}{Xinru Liao}, \bibinfo{person}{Quan Liu}, \bibinfo{person}{Jian Xu}, {and} \bibinfo{person}{Bo Zheng}.} \bibinfo{year}{2022}\natexlab{a}.
\newblock \showarticletitle{Leaving no one behind: A multi-scenario multi-task meta learning approach for advertiser modeling}. In \bibinfo{booktitle}{\emph{Proceedings of the Fifteenth ACM International Conference on Web Search and Data Mining}}. \bibinfo{pages}{1368--1376}.
\newblock


\bibitem[Zhang and Balog(2020)]%
        {zhang2020evaluating}
\bibfield{author}{\bibinfo{person}{Shuo Zhang} {and} \bibinfo{person}{Krisztian Balog}.} \bibinfo{year}{2020}\natexlab{}.
\newblock \showarticletitle{Evaluating conversational recommender systems via user simulation}. In \bibinfo{booktitle}{\emph{Proceedings of the 26th acm sigkdd international conference on knowledge discovery \& data mining}}. \bibinfo{pages}{1512--1520}.
\newblock


\bibitem[Zhang et~al\mbox{.}(2021)]%
        {zhang2021deep}
\bibfield{author}{\bibinfo{person}{Weinan Zhang}, \bibinfo{person}{Jiarui Qin}, \bibinfo{person}{Wei Guo}, \bibinfo{person}{Ruiming Tang}, {and} \bibinfo{person}{Xiuqiang He}.} \bibinfo{year}{2021}\natexlab{}.
\newblock \showarticletitle{Deep learning for click-through rate estimation}.
\newblock \bibinfo{journal}{\emph{arXiv preprint arXiv:2104.10584}} (\bibinfo{year}{2021}).
\newblock


\bibitem[Zhang et~al\mbox{.}(2020)]%
        {zhang2020conversational}
\bibfield{author}{\bibinfo{person}{Xiaoying Zhang}, \bibinfo{person}{Hong Xie}, \bibinfo{person}{Hang Li}, {and} \bibinfo{person}{John CS~Lui}.} \bibinfo{year}{2020}\natexlab{}.
\newblock \showarticletitle{Conversational contextual bandit: Algorithm and application}. In \bibinfo{booktitle}{\emph{Proceedings of the web conference 2020}}. \bibinfo{pages}{662--672}.
\newblock


\bibitem[Zhang et~al\mbox{.}(2022c)]%
        {zhang2022deepvt}
\bibfield{author}{\bibinfo{person}{Xuanyu Zhang}, \bibinfo{person}{Qing Yang}, {and} \bibinfo{person}{Dongliang Xu}.} \bibinfo{year}{2022}\natexlab{c}.
\newblock \showarticletitle{Deepvt: Deep view-temporal interaction network for news recommendation}. In \bibinfo{booktitle}{\emph{Proceedings of the 31st ACM International Conference on Information \& Knowledge Management}}. \bibinfo{pages}{2640--2650}.
\newblock


\bibitem[Zhang et~al\mbox{.}(2018)]%
        {zhang2018towards}
\bibfield{author}{\bibinfo{person}{Yongfeng Zhang}, \bibinfo{person}{Xu Chen}, \bibinfo{person}{Qingyao Ai}, \bibinfo{person}{Liu Yang}, {and} \bibinfo{person}{W~Bruce Croft}.} \bibinfo{year}{2018}\natexlab{}.
\newblock \showarticletitle{Towards conversational search and recommendation: System ask, user respond}. In \bibinfo{booktitle}{\emph{Proceedings of the 27th acm international conference on information and knowledge management}}. \bibinfo{pages}{177--186}.
\newblock


\bibitem[Zhang et~al\mbox{.}(2023a)]%
        {zhang2023collm}
\bibfield{author}{\bibinfo{person}{Yang Zhang}, \bibinfo{person}{Fuli Feng}, \bibinfo{person}{Jizhi Zhang}, \bibinfo{person}{Keqin Bao}, \bibinfo{person}{Qifan Wang}, {and} \bibinfo{person}{Xiangnan He}.} \bibinfo{year}{2023}\natexlab{a}.
\newblock \showarticletitle{Collm: Integrating collaborative embeddings into large language models for recommendation}.
\newblock \bibinfo{journal}{\emph{arXiv preprint arXiv:2310.19488}} (\bibinfo{year}{2023}).
\newblock


\bibitem[Zhang et~al\mbox{.}(2022b)]%
        {zhang2022scenario}
\bibfield{author}{\bibinfo{person}{Yuanliang Zhang}, \bibinfo{person}{Xiaofeng Wang}, \bibinfo{person}{Jinxin Hu}, \bibinfo{person}{Ke Gao}, \bibinfo{person}{Chenyi Lei}, {and} \bibinfo{person}{Fei Fang}.} \bibinfo{year}{2022}\natexlab{b}.
\newblock \showarticletitle{Scenario-Adaptive and Self-Supervised Model for Multi-Scenario Personalized Recommendation}. In \bibinfo{booktitle}{\emph{Proceedings of the 31st ACM International Conference on Information \& Knowledge Management}}. \bibinfo{pages}{3674--3683}.
\newblock


\bibitem[Zhao et~al\mbox{.}(2018)]%
        {zhao2018recommendations}
\bibfield{author}{\bibinfo{person}{Xiangyu Zhao}, \bibinfo{person}{Liang Zhang}, \bibinfo{person}{Zhuoye Ding}, \bibinfo{person}{Long Xia}, \bibinfo{person}{Jiliang Tang}, {and} \bibinfo{person}{Dawei Yin}.} \bibinfo{year}{2018}\natexlab{}.
\newblock \showarticletitle{Recommendations with negative feedback via pairwise deep reinforcement learning}. In \bibinfo{booktitle}{\emph{Proceedings of the 24th ACM SIGKDD international conference on knowledge discovery \& data mining}}. \bibinfo{pages}{1040--1048}.
\newblock


\bibitem[Zhao et~al\mbox{.}(2013)]%
        {zhao2013interactive}
\bibfield{author}{\bibinfo{person}{Xiaoxue Zhao}, \bibinfo{person}{Weinan Zhang}, {and} \bibinfo{person}{Jun Wang}.} \bibinfo{year}{2013}\natexlab{}.
\newblock \showarticletitle{Interactive collaborative filtering}. In \bibinfo{booktitle}{\emph{Proceedings of the 22nd ACM international conference on Information \& Knowledge Management}}. \bibinfo{pages}{1411--1420}.
\newblock


\bibitem[Zhao et~al\mbox{.}(2020)]%
        {zhao2020jointly}
\bibfield{author}{\bibinfo{person}{Xiangyu Zhao}, \bibinfo{person}{Xudong Zheng}, \bibinfo{person}{Xiwang Yang}, \bibinfo{person}{Xiaobing Liu}, {and} \bibinfo{person}{Jiliang Tang}.} \bibinfo{year}{2020}\natexlab{}.
\newblock \showarticletitle{Jointly learning to recommend and advertise}. In \bibinfo{booktitle}{\emph{Proceedings of the 26th ACM SIGKDD International Conference on Knowledge Discovery \& Data Mining}}. \bibinfo{pages}{3319--3327}.
\newblock


\bibitem[Zhou et~al\mbox{.}(2020a)]%
        {zhou2020can}
\bibfield{author}{\bibinfo{person}{Guorui Zhou}, \bibinfo{person}{Weijie Bian}, \bibinfo{person}{Kailun Wu}, \bibinfo{person}{Lejian Ren}, \bibinfo{person}{Qi Pi}, \bibinfo{person}{Yujing Zhang}, \bibinfo{person}{Can Xiao}, \bibinfo{person}{Xiang-Rong Sheng}, \bibinfo{person}{Na Mou}, \bibinfo{person}{Xinchen Luo}, {et~al\mbox{.}}} \bibinfo{year}{2020}\natexlab{a}.
\newblock \showarticletitle{CAN: revisiting feature co-action for click-through rate prediction}.
\newblock \bibinfo{journal}{\emph{arXiv preprint arXiv:2011.05625}} (\bibinfo{year}{2020}).
\newblock


\bibitem[Zhou et~al\mbox{.}(2018)]%
        {zhou2018deep}
\bibfield{author}{\bibinfo{person}{Guorui Zhou}, \bibinfo{person}{Xiaoqiang Zhu}, \bibinfo{person}{Chenru Song}, \bibinfo{person}{Ying Fan}, \bibinfo{person}{Han Zhu}, \bibinfo{person}{Xiao Ma}, \bibinfo{person}{Yanghui Yan}, \bibinfo{person}{Junqi Jin}, \bibinfo{person}{Han Li}, {and} \bibinfo{person}{Kun Gai}.} \bibinfo{year}{2018}\natexlab{}.
\newblock \showarticletitle{Deep interest network for click-through rate prediction}. In \bibinfo{booktitle}{\emph{Proceedings of the 24th ACM SIGKDD international conference on knowledge discovery \& data mining}}. \bibinfo{pages}{1059--1068}.
\newblock


\bibitem[Zhou et~al\mbox{.}(2023)]%
        {zhou2023hinet}
\bibfield{author}{\bibinfo{person}{Jie Zhou}, \bibinfo{person}{Xianshuai Cao}, \bibinfo{person}{Wenhao Li}, \bibinfo{person}{Lin Bo}, \bibinfo{person}{Kun Zhang}, \bibinfo{person}{Chuan Luo}, {and} \bibinfo{person}{Qian Yu}.} \bibinfo{year}{2023}\natexlab{}.
\newblock \showarticletitle{HiNet: Novel Multi-Scenario \& Multi-Task Learning with Hierarchical Information Extraction}.
\newblock \bibinfo{journal}{\emph{arXiv preprint arXiv:2303.06095}} (\bibinfo{year}{2023}).
\newblock


\bibitem[Zhou et~al\mbox{.}(2020b)]%
        {zhou2020towards}
\bibfield{author}{\bibinfo{person}{Kun Zhou}, \bibinfo{person}{Yuanhang Zhou}, \bibinfo{person}{Wayne~Xin Zhao}, \bibinfo{person}{Xiaoke Wang}, {and} \bibinfo{person}{Ji-Rong Wen}.} \bibinfo{year}{2020}\natexlab{b}.
\newblock \showarticletitle{Towards topic-guided conversational recommender system}.
\newblock \bibinfo{journal}{\emph{arXiv preprint arXiv:2010.04125}} (\bibinfo{year}{2020}).
\newblock


\bibitem[Zhu et~al\mbox{.}(2021)]%
        {zhu2021open}
\bibfield{author}{\bibinfo{person}{Jieming Zhu}, \bibinfo{person}{Jinyang Liu}, \bibinfo{person}{Shuai Yang}, \bibinfo{person}{Qi Zhang}, {and} \bibinfo{person}{Xiuqiang He}.} \bibinfo{year}{2021}\natexlab{}.
\newblock \showarticletitle{Open benchmarking for click-through rate prediction}. In \bibinfo{booktitle}{\emph{Proceedings of the 30th ACM international conference on information \& knowledge management}}. \bibinfo{pages}{2759--2769}.
\newblock


\bibitem[Zhu et~al\mbox{.}(2023)]%
        {zhu2023collaborative}
\bibfield{author}{\bibinfo{person}{Yaochen Zhu}, \bibinfo{person}{Liang Wu}, \bibinfo{person}{Qilnli Guo}, \bibinfo{person}{Liangjie Hong}, {and} \bibinfo{person}{Jundong Li}.} \bibinfo{year}{2023}\natexlab{}.
\newblock \showarticletitle{Collaborative Large Language Model for Recommender Systems}.
\newblock \bibinfo{journal}{\emph{ArXiv}}  \bibinfo{volume}{abs/2311.01343} (\bibinfo{year}{2023}).
\newblock


\bibitem[Zhuang et~al\mbox{.}(2018)]%
        {zhuang2018globally}
\bibfield{author}{\bibinfo{person}{Tao Zhuang}, \bibinfo{person}{Wenwu Ou}, {and} \bibinfo{person}{Zhirong Wang}.} \bibinfo{year}{2018}\natexlab{}.
\newblock \showarticletitle{Globally optimized mutual influence aware ranking in e-commerce search}.
\newblock \bibinfo{journal}{\emph{arXiv preprint arXiv:1805.08524}} (\bibinfo{year}{2018}).
\newblock


\bibitem[Zou et~al\mbox{.}(2020)]%
        {zou2020pseudo}
\bibfield{author}{\bibinfo{person}{Lixin Zou}, \bibinfo{person}{Long Xia}, \bibinfo{person}{Pan Du}, \bibinfo{person}{Zhuo Zhang}, \bibinfo{person}{Ting Bai}, \bibinfo{person}{Weidong Liu}, \bibinfo{person}{Jian-Yun Nie}, {and} \bibinfo{person}{Dawei Yin}.} \bibinfo{year}{2020}\natexlab{}.
\newblock \showarticletitle{Pseudo Dyna-Q: A reinforcement learning framework for interactive recommendation}. In \bibinfo{booktitle}{\emph{Proceedings of the 13th International Conference on Web Search and Data Mining}}. \bibinfo{pages}{816--824}.
\newblock


\end{thebibliography}

\end{document}